\documentclass[11pt]{cernrep}
\usepackage{graphicx}
\usepackage{here}
\usepackage{url}
\usepackage{epsfig}
\usepackage{axodraw}
\usepackage{subfigure}
\usepackage{amssymb}
\catcode `@ 11
\newcommand\labelsection[1]{{\def\@currentlabel{\@arabic\c@section}
                            \label{#1}}}
\catcode `@ 12
\newcommand     \mub            {\mu{\rm b}}

\newcommand     \ba             {\begin{eqnarray}}
\newcommand     \be             {\begin{equation}}

\newcommand     \ea             {\end{eqnarray}}
\newcommand     \ee             {\end{equation}}
\newcommand     \ep             {\epsilon}

\newcommand     \epem           {\ifmmode{e^+e^-}\else{$e^+e^-$}\fi}

\newcommand     \lambdamsb     {\ifmmode
          \Lambda_5^{\rm \scriptscriptstyle \overline{MS}} \else
         $\Lambda_5^{\rm \scriptscriptstyle \overline{MS}}$ \fi}
\newcommand     \Lambdamsb      \lambdamsb
\newcommand     \LambdaQCD     {\ifmmode
          \Lambda_{\rm \scriptscriptstyle QCD} \else
         $\Lambda_{\rm \scriptscriptstyle QCD}$ \fi}

\newcommand     \MSB            {\ifmmode {\overline{\rm MS}} \else
                                 $\overline{\rm MS}$  \fi}
\newcommand     \muF            {\mu_{\rm F}}
\newcommand     \muR            {\mu_{\rm R}}

\newcommand     \ptmin     {\ifmmode p_{\scriptscriptstyle T}^{\sss min} \else
                           $p_{\scriptscriptstyle T}^{\sss min}$ \fi}

\newcommand \abs[1]{\left| #1 \right|}

\newcommand\sss{\scriptscriptstyle\rm}

\newcommand\as{\alpha_{\sss S}}

\def \pt   {p_{\scriptscriptstyle T}}

\def \to   {\mbox{$\rightarrow$}}

\newcommand\gsim{\mathop{\mbox{\vbox{\hbox{$>$} \vskip -9pt \hbox{$\sim$}
             \vskip -3pt  }}}}

\newcommand\MRSTH{MRST$(g{\scriptstyle\uparrow})$}
\newcommand\MRSTL{MRST$(g{\scriptstyle\downarrow})$}
\newcommand\MRSTM{MRST$(\as{\scriptstyle\downarrow})$}
\newcommand\MRSTP{MRST$(\as{\scriptstyle\uparrow})$}
\newcommand{\Jpsitomm}{\mbox{$J/\psi \rightarrow \mu^+ \mu^-$}}
\newcommand{\bJpsi}{\mbox{$b \rightarrow J/\psi X$}}
\newcommand{\bJpsimm}{\mbox{$ b \rightarrow J/\psi(\mu\mu) X\ $}}
\newcommand{\bbJpsi}{\mbox{$ b \overline{b} \rightarrow J/\psi X\ $}}
\newcommand{\bbJpsimm}{\mbox{$ b \overline{b} \rightarrow J/\psi(\mu\mu) X\ $}}
\newcommand{\Jpsitoee}{\mbox{$J/\psi \rightarrow e^{+} e^{-}\ $}}
\newcommand{\bbJpsiee}{\mbox{$ b \overline{b} \rightarrow J/\psi(ee) X\ $}}
\newcommand{\alphas}{\alpha_{\mathrm{s}}}
\newcommand{\xF}{x_{\mathrm{F}}}
\newcommand{\pperp}{p_{\perp}}
\newcommand{\kperp}{k_{\perp}}

\renewcommand{\b}{\mathrm{b}}
\renewcommand{\c}{\mathrm{c}}
\renewcommand{\d}{\mathrm{d}}
\newcommand{\e}{\mathrm{e}}
\newcommand{\g}{\mathrm{g}}
\newcommand{\p}{\mathrm{p}}
\newcommand{\q}{\mathrm{q}}
\newcommand{\s}{\mathrm{s}}
\renewcommand{\u}{\mathrm{u}}
\newcommand{\D}{\mathrm{D}}

\newcommand{\Q}{\mathrm{Q}}
\newcommand{\V}{\mathrm{V}}
\newcommand{\bbar}{\overline{\mathrm{b}}}
\newcommand{\cbar}{\overline{\mathrm{c}}}
\newcommand{\dbar}{\overline{\mathrm{d}}}
\newcommand{\qbar}{\overline{\mathrm{q}}}
\newcommand{\sbar}{\overline{\mathrm{s}}}
\newcommand{\ubar}{\overline{\mathrm{u}}}
\newcommand{\Bbar}{\overline{\mathrm{B}}}
\newcommand{\Dbar}{\overline{\mathrm{D}}}
\newcommand{\Qbar}{\overline{\mathrm{Q}}}
\newcommand{\Py}{{\sc{Pythia}}}
\def\simgt{\rlap{\lower 3.5 pt \hbox{$\mathchar \sim$}} \raise 1pt \hbox {$>$}}
\def\simlt{\rlap{\lower 3.5 pt \hbox{$\mathchar \sim$}} \raise 1pt \hbox {$<$}}
\def\bb{b\overline{b}}
\def\pthat{\hat{P_{t}}}
\newcommand\pthatmin{\hat{p}_T^{\rm min}}
\def\herw{{\small HERWIG}}
\def\beq{\begin{equation}}
\def\eeq{\end{equation}}

\def\beqn{\begin{eqnarray}}
\def\eeqn{\end{eqnarray}}
\newcommand\AP{{\rm\bf A}}
\newcommand\BP{{\rm\bf B}}

\newcommand\ibid[3]{ibid. {\bf #1}, #3 (#2)}
\newcommand\hepph[1]{hep-ph/#1}
\newcommand\hepex[1]{hep-ex/#1}
\newcommand\heplat[1]{hep-lat/#1}
\newcommand\npb[3]{Nucl. Phys. {\bf B#1},  #3 (#2) }
\newcommand\plb[3]{Phys. Lett. {\bf B#1}, #3 (#2)}
\newcommand\prd[3]{Phys. Rev. {\bf D#1}, #3 (#2)}
\newcommand\sjnp[3]{Sov. J. Nucl. Phys. {\bf #1}, #3 (#2)}
\newcommand\spj[3]{Sov. Phys. JETP  {\bf #1}, #3 (#2) }

\begin{document}
\title{BOTTOM PRODUCTION}
\author{{\bf Convenors}: P.~Nason, G.~Ridolfi, O.~Schneider
 G.F.~Tartarelli and P.~Vikas \\
  {\bf Contributing authors}:
J.~Baines,
S.P.~Baranov,
P.~Bartalini,
A.~Bay,
E.~Bouhova,
M.~Cacciari,
A.~Caner,
Y.~Coadou,
G.~Corti,
J.~Damet,
R.~Dell'Orso,
J.R.T.~De Mello Neto,
J.L.~Domenech,
V.~Drollinger,
P.~Eerola,
N.~Ellis,
B.~Epp,
S.~Frixione, 
S.~Gadomski,
I.~Gavrilenko,
S.~Gennai,
S.~George,
V.M.~Ghete,
L.~Guy,
Y.~Hasegawa,
P.~Iengo,
A.~Jacholkowska,
R.~Jones,
A.~Kharchilava,
E.~Kneringer,
P.~Koppenburg,
H.~Korsmo,
M.~Kr\"amer,
N.~Labanca,
M.~Lehto,
F.~Maltoni,
M.L.~Mangano,
S.~Mele,
A.M.~Nairz,
T.~Nakada,
N.~Nikitin,
A.~Nisati,
E.~Norrbin,
F.~Palla,
F.~Rizatdinova,
S.~Robins,
D.~Rousseau,
M.A.~Sanchis-Lozano,
M.~Shapiro,
P.~Sherwood,
L.~Smirnova,
M.~Smizanska,
A.~Starodumov,
N.~Stepanov,
R.~Vogt
}
\institute{~}
\maketitle
\begin{abstract}
We review the prospects for bottom production physics at the LHC.
\end{abstract}

\section{INTRODUCTION}
In the context of the LHC experiments, the physics of bottom
flavoured hadrons enters in different contexts. It can be used
for QCD tests, it affects the possibilities of $B$ decays studies,
and it is an important source of background for several processes
of interest.

The physics of $b$ production at hadron
colliders has a rather long story, dating back to its first observation
in the UA1 experiment.
Subsequently, $b$ production has been studied at the Tevatron.
Besides the transverse momentum spectrum of a single $b$, it has
also become possible, in recent time, to study correlations
in the production characteristics of the $b$ and the $\overline{b}$.

At the LHC new opportunities will be offered by the high statistics
and the high energy reach. One expects to be able to study the transverse
momentum spectrum at higher transverse momenta, and also to
exploit the large statistics to perform more accurate
studies of correlations.

This chapter is organized as follows.

Section \ref{bprod:theory} is mostly theoretical. Its goal is to
provide benchmark cross sections and distributions for the LHC,
including rates relevant for the trigger requirements of
the experiments.  Furthermore, a discussion of the present status of
$b$ production phenomenology at hadron colliders is given. In this
context, one cannot forget that the theoretical status is a mixed
success.  On one side, the shape of distributions and
correlations are reasonably well explained by perturbative QCD. On the
other side, however, the observed cross sections at the Tevatron are
larger than QCD predictions. It is hoped that further studies may help
to understand the nature of the discrepancy.  As of now, we see two
possible explanations: either the absolute normalization of the cross
section is not correctly predicted due to the presence of large
higher order terms, or the shape of the distributions is distorted by
some perturbative or non-perturbative effects (like, for example,
fragmentation effects). With the wide $\pt$ range covered by the LHC
experiments, and perhaps also with the possibility of performing more
accurate studies of correlations, these two possibilities may be
distinguished. The problem of fragmentation effects has been studied
in this workshop also from the point of view of hadronization models
in Monte Carlo programs, in Section \ref{bprod:MLMFrix}.  This study
deals with the hadronization model in the HERWIG Monte Carlo program.
Its aim was to understand whether, in simple realistic models of
hadronization, the usual assumption of QCD factorization is really at
work. In general, the problem of studying how realistic is the heavy
flavour production mechanism implemented in shower Monte Carlo's is
quite important, and probably will require a considerable effort. Along
this line, in Section \ref{bprod:gennai}, a problem in the heavy
flavour production mechanisms implemented in \Py\ is examined.

Further self-contained theoretical topics are dealt with in Sections
\ref{bprod:asymmetry} and \ref{bprod:quarkonium}.
Section~\ref{bprod:asymmetry} deals with the charge asymmetry
in $b$ production in $pp$ collisions. In this context,
QCD is not of great help, since in perturbative
QCD charge asymmetries turn out to be extremely small. Instead, studies
are made within specific hadronization models, that are parametrized in
such a way that they fit charm asymmetry data. This topic, besides
being interesting in its own, since it deals with a phenomenon which is
dominated by non-perturbative physics, has also an impact on $CP$
violation studies in $B$ decays.

In Section~\ref{bprod:quarkonium} quarkonium production is discussed.
This subject has been intensively studied in recent years, following an
initial CDF observation of a $J/\Psi$ production rate much higher than
theoretical predictions.  This has triggered, from the theoretical
side, the understanding that the fragmentation process is the dominant
mechanism in quarkonium production.  Besides this, a novel branch of
applications of perturbative QCD, the NRQCD approach, has emerged,
that may be useful to explain the production process.

In Section~\ref{bprod:smizanska}, the prospects for $b$ detection
are discussed. It is shown that there is a complementarity between
ATLAS/CMS and the LHCb experiment, with a certain region of overlap.
In particular, the LHCb experiment can detect very low momentum heavy
quarks, while the other experiments can reach the very high transverse
momentum region. Some results on correlations measurements are also
given, exploring the possibility of looking at one $b$ decaying into a
$J/\Psi$, and the other decaying semileptonically. Double heavy flavour
production, charge asymmetry, polarization effects, and doubly-heavy
meson production are also discussed.

In Section~\ref{bprod:bartalini} the tuning of the multiple
interaction parameters in \Py\ is illustrated.
The correct treatment of multiple interactions is
important to model the multiplicity observables in both minimum-bias and
heavy flavour events.

\section{BENCHMARK CROSS SECTIONS\protect\footnote{Section
    coordinator: P.~Nason and G.~Ridolfi}}
\labelsection{bprod:theory}
\subsection{Total cross sections}
\begin{figure}[htb]
\begin{center}
    \includegraphics[width=0.2\textwidth,clip]{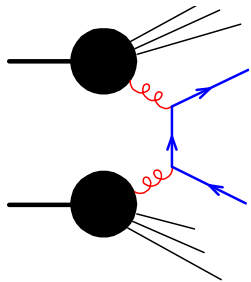}
    \caption{Typical diagram for heavy flavour production}
    \label{fig:hadroprod}
\end{center}
\end{figure}

It is assumed that heavy flavour production in hadronic collisions
can be described in the usual improved parton model approach, where
light partons in the incoming hadrons collide and produce a heavy
quark-antiquark pair via elementary strong interaction vertices, like, for
example, in the diagram of fig.~\ref{fig:hadroprod}.
This description is appropriate for all hard processes in hadronic
collisions, and thus, in the case of heavy flavours, is applicable as long as
the mass of the heavy flavour can be considered sufficiently large.
The perturbative QCD cross section for heavy flavour production
has been computed to next-to-leading order accuracy (i.e. ${\cal O}(\as^3)$)
a long time ago
\cite{bprod:NDE1} \cite{bprod:Beenakker1} \cite{bprod:NDE2} \cite{bprod:Beenakker2}
\cite{bprod:MNR} \cite{bprod:Ellis89} \cite{bprod:Smith92} \cite{bprod:FMNR94}
\cite{bprod:Laenen93} \cite{bprod:Laenen93a} \cite{bprod:Harris95},
and a large amount of experimental and theoretical work has been
done in this field. A relatively recent account of the status
of this field can be found in ref.~\cite{bprod:Buras2}.
It can be said that qualitatively the QCD description of heavy flavour
production seems to be adequate also for charm production,
while quantitatively large uncertainties are present in the
calculation of the charm and bottom cross section. Only for a quark as heavy
as the top quark the perturbative calculation seems (up to now)
to predict the cross section with a good accuracy.

Large uncertainties are also found in the calculation of the
bottom production cross section at the LHC.
The largest uncertainty is due to unknown higher order effects,
and it is traditionally quantified by estimating the scale dependence
of the cross section when the renormalization and factorization scales
are varied by a factor of 2 above and below their central value,
which is usually taken equal to the heavy quark mass.
Since this uncertainty is due to a limitation in our current theoretical
knowledge, it is hard to overcome. Other sources of uncertainty
are related to theoretical and experimental errors in the parameters
that enter the perturbative calculation: the value of the strong
coupling constant, the heavy quark mass, and the parton density functions.

We present here a benchmark study of $b$ total cross sections at the LHC,
using the FMNR package for heavy flavour cross sections \cite{bprod:MNR}
\cite{bprod:FMNR94} (the code for this package is
available upon request to the authors). In the study we consider
\begin{itemize}
\item
The dependence of the total cross section on the choice of
the factorization and renormalization scales. We will use the values
$\mu=m_b,2m_b,m_b/2$.
\item
The dependence on the parton density parametrization.
We will use the sets MRST~\cite{bprod:Martin:1998sq},
\MRSTH, \MRSTL, \MRSTM\ and \MRSTP.
The first set is used as reference set. \MRSTH\ and \MRSTL\ have extreme
gluon densities, \MRSTM-\MRSTP\ have extreme values of the 
strong coupling constant: $\Lambda_5=220\;$MeV for MRST, 164~MeV for \MRSTM,
280~MeV for \MRSTP.
Cross section values obtained with the CTEQ4 \cite{bprod:cteq4}
set are very similar to the MRST
set. We have preferred to use the MRST sets because they gave us
the possibility to perform a study of sensitivity to $\Lambda$
and to variations in the gluon density.
\item
The dependence on the $b$ quark mass: $m_b=4.75\pm0.25\;$GeV.
\end{itemize}
Factorization and renormalization scale dependence of the total
cross section at $\sqrt{S}=14\;$TeV is reported in table~\ref{tab:sigma0},
where we have used the 
MRST parton densities, with $\Lambda_5=220\;$MeV, and we
have fixed the $b$ mass at the value $m_b=4.75\;$GeV.
\begin{table}[t]
\begin{center}
\caption{Dependence of the $b$ cross section on scale choices.}
\label{tab:sigma0}
\vskip0.2cm
\begin{tabular}{|c|c|c|c|} \hline
{ $\muF/m_b$}&{ $\muR/m_b$}&{ Total ($\mu b$)}
&{ Born ($\mu b$)}\\ \hline  
 0.50 & 1.00 & 0.2779 $10^3$ & 0.6465 $10^2$ \\ \hline
 1.00 & 1.00 & 0.4960 $10^3$ & 0.1796 $10^3$ \\ \hline 
 2.00 & 1.00 & 0.6453 $10^3$ & 0.3253 $10^3$ \\ \hline 
 0.50 & 0.50 & 0.5126 $10^3$ & 0.1078 $10^3$ \\ \hline 
 1.00 & 0.50 & 0.8289 $10^3$ & 0.2995 $10^3$ \\ \hline 
 2.00 & 0.50 & { 0.9538 $10^3$} & { 0.5426 $10^3$} \\ \hline 
 0.50 & 2.00 & { 0.1758 $10^3$} & { 0.4355 $10^2$} \\ \hline 
 1.00 & 2.00 & 0.3353 $10^3$ & 0.1209 $10^3$ \\ \hline 
 2.00 & 2.00 & 0.4669 $10^3$ & 0.2191 $10^3$ \\ \hline
\end{tabular}
\end{center}
\end{table}
Notice that:
\begin{itemize}
\item
If we keep $\muF=\muR$, the full cross section variation
is small (467 to 512 $\mu b$).
\item
The largest cross section corresponds to large $\muF$ and small $\muR$
\item
The smallest cross section corresponds to small $\muF$ and large $\muR$
\end{itemize}
This is understood since, at small $x$, the gluon density $g(x)$
grows with the scale, and $\as$ decreases with the scale.

The dependence on the choice of parton density parametrization is shown
in table~\ref{tab:pdf}.
\begin{table}[ht]
\begin{center}
\caption{Parton density dependence of total cross sections (in $\mu b$).}
\label{tab:pdf}
\vskip 0.2cm
\begin{tabular}{|c|c|c|c|} \hline
& central & lowest & highest \\ \hline
MRST & 0.4960 $10^3$ & 0.1758 $10^3$ & 0.9538 $10^3$ \\ \hline
\MRSTH & 0.4866 $10^3$ & 0.1727 $10^3$ & 0.9337 $10^3$ \\ \hline
\MRSTL & 0.4992 $10^3$ & 0.1751 $10^3$ & 0.9610 $10^3$ \\ \hline
\MRSTM & 0.4487 $10^3$ & 0.1799 $10^3$ & { 0.7878 $10^3$} \\ \hline
\MRSTP & 0.6001 $10^3$ & 0.1894 $10^3$ & { 0.1267 $10^4$} \\ \hline
\end{tabular}
\end{center}
\end{table}
As one can see, the sensitivity to the variation of the gluon density
is small. Apparently, the constraints from HERA data are strong
enough in the $x$ region where most of the $b$ production takes place.
The dependence upon the strong coupling constant is instead larger,
and can increase the upper limit of the cross section by about 40\%.
The last two sets have $\Lambda_5=164$ and 288 MeV respectively,
corresponding to $\as(M_Z)=0.1125$ and 0.1225, which is a reasonably large
range.

\begin{figure}[htb]
\begin{center}
    \includegraphics[width=0.6\textwidth,clip]{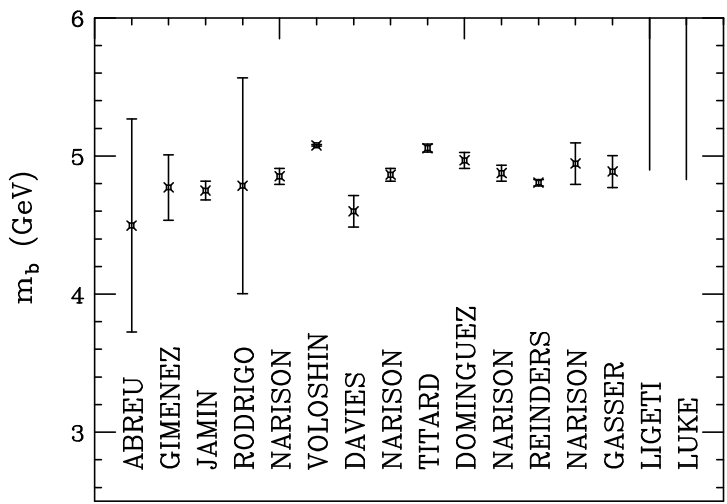}
    \caption{Different determinations of the $b$ quark pole mass.}
    \label{fig:bmass}
\end{center}
\end{figure}

Mass uncertainties are quite important, especially if $m_b$ is allowed to take
very small values. This can be seen from table~\ref{tab:mass}.
\begin{table}[ht]
\begin{center}
\caption{Mass dependence of total cross sections (in $\mu b$).}
\label{tab:mass}
\vskip 0.2cm
\begin{tabular}{|c|c|c|c|} \hline
$m_b$ (GeV) & central & lowest & highest \\ \hline
4     & { 0.7957 $10^3$} & { 0.2336 $10^3$} & { 0.1706 $10^4$} \\ \hline
4.5   & 0.5789 $10^3$ & 0.1945 $10^3$ & 0.1138 $10^4$ \\ \hline
5     & 0.4313 $10^3$ & 0.1609 $10^3$ & 0.8087 $10^3$ \\ \hline
\end{tabular}
\end{center}
\end{table}
We see that lowering the $b$ mass from 4.5 down to 4~GeV raises the upper
limit of the cross section by about 50\%.
It is however unlikely that such small values are viable.
A rough view of the status of the bottom mass determination
is given in fig.~\ref{fig:bmass},
which we obtained by taking the various determinations of the
$\MSB$ bottom mass from the Particle Data Book, and rescale them
by a factor of (1+0.09+0.06), to account for the two-loop
correction needed to translate the $\MSB$ mass into the pole mass.
As one can see, not all determinations are consistent among
each other. A critical review of all determinations is beyond the
scope of this workshop. We should however point out that recent progress
has been made in the bottom mass determination.
The reader can find a summary of these issues
and further references in ref.~\cite{bprod:Beneke-hf8}.
It is argued there
that the bottom mass is determined with higher precision in processes
where it is probed at short distances, like in the
$\Upsilon$ mass, or via sum rules applied to the bottom
vector current spectral function in $e^+e^-$ annihilation.
The mass extracted in this way can be reliably related to
the so called $\MSB$ mass. The relation of the $\MSB$ mass to the
pole mass is instead not so precise, because the perturbative expansion
that relates the two quantities is not convergent.
In ref.~\cite{bprod:Beneke-hf8} a preferred value
of $\bar{m}_b(m_b)=4.23\pm 0.08$
is given, where $\bar{m}_b(m_b)$ is the \MSB\ bottom mass at the
scale of the bottom mass itself. The corresponding pole mass,
obtained using the newly computed 3-loop relation
between the $\MSB$ and pole mass \cite{bprod:Chetyrkin99} \cite{bprod:Melnikov99},
is $4.98\pm 0.09$ GeV.
If one wanted to account for the uncertainties due to the lack of convergence
of the perturbative expansion, the range obtained in this way should
be enlarged by some amount, of the order of 100 MeV.
The question arises whether it would be possible to eliminate this
uncertainty by expressing the hadroproduction
cross section in terms of the \MSB\ mass. In our view, the answer
is most likely no, since the bottom hadroproduction cross section
does not have the same inclusive
character as the sum rules applied to the $e^+e^-$ bottom spectral function.

In the present work we thus used the traditional range
$4.5$~GeV~$<m_b<5$~GeV for the
bottom pole mass in the hadroproduction process, keeping in mind that recent
determinations seem to favour the upper region of this range.
The sensitivity of the cross section to the bottom mass in this range
is at most of $\pm 10$\%, and it becomes much smaller if transverse
momentum cuts are applied. Thus, as far as the LHC is concerned,
this uncertainty is not very important.

\begin{figure}[htb]
\begin{center}
    \includegraphics[width=0.8\textwidth,clip]{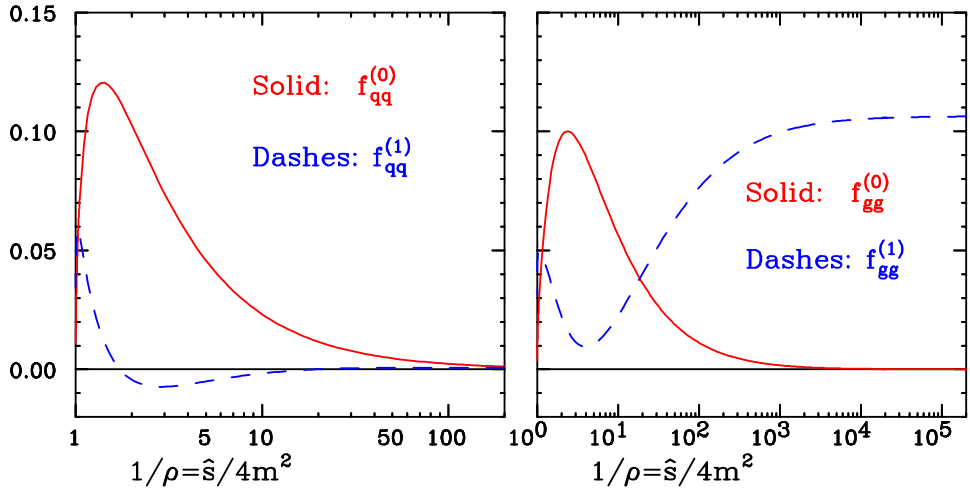}
    \caption{Partonic cross section for the $q\bar{q}$ and $gg$ subprocesses.}
    \label{fig:f1qqgg}
\end{center}
\end{figure}

The largest uncertainy in the cross section comes from the scale
uncertainty, which is a (rather arbitrary) method to assess the
possible impact of unknown higher order corrections. In the following we
report a brief discussion of the origin of these large corrections. 
Radiative corrections for the total cross section are usually parametrized as
follows. The total cross sections
$\sigma_{ij}$ for the various parton
subprocesses (${\overline q}q, qg,gg$) have a perturbative expansions given by
\begin{equation}
\sigma_{ij}=\frac{\as^2(\mu)}{m^2}\left[
                 f_{ij}^{(0)}(\rho) + 4\pi\as\left(f_{ij}^{(1)}(\rho)
                 +\bar{f}^{(1)}(\rho)\log\frac{\mu^2}{m^2}\right)\right],
\end{equation}
where $\rho=4m_b^2/\hat{s}$ and $\hat{s}$ is the squared partonic
center-of-mass energy.
The functions $f_{ij}^{(0,1)}$ for the ${\overline q}q$ and $gg$ subprocesses
are displayed in fig.~\ref{fig:f1qqgg}.
Notice the behaviour near threshold
\begin{eqnarray}
    && f_{q\overline{q}}^{(1)} \rightarrow
       \frac{f^{(0)}_{q\bar{q}}(\rho)}{8\pi^2}
    \Bigg[ - \frac{ \pi^2 }{6 \beta } 
    + \frac{16}{3 } \ln ^2 \big(8 \beta^2 \big) 
    - \frac{82}{3} \ln(8 \beta^2) \Bigg] \nonumber \\
    && f_{gg}^{(1)} \rightarrow \frac{f_{gg}^{(0)}(\rho)}{8\pi^2}
    \Bigg[  \frac{ 11 \pi^2 }{42\beta } + 
    12 \ln ^2 \big(8 \beta^2 \big) 
    - \frac{366}{7} \ln( 8 \beta^2) \Bigg] \nonumber
\end{eqnarray}
due to Coulomb $1/\beta$ singularities
and to Sudakov double logarithms. Near threshold,
these terms may require special treatement, such as 
resummation to all orders.
Notice also the constant asymptotic behaviour of $f^{(1)}_{gg}$, which
may cause problems far above threshold.

Plotting the cross section as a function of the partonic energy $\hat{s}$
may help to understand the origin of large corrections.
\begin{figure}[htb]
\begin{center}
    \includegraphics[width=0.6\textwidth,clip]{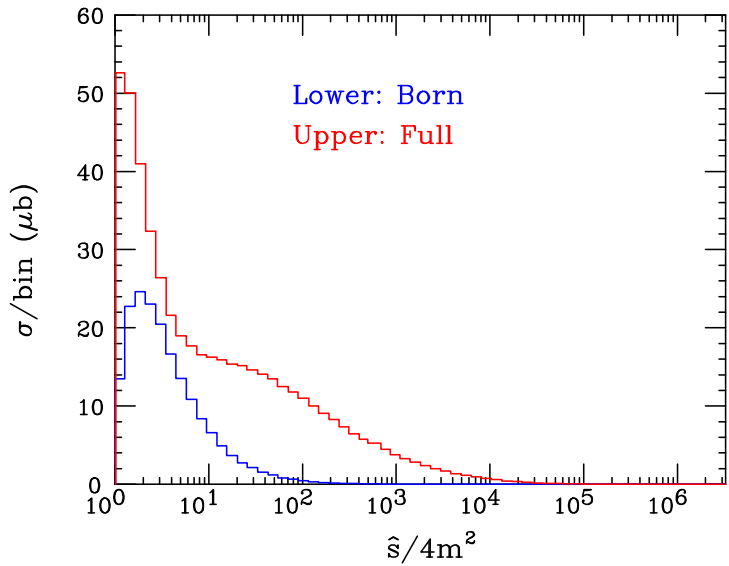}
    \caption{Partonic cross section as a function of $\hat{s}/4m^2$}.
    \label{fig:roplot}
\end{center}
\end{figure}
We find that radiative corrections are large near the production
threshold. This problem becomes more and more severe as we approach
the production threshold. Thus, it is more important for production of
$b$ at fixed target energies, or for production of $t\bar{t}$ pairs at
colliders.  Techniques to resum these large corrections to all orders
of perturbation theory, at the NLO level are available
\cite{bprod:Bonciani}, but it is found that little improvement is
achieved for the bottom cross section at collider energies.  Large
corrections are also found far above threshold.  This effect is bound
to become more and more pronounced in the high energy limit. In order
to reduce the scale uncertainties coming from these corrections, one
should resum them at the next-to-leading order level. This problem has
been discussed in the literature, so far, only at the leading
logarithmic level
\cite{bprod:smallx1} \cite{bprod:smallx2} \cite{bprod:smallx3}
\cite{bprod:smallx4}.
At the time of the completion
of this workshop, no further progress has been achieved in this field.

\begin{figure}[htb]
\begin{center}
    \includegraphics[width=0.8\textwidth,clip]{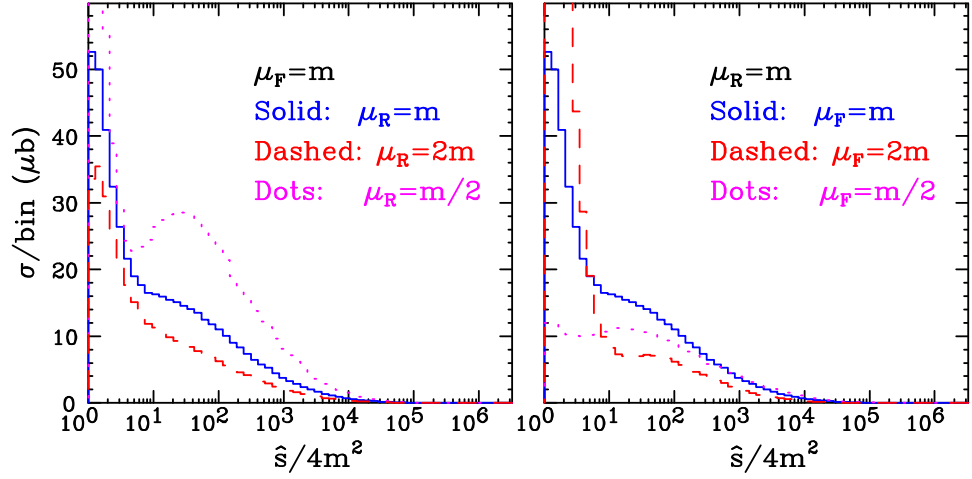}
    \caption{Scale dependence of the total cross section.}
    \label{fig:roplotscdep}
\end{center}
\end{figure}

In fig.~\ref{fig:roplotscdep} we present a study
of the scale dependence of the total cross section as a function of $\rho$.
We find a large scale dependence near threshold, due to both renormalization
and factorization scale variation, and a
large scale dependence far from threshold. Here, the renormalization
scale dependence  plays a dominant role.
Renormalization scale variations are mainly due to the large
variation of the coupling constant in the ${\cal O}(\as^3)$
terms.  Where radiative corrections are small, a reasonable scale
compensation takes place. Thus, both the threshold and the high
energy regions, where corrections are large, are strongly affected.
Factorization scale variation has a strong impact on threshold
corrections, while in the high energy region we observe
some compensation. In fact, the cross section near threshold increases
with $\muF$ near threshold, while above threshold the $\muF=m$
value is above both the $\muF=m/2$ and the $\muF=2 m$ curves, indicating
the presence of some sort of compensation. As of now, it appears therefore
that a better understanding of the high energy region will not
strongly reduce the scale uncertainty, although it might, of course,
improve our confidence in the error band we quote.

\begin{figure}[t]
\begin{center}
    \includegraphics[width=0.8\textwidth,clip]{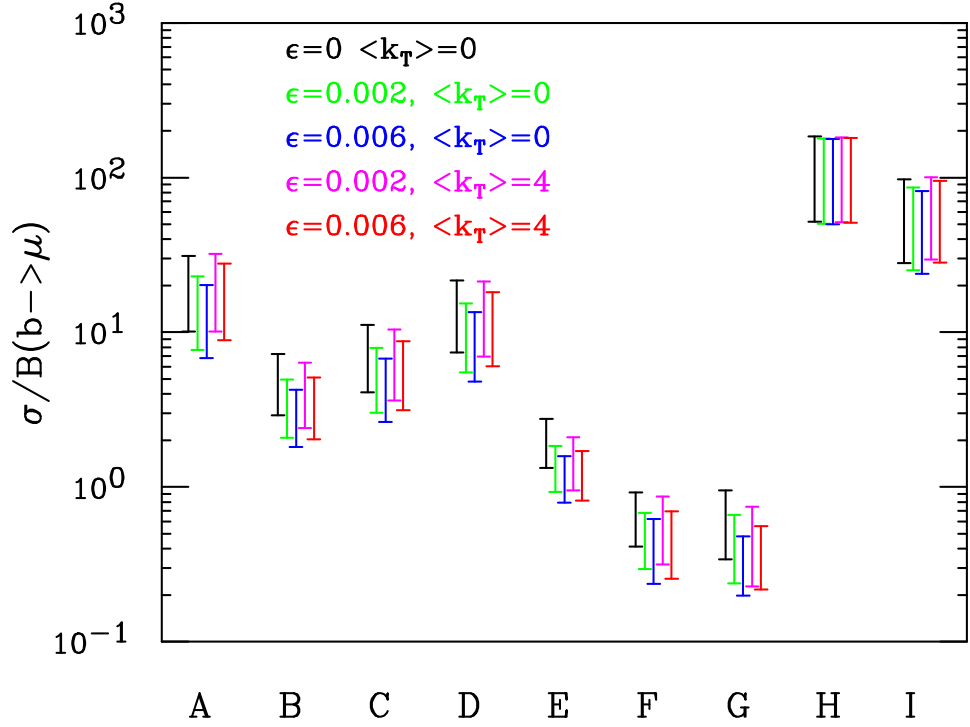}
    \caption{Cross sections as in table \ref{tab:triggerrates}. The four bars
in each group correspond to different choices of $\epsilon$ and
$\langle k_T\rangle$, in the order reported in the figure.}
    \label{fig:triggerrates}
\end{center}
\end{figure}

The study given here deals with total cross sections.  It should be
repeated with appropriate rapidity cuts, since this may reduce large
effects due to the high energy limit.  In general, we may expect that
the cross section with rapidity and transverse momentum cuts may have
smaller error bars than the total.
It is particularly interesting to investigate directly cross sections
for muons originating from $B$ decays, since muons are often
used as trigger objects for $B$ physics. We have performed this study using
a simple implementation of the $B$ semileptonic decay in the FMNR
program, that will be described in more details in the following
subsections.
The results are shown in table~\ref{tab:triggerrates}.
\begin{table}[t]
\begin{center}
\caption{Cross sections (in $\mu b$) for $b\to \mu+X$ production, with a muon,
         or both,
         satisfying appropriate cuts. Only muons coming directly
         from $B$ decays are included here. The calculation was
         performed using the CTQ4M parton densities.
	 The upper number are the maximum, and the lower number the minimum
         of the values obtained by varying the scales in the usual way.
         The corresponding
         total cross sectios are 165 to 864 $\mub$
         The $B\to \mu$ branching fraction was taken equal to 10.5\%.
         Different values for the $\epsilon$ parameter of the
         Peterson fragmentation function are assumed.
         The last two column show the impact of a rather large
         intrinsic transverse momentum of the incoming partons.}
\label{tab:triggerrates}
\vskip0.2cm
\begin{tabular}{|l|c|c|c|c|c|} \hline
 $\epsilon$ & 0 & 0.002 & 0.006 & 0.002 & 0.006 \\ \hline
 $\langle k_T \rangle$ (GeV) & 0 & 0 & 0 & 4 & 4 \\ \hline\hline
A: $b\bar{b}\to\mu(|\eta|<2.4,\; p_T\ge 6)$\rule[-3mm]{0mm}{8mm}
  & $   3.3\atop   1.06$
  & $   2.41\atop   0.81$
  & $   2.12\atop   0.72$
  & $   3.4\atop   1.06$
  & $   2.91\atop   0.94$
  \\ \hline
B: $b\bar{b}\to\mu(p_T>6)\;\mu(p_T>3)$\rule[-3mm]{0mm}{8mm}
  & $   0.76\atop   0.304$
  & $   0.52\atop   0.219$
  & $   0.45\atop   0.19$
  & $   0.67\atop   0.252$
  & $   0.54\atop   0.214$
  \\ \hline
C: $b\bar{b}\to\mu(p_T>6)\; e(p_T>2)$\rule[-3mm]{0mm}{8mm}
  & $   1.18\atop   0.43$
  & $   0.83\atop   0.32$
  & $   0.71\atop   0.277$
  & $   1.1\atop   0.38$
  & $   0.92\atop   0.33$
  \\ \hline
D: $b\bar{b}\to\mu(p_T > 7,\;|\eta|<2.4)$\rule[-3mm]{0mm}{8mm}
  & $   2.26\atop   0.78$
  & $   1.62\atop   0.58$
  & $   1.41\atop   0.5$
  & $   2.23\atop   0.73$
  & $   1.9\atop   0.63$
  \\ \hline
E: $b\bar{b}\to\mu( p_T > 7 ,\;  |\eta| < 2.4)$&&&&&\\
\phantom{E: $b\bar{b}\to$}
    $\mu(p_T > 4.5 ,\;\;0<|\eta| < 1.5)$\rule[-3mm]{0mm}{3mm}
  & $   0.0304\atop   0.0146$
  & $   0.0203\atop   0.0102$
  & $   0.0174\atop   0.0087$
  & $   0.0232\atop   0.0105$
  & $   0.0188\atop   0.009$
  \\ \hline
F: $b\bar{b}\to\mu( p_T > 7 ,\;  |\eta| < 2.4)$&&&&&\\
\phantom{F: $b\bar{b}\to$}
    $\mu(p_T > 3.6 ,\;  1.5 < |\eta| < 2)$\rule[-3mm]{0mm}{3mm}
  & $   0.0101\atop   0.0045$
  & $   0.0075\atop   0.0032$
  & $   0.0068\atop   0.0026$
  & $   0.0096\atop   0.0035$
  & $   0.0076\atop   0.00281$
  \\ \hline
G: $b\bar{b}\to\mu( p_T > 7 ,\;  |\eta| < 2.4)$ &&&&&\\
\phantom{G: $b\bar{b}\to$}
     $\mu(p_T > 2.6 ,\; 2 < |\eta| < 2.4)$\rule[-3mm]{0mm}{3mm}
  & $   0.0105\atop   0.0038$
  & $   0.0073\atop   0.00263$
  & $   0.0053\atop   0.00219$
  & $   0.0082\atop   0.00251$
  & $   0.0062\atop   0.0024$
  \\ \hline
H: $b\bar{b}\to\mu(p_T > 1,\; 2<|\eta|<6)$\rule[-3mm]{0mm}{8mm}
  & $   19.3\atop   5.4$
  & $   18.8\atop   5.3$
  & $   18.6\atop   5.2$
  & $   19.1\atop   5.4$
  & $   18.9\atop   5.3$
  \\ \hline
I: $b\bar{b}\to\mu(p_T > 2,\; 2<|\eta|<6)$\rule[-3mm]{0mm}{8mm}
  & $   10.2\atop   2.94$
  & $   9.1\atop   2.65$
  & $   8.6\atop   2.51$
  & $   10.6\atop   3.11$
  & $   10.\atop   2.96$
  \\ \hline
\end{tabular}
\end{center}
\end{table}
The same results are also reported in fig.~\ref{fig:triggerrates}, since
several features become more apparent there.
First of all, we point out that, as expected, there is a considerable reduction
in the scale dependence in these muon rates. This is mostly due to the presence
of cuts in the transverse momentum of the muon, that increases the total
transverse energy that characterizes the cross section. Thus, while the ratio
of the upper to the lower limit of the cross section is above a factor
of 5 in the total rate, it is between a factor of 2 and 4 in the muon rates.
The smallest values are achieved for the highest momentum cuts.
A non perturbative fragmentation function of the Peterson form
was also included in the calculation, with $\epsilon$ parameter
taking the values $0$ (i.e., no fragmentation), $0.002$ and $0.006$.
More details on its implementation are given in the following subsections.
Observe that for softer fragmentation functions (i.e. larger $\ep$
parameter) the uncertainty is
reduced, since they imply higher quark momenta.
The reduction in the scale uncertainty is obtained at the price of introducing
a sensitivity to the fragmentation function parameter.
We considered as realistic values of $\epsilon$ between 0.002 and 0.006.
The corresponding variation of the cross section is not large.
The impact of an intrinsic transverse momentum of the incoming partons
(see the following subsections)
is also studied. We have chosen the unrealistically large value
$\langle k_T\rangle =4$~GeV just to show that its effect is in all cases
not a dramatic one.

\subsection{Transverse momentum spectrum}
\subsubsection{Benchmark single-inclusive distributions}
\begin{figure}[t]
\begin{center}
    \includegraphics[width=0.65\textwidth,clip]{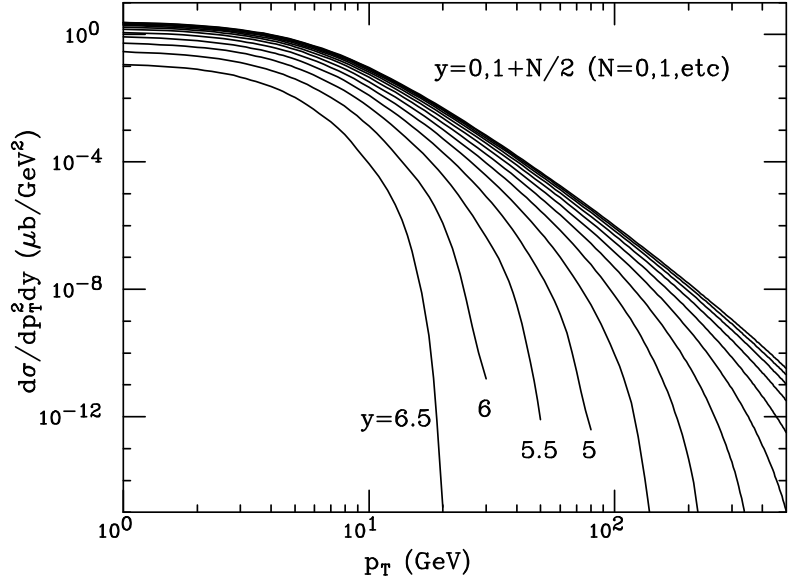}
    \caption{Differential cross section for heavy flavour production
             vs. $p_T$, for different rapidities.}
    \label{bprod:fig:diffpt}
\end{center}
\end{figure}

\begin{figure}[t]
\begin{center}
    \includegraphics[width=0.65\textwidth,clip]{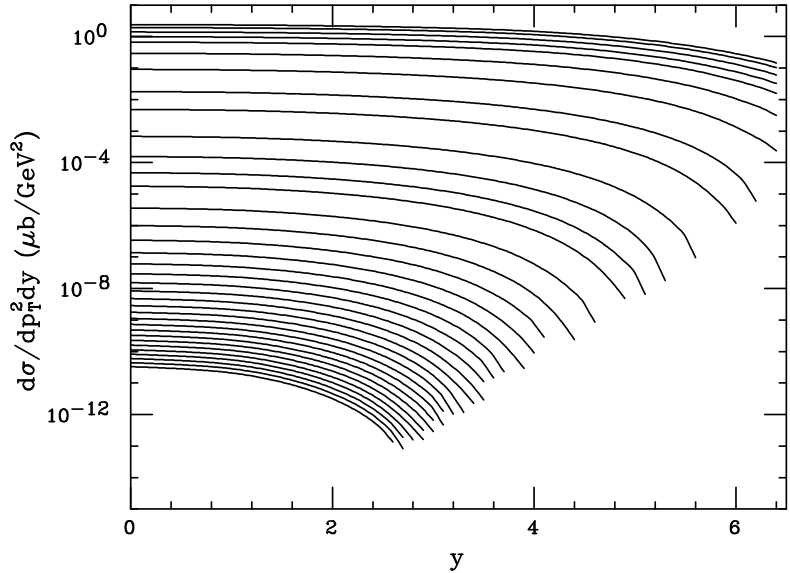}
    \caption{Differential cross section for heavy flavour production
             vs. $y$ for different $p_T$ values, as
             given in tables~\ref{bprod:tab:diff1}-\ref{bprod:tab:diff4}.}
    \label{bprod:fig:diffy}
\end{center}
\end{figure}

\begin{figure}[t]
\begin{center}
    \includegraphics[width=0.65\textwidth,clip]{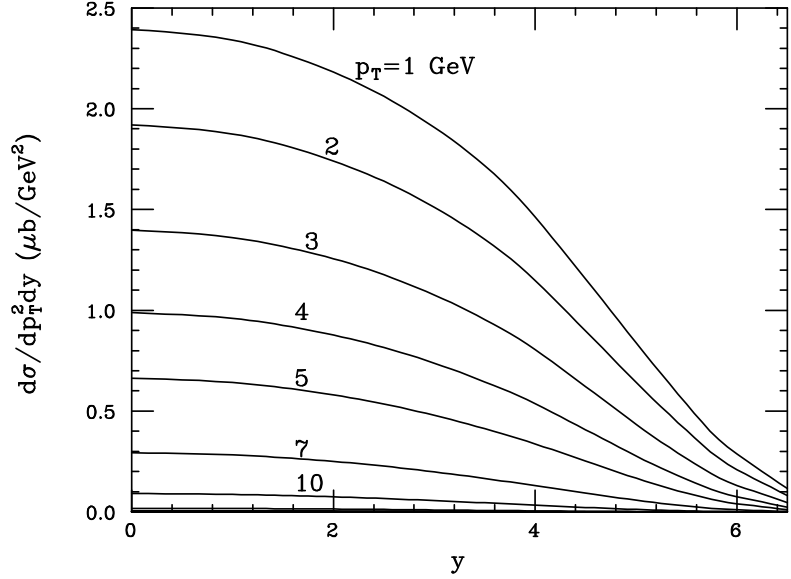}
    \caption{Differential cross section for heavy flavour production
             vs. $y$ for different $p_T$ values.}
    \label{bprod:fig:diffy1}
\end{center}
\end{figure}

\begin{figure}[htb]
\begin{center}
    \includegraphics[width=0.65\textwidth,clip]{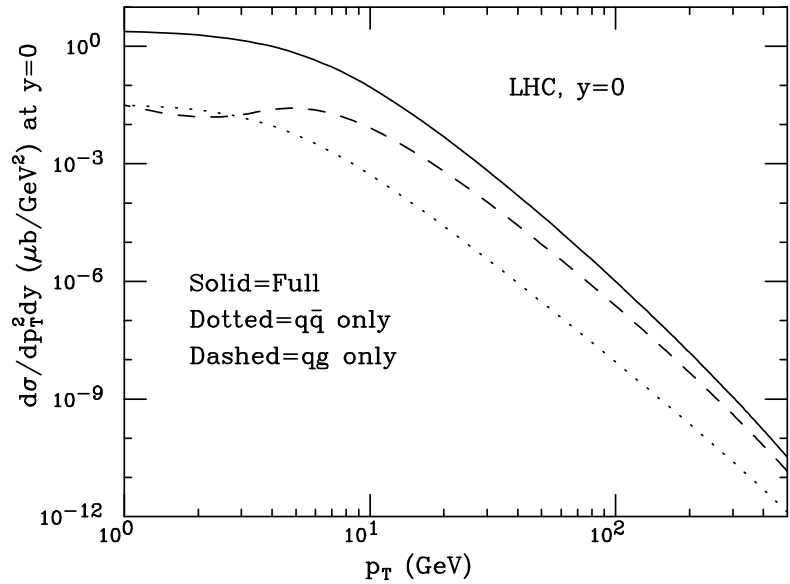}
    \caption{The total, $qq$ and $qg$ components of the differential
             cross section for heavy flavour production.}
    \label{bprod:fig:totqqqg}
\end{center}
\end{figure}

The fixed-order, NLO result for single-inclusive $b$ production
has several limitations in different regions of the phase space.
In particular, one should be aware of the high-energy limit
problem when $p_T$ is small compared to the incoming energy,
of the logarithms of $m_b/p_T$ for high transverse momenta,
and of further problems when approaching the threshold region.
All these issues will be discussed in some detail in the next Sections.
However, the fixed-order calculation at NLO provides a useful starting
point for estimating the differential cross section. At this time, it
is probably not useful to perform a cross section study with different
sets of parton densities, and for different values of the $b$ mass. We
limit ourselves to the MRST set, and we only study the scale
dependence of the cross section.  We do not include, at this stage,
fragmentation effects, which, as shown in the following Sections, can
be easily accounted for.  In
tables~\ref{bprod:tab:diff1}-\ref{bprod:tab:diff4} we collect the
results of this study.
\begin{table}[htb]
\begin{center}
\caption{Differential cross section $d\sigma/dp_T^2 dy$ for single
inclusive $b$ production at the LHC, for $p_T$ from 1 to 80~GeV and
$y$ from 0 to 4. The table was computed with the $MRST$ parton
density set, for $m_b=4.75\;$GeV. The central value was obtained with
the factorization and renormalization scale set to $\sqrt{p_T^2+m^2}$.
The upper and lower values give the maximal variation when varying the
scales independently by a factor of 2 above and below the central
value. Cross sections are in $\mu b$; each element in a row should be
multiplied by the common scale factor in the left column.}
\vskip0.2cm
\label{bprod:tab:diff1}
\begin{tabular}{|l|c|c|c|c|c|c|c|c|}
\hline
$y$ & 0& 1&1.5& 2&2.5& 3&3.5& 4\\
\hline
 $p_T=1 $
 & $7.7 $
 & $7.6 $
 & $7.5 $
 & $7.3 $
 & $7 $
 & $6.6 $
 & $6.1 $
 & $5.4 $
 \\
 \hfill $10^{ -1}\times$
 & $23.9 $
 & $23.4 $
 & $22.7 $
 & $21.8 $
 & $20.6 $
 & $19.1 $
 & $17.2 $
 & $14.6 $
 \\
 & $41.3 $
 & $40.5 $
 & $39.5 $
 & $38.2 $
 & $36.4 $
 & $34.1 $
 & $31 $
 & $26.1 $
 \\
 \hline
 $p_T=2 $
 & $6.7 $
 & $6.6 $
 & $6.5 $
 & $6.3 $
 & $6 $
 & $5.7 $
 & $5.2 $
 & $4.56 $
 \\
 \hfill $10^{ -1}\times$
 & $19.2 $
 & $18.8 $
 & $18.2 $
 & $17.4 $
 & $16.4 $
 & $15.1 $
 & $13.5 $
 & $11.5 $
 \\
 & $30.3 $
 & $29.7 $
 & $29 $
 & $28 $
 & $26.7 $
 & $24.9 $
 & $22.6 $
 & $19.2 $
 \\
 \hline
 $p_T=3 $
 & $5.4 $
 & $5.3 $
 & $5.2 $
 & $5 $
 & $4.76 $
 & $4.43 $
 & $4.01 $
 & $3.48 $
 \\
 \hfill $10^{ -1}\times$
 & $14 $
 & $13.6 $
 & $13.2 $
 & $12.6 $
 & $11.8 $
 & $10.8 $
 & $9.5 $
 & $8 $
 \\
 & $21.8 $
 & $21.3 $
 & $20.7 $
 & $19.9 $
 & $18.8 $
 & $17.4 $
 & $15.6 $
 & $13.2 $
 \\
 \hline
 $p_T=4 $
 & $4.04 $
 & $3.96 $
 & $3.85 $
 & $3.7 $
 & $3.48 $
 & $3.21 $
 & $2.87 $
 & $2.45 $
 \\
 \hfill $10^{ -1}\times$
 & $9.9 $
 & $9.6 $
 & $9.3 $
 & $8.8 $
 & $8.2 $
 & $7.4 $
 & $6.5 $
 & $5.4 $
 \\
 & $15.9 $
 & $15.5 $
 & $15 $
 & $14.2 $
 & $13.3 $
 & $12.1 $
 & $10.6 $
 & $8.9 $
 \\
 \hline
 $p_T=5 $
 & $2.91 $
 & $2.84 $
 & $2.75 $
 & $2.62 $
 & $2.45 $
 & $2.24 $
 & $1.97 $
 & $1.64 $
 \\
 \hfill $10^{ -1}\times$
 & $6.6 $
 & $6.4 $
 & $6.2 $
 & $5.8 $
 & $5.3 $
 & $4.79 $
 & $4.13 $
 & $3.37 $
 \\
 & $11.1 $
 & $10.7 $
 & $10.3 $
 & $9.7 $
 & $9 $
 & $8 $
 & $6.9 $
 & $5.7 $
 \\
 \hline
 $p_T=7 $
 & $14 $
 & $13.6 $
 & $13.1 $
 & $12.3 $
 & $11.3 $
 & $10.1 $
 & $8.6 $
 & $6.9 $
 \\
 \hfill $10^{ -2}\times$
 & $29.4 $
 & $28.2 $
 & $26.9 $
 & $25 $
 & $22.7 $
 & $19.8 $
 & $16.6 $
 & $13 $
 \\
 & $ 51 $
 & $48.5 $
 & $46 $
 & $42.7 $
 & $38.5 $
 & $33.5 $
 & $27.9 $
 & $21.7 $
 \\
 \hline
 $p_T=10 $
 & $4.72 $
 & $4.54 $
 & $4.31 $
 & $3.99 $
 & $3.58 $
 & $3.09 $
 & $2.52 $
 & $1.9 $
 \\
 \hfill $10^{ -2}\times$
 & $9.1 $
 & $8.7 $
 & $8.2 $
 & $7.5 $
 & $6.6 $
 & $5.6 $
 & $4.49 $
 & $3.32 $
 \\
 & $15.6 $
 & $14.8 $
 & $13.9 $
 & $12.6 $
 & $11.1 $
 & $9.3 $
 & $7.4 $
 & $5.4 $
 \\
 \hline
 $p_T=15 $
 & $10.3 $
 & $9.8 $
 & $9.1 $
 & $8.2 $
 & $7.1 $
 & $5.9 $
 & $4.49 $
 & $3.11 $
 \\
 \hfill $10^{ -3}\times$
 & $17.8 $
 & $16.8 $
 & $15.6 $
 & $14 $
 & $12 $
 & $9.7 $
 & $7.3 $
 & $4.91 $
 \\
 & $29.3 $
 & $27.5 $
 & $25.4 $
 & $22.6 $
 & $19.2 $
 & $15.4 $
 & $11.4 $
 & $7.6 $
 \\
 \hline
 $p_T=20 $
 & $2.99 $
 & $2.81 $
 & $2.6 $
 & $2.3 $
 & $1.94 $
 & $1.53 $
 & $1.1 $
 & $0.7 $
 \\
 \hfill $10^{ -3}\times$
 & $4.86 $
 & $4.55 $
 & $4.16 $
 & $3.65 $
 & $3.04 $
 & $2.36 $
 & $1.67 $
 & $1.04 $
 \\
 & $7.7 $
 & $7.1 $
 & $6.5 $
 & $5.7 $
 & $4.69 $
 & $3.61 $
 & $2.52 $
 & $1.55 $
 \\
 \hline
 $p_T=30 $
 & $4.53 $
 & $4.2 $
 & $3.79 $
 & $3.25 $
 & $2.61 $
 & $1.91 $
 & $1.25 $
 & $0.68 $
 \\
 \hfill $10^{ -4}\times$
 & $6.8 $
 & $6.3 $
 & $5.6 $
 & $4.77 $
 & $3.78 $
 & $2.74 $
 & $1.75 $
 & $0.93 $
 \\
 & $10.1 $
 & $9.3 $
 & $8.3 $
 & $7 $
 & $5.5 $
 & $3.94 $
 & $2.49 $
 & $1.3 $
 \\
 \hline
 $p_T=40 $
 & $10.9 $
 & $10 $
 & $8.8 $
 & $7.4 $
 & $5.7 $
 & $3.9 $
 & $2.31 $
 & $1.08 $
 \\
 \hfill $10^{ -5}\times$
 & $15.5 $
 & $14.1 $
 & $12.5 $
 & $10.3 $
 & $7.8 $
 & $5.3 $
 & $3.1 $
 & $1.41 $
 \\
 & $22.3 $
 & $20.2 $
 & $17.8 $
 & $14.6 $
 & $11 $
 & $7.4 $
 & $4.23 $
 & $1.92 $
 \\
 \hline
 $p_T=50 $
 & $3.47 $
 & $3.13 $
 & $2.72 $
 & $2.21 $
 & $1.63 $
 & $1.06 $
 & $0.57 $
 & $0.218 $
 \\
 \hfill $10^{ -5}\times$
 & $4.77 $
 & $4.28 $
 & $3.71 $
 & $2.99 $
 & $2.18 $
 & $1.4 $
 & $0.74 $
 & $0.287 $
 \\
 & $6.7 $
 & $6 $
 & $5.2 $
 & $4.12 $
 & $2.98 $
 & $1.89 $
 & $1.01 $
 & $0.395 $
 \\
 \hline
 $p_T= 60 $
 & $13.2 $
 & $11.8 $
 & $10.1 $
 & $8 $
 & $5.7 $
 & $3.48 $
 & $1.66 $
 & $0.53 $
 \\
 \hfill $10^{ -6}\times$
 & $17.8 $
 & $15.8 $
 & $13.5 $
 & $10.6 $
 & $7.5 $
 & $4.52 $
 & $2.18 $
 & $0.71 $
 \\
 & $24.3 $
 & $21.5 $
 & $18.3 $
 & $14.3 $
 & $10 $
 & $6.1 $
 & $2.97 $
 & $0.98 $
 \\
 \hline
 $p_T= 80 $
 & $2.77 $
 & $2.42 $
 & $2.01 $
 & $1.5 $
 & $0.98 $
 & $0.53 $
 & $0.207 $
 & $0.0449 $
 \\
 \hfill $10^{ -6}\times$
 & $3.59 $
 & $3.12 $
 & $2.59 $
 & $1.95 $
 & $1.28 $
 & $0.69 $
 & $0.273 $
 & $0.06 $
 \\
 & $4.76 $
 & $4.14 $
 & $3.45 $
 & $2.6 $
 & $1.72 $
 & $0.94 $
 & $0.376 $
 & $0.083 $
 \\
 \hline
\end{tabular}
\end{center}
\end{table}
\begin{table}[htb]
\begin{center}
\caption{As in table \ref{bprod:tab:diff1}, for $y$ from 4.5 to 6.5.
}
\label{bprod:tab:diff2}
\vskip0.2cm
\begin{tabular}{|l|c|c|c|c|c|}
\hline
$y$ &4.5& 5&5.5& 6&6.5\\
\hline
 $p_T=1 $
 & $4.51 $
 & $3.46 $
 & $2.36 $
 & $1.36 $
 & $0.59 $
 \\
 \hfill $10^{ -1}\times$
 & $11.6 $
 & $8.4 $
 & $5.3 $
 & $2.87 $
 & $1.15 $
 \\
 & $20 $
 & $14.2 $
 & $8.5 $
 & $4.47 $
 & $1.8 $
 \\
 \hline
 $p_T=2 $
 & $3.74 $
 & $2.82 $
 & $1.88 $
 & $1.04 $
 & $0.431 $
 \\
 \hfill $10^{ -1}\times$
 & $9 $
 & $6.4 $
 & $4.04 $
 & $2.1 $
 & $0.8 $
 \\
 & $14.6 $
 & $10.2 $
 & $6.3 $
 & $3.2 $
 & $1.23 $
 \\
 \hline
 $p_T=3 $
 & $2.8 $
 & $2.06 $
 & $1.32 $
 & $0.69 $
 & $0.261 $
 \\
 \hfill $10^{ -1}\times$
 & $6.2 $
 & $4.35 $
 & $2.65 $
 & $1.31 $
 & $0.458 $
 \\
 & $10.1 $
 & $6.8 $
 & $4.07 $
 & $1.97 $
 & $0.7 $
 \\
 \hline
 $p_T=4 $
 & $1.93 $
 & $1.37 $
 & $0.84 $
 & $0.415 $
 & $0.137 $
 \\
 \hfill $10^{ -1}\times$
 & $4.08 $
 & $2.78 $
 & $1.62 $
 & $0.75 $
 & $0.231 $
 \\
 & $6.7 $
 & $4.44 $
 & $2.53 $
 & $1.15 $
 & $0.35 $
 \\
 \hline
 $p_T=5 $
 & $1.27 $
 & $0.87 $
 & $0.51 $
 & $0.233 $
 & $0.065 $
 \\
 \hfill $10^{ -1}\times$
 & $2.52 $
 & $1.67 $
 & $0.93 $
 & $0.404 $
 & $0.105 $
 \\
 & $4.19 $
 & $2.71 $
 & $1.5 $
 & $0.63 $
 & $0.16 $
 \\
 \hline
 $p_T=7 $
 & $5.1 $
 & $3.24 $
 & $1.71 $
 & $0.65 $
 & $0.12 $
 \\
 \hfill $10^{ -2}\times$
 & $9.3 $
 & $5.8 $
 & $2.93 $
 & $1.06 $
 & $0.178 $
 \\
 & $15.4 $
 & $9.4 $
 & $4.66 $
 & $1.63 $
 & $0.268 $
 \\
 \hline
 $p_T=10 $
 & $1.29 $
 & $0.75 $
 & $0.333 $
 & $0.09 $
 & $6.4\,10^{ -3} $
 \\
 \hfill $10^{ -2}\times$
 & $2.2 $
 & $1.23 $
 & $0.53 $
 & $0.134 $
 & $8.8\,10^{ -3} $
 \\
 & $3.54 $
 & $1.95 $
 & $0.81 $
 & $0.2 $
 & $0.0128 $
 \\
 \hline
 $p_T=15 $
 & $1.86 $
 & $0.88 $
 & $0.279 $
 & $0.0324 $
 & $1.2\,10^{ -4} $
 \\
 \hfill $10^{ -3}\times$
 & $2.86 $
 & $1.31 $
 & $0.395 $
 & $0.0431 $
 & $1.57\,10^{ -4} $
 \\
 & $4.36 $
 & $1.95 $
 & $0.57 $
 & $0.061 $
 & $2.25\,10^{ -4} $
 \\
 \hline
 $p_T=20 $
 & $0.371 $
 & $0.143 $
 & $0.0288 $
 & $8.8\,10^{ -4} $
 & $1.17\,10^{ -12} $
 \\
 \hfill $10^{ -3}\times$
 & $0.54 $
 & $0.199 $
 & $0.0379 $
 & $1.18\,10^{ -3} $
 & $1.44\,10^{ -12} $
 \\
 & $0.78 $
 & $0.281 $
 & $0.053 $
 & $1.69\,10^{ -3} $
 & $1.57\,10^{ -12} $
 \\
 \hline
 $p_T=30 $
 & $0.278 $
 & $0.062 $
 & $3.28\,10^{ -3} $
 & $1.14\,10^{ -7} $
 & $0 $
 \\
 \hfill $10^{ -4}\times$
 & $0.369 $
 & $0.082 $
 & $4.39\,10^{ -3} $
 & $1.55\,10^{ -7} $
 & $0 $
 \\
 & $0.5 $
 & $0.114 $
 & $6.2\,10^{ -3} $
 & $2.02\,10^{ -7} $
 & $0 $
 \\
 \hline
 $p_T=40 $
 & $0.316 $
 & $0.036 $
 & $2.48\,10^{ -4} $
 & $0 $
 & $0 $
 \\
 \hfill $10^{ -5}\times$
 & $0.418 $
 & $0.0481 $
 & $3.24\,10^{ -4} $
 & $0 $
 & $0 $
 \\
 & $0.58 $
 & $0.067 $
 & $4.43\,10^{ -4} $
 & $0 $
 & $0 $
 \\
 \hline
 $p_T=50 $
 & $0.0457 $
 & $2.25\,10^{ -3} $
 & $6.4\,10^{ -8} $
 & $0 $
 & $0 $
 \\
 \hfill $10^{ -5}\times$
 & $0.061 $
 & $3.02\,10^{ -3} $
 & $8.2\,10^{ -8} $
 & $0 $
 & $0 $
 \\
 & $0.085 $
 & $4.2\,10^{ -3} $
 & $1.07\,10^{ -7} $
 & $0 $
 & $0 $
 \\
 \hline
 $p_T= 60 $
 & $0.076 $
 & $1.23\,10^{ -3} $
 & $0 $
 & $0 $
 & $0 $
 \\
 \hfill $10^{ -6}\times$
 & $0.102 $
 & $1.64\,10^{ -3} $
 & $0 $
 & $0 $
 & $0 $
 \\
 & $0.143 $
 & $2.25\,10^{ -3} $
 & $0 $
 & $0 $
 & $0 $
 \\
 \hline
 $p_T= 80 $
 & $2.48\,10^{ -3} $
 & $2.93\,10^{ -7} $
 & $0 $
 & $0 $
 & $0 $
 \\
 \hfill $10^{ -6}\times$
 & $3.29\,10^{ -3} $
 & $3.9\,10^{ -7} $
 & $0 $
 & $0 $
 & $0 $
 \\
 & $4.56\,10^{ -3} $
 & $4.78\,10^{ -7} $
 & $0 $
 & $0 $
 & $0 $
 \\
 \hline
\end{tabular}
\end{center}
\end{table}
\begin{table}[htb]
\begin{center}
\caption{As in table \ref{bprod:tab:diff1}, for $p_T$ from 100 to 300.}
\label{bprod:tab:diff3}
\vskip0.2cm
\begin{tabular}{|l|c|c|c|c|c|c|c|c|c|}
\hline
$y$ & 0& 1&1.5& 2&2.5& 3
 &3.5& 4&4.5\\
\hline
 $p_T= 100 $
 & $7.8 $
 & $6.6 $
 & $5.3 $
 & $3.82 $
 & $2.33 $
 & $1.11 $
 & $0.354 $
 & $0.0484 $
 & $6.9\,10^{ -4} $
 \\
 \hfill $10^{ -7}\times$
 & $10 $
 & $8.5 $
 & $6.9 $
 & $4.97 $
 & $3.05 $
 & $1.47 $
 & $0.469 $
 & $0.064 $
 & $9 \,10^{ -4} $
 \\
 & $13.3 $
 & $11.4 $
 & $9.3 $
 & $6.7 $
 & $4.14 $
 & $2.02 $
 & $0.65 $
 & $0.09 $
 & $1.23\,10^{ -3} $
 \\
 \hline
 $p_T= 120 $
 & $2.66 $
 & $2.22 $
 & $1.75 $
 & $1.2 $
 & $0.69 $
 & $0.291 $
 & $0.073 $
 & $5.8\,10^{ -3} $
 & $9.5\,10^{ -6} $
 \\
 \hfill $10^{ -7}\times$
 & $3.44 $
 & $2.89 $
 & $2.28 $
 & $1.57 $
 & $0.9 $
 & $0.387 $
 & $0.098 $
 & $7.7\,10^{ -3} $
 & $1.11\,10^{ -5} $
 \\
 & $4.59 $
 & $3.87 $
 & $3.06 $
 & $2.13 $
 & $1.23 $
 & $0.53 $
 & $0.135 $
 & $0.0106 $
 & $1.54\,10^{ -5} $
 \\
 \hline
 $p_T= 140 $
 & $10.5 $
 & $8.7 $
 & $6.7 $
 & $4.41 $
 & $2.35 $
 & $0.88 $
 & $0.172 $
 & $7 \,10^{ -3} $
 & $4.61\,10^{ -8} $
 \\
 \hfill $10^{ -8}\times$
 & $13.7 $
 & $11.3 $
 & $8.7 $
 & $5.8 $
 & $3.1 $
 & $1.17 $
 & $0.229 $
 & $9.1\,10^{ -3} $
 & $5.4\,10^{ -8} $
 \\
 & $18.4 $
 & $15.2 $
 & $11.8 $
 & $7.9 $
 & $4.24 $
 & $1.62 $
 & $0.317 $
 & $0.0125 $
 & $6.1\,10^{ -8} $
 \\
 \hline
 $p_T= 160 $
 & $4.68 $
 & $3.79 $
 & $2.84 $
 & $1.81 $
 & $0.9 $
 & $0.296 $
 & $0.0433 $
 & $7.6\,10^{ -4} $
 & $0 $
 \\
 \hfill $10^{ -8}\times$
 & $6.1 $
 & $4.96 $
 & $3.73 $
 & $2.38 $
 & $1.19 $
 & $0.396 $
 & $0.058 $
 & $9.9\,10^{ -4} $
 & $0 $
 \\
 & $8.2 $
 & $6.7 $
 & $5.1 $
 & $3.24 $
 & $1.63 $
 & $0.55 $
 & $0.08 $
 & $1.33\,10^{ -3} $
 & $0 $
 \\
 \hline
 $p_T= 180 $
 & $22.6 $
 & $18 $
 & $13.2 $
 & $8.1 $
 & $3.74 $
 & $1.08 $
 & $0.115 $
 & $6.8\,10^{ -4} $
 & $0 $
 \\
 \hfill $10^{ -9}\times$
 & $29.5 $
 & $23.6 $
 & $17.4 $
 & $10.7 $
 & $4.96 $
 & $1.43 $
 & $0.152 $
 & $8.7\,10^{ -4} $
 & $0 $
 \\
 & $39.8 $
 & $31.9 $
 & $23.6 $
 & $14.6 $
 & $6.8 $
 & $1.99 $
 & $0.21 $
 & $1.13\,10^{ -3} $
 & $0 $
 \\
 \hline
 $p_T= 200 $
 & $11.7 $
 & $9.2 $
 & $6.6 $
 & $3.87 $
 & $1.67 $
 & $0.415 $
 & $0.0312 $
 & $3.29\,10^{ -5} $
 & $0 $
 \\
 \hfill $10^{ -9}\times$
 & $15.3 $
 & $12.1 $
 & $8.7 $
 & $5.1 $
 & $2.21 $
 & $0.55 $
 & $0.0413 $
 & $4.68\,10^{ -5} $
 & $0 $
 \\
 & $20.6 $
 & $16.3 $
 & $11.8 $
 & $7 $
 & $3.03 $
 & $0.76 $
 & $0.056 $
 & $5.4\,10^{ -5} $
 & $0 $
 \\
 \hline
 $p_T= 220 $
 & $6.4 $
 & $4.93 $
 & $3.46 $
 & $1.95 $
 & $0.78 $
 & $0.167 $
 & $8.4\,10^{ -3} $
 & $4.96\,10^{ -7} $
 & $0 $
 \\
 \hfill $10^{ -9}\times$
 & $8.4 $
 & $6.5 $
 & $4.58 $
 & $2.59 $
 & $1.04 $
 & $0.223 $
 & $0.0111 $
 & $6.5\,10^{ -7} $
 & $0 $
 \\
 & $11.3 $
 & $8.8 $
 & $6.2 $
 & $3.54 $
 & $1.43 $
 & $0.307 $
 & $0.015 $
 & $7.1\,10^{ -7} $
 & $0 $
 \\
 \hline
 $p_T= 240 $
 & $3.66 $
 & $2.79 $
 & $1.91 $
 & $1.04 $
 & $0.385 $
 & $0.07 $
 & $2.24\,10^{ -3} $
 & $2.82\,10^{ -10} $
 & $0 $
 \\
 \hfill $10^{ -9}\times$
 & $4.8 $
 & $3.67 $
 & $2.52 $
 & $1.37 $
 & $0.51 $
 & $0.092 $
 & $2.92\,10^{ -3} $
 & $8 \,10^{ -10} $
 & $0 $
 \\
 & $6.5 $
 & $4.99 $
 & $3.45 $
 & $1.88 $
 & $0.7 $
 & $0.127 $
 & $3.91\,10^{ -3} $
 & $9.7\,10^{ -10} $
 & $0 $
 \\
 \hline
 $p_T= 260 $
 & $21.8 $
 & $16.3 $
 & $11 $
 & $5.7 $
 & $1.96 $
 & $0.298 $
 & $5.7\,10^{ -3} $
 & $0 $
 & $0 $
 \\
 \hfill $10^{ -10}\times$
 & $28.7 $
 & $21.6 $
 & $14.5 $
 & $7.6 $
 & $2.61 $
 & $0.396 $
 & $7.4\,10^{ -3} $
 & $0 $
 & $0 $
 \\
 & $38.8 $
 & $29.3 $
 & $19.8 $
 & $10.3 $
 & $3.58 $
 & $0.54 $
 & $9.6\,10^{ -3} $
 & $0 $
 & $0 $
 \\
 \hline
 $p_T= 280 $
 & $13.4 $
 & $9.9 $
 & $6.5 $
 & $3.24 $
 & $1.02 $
 & $0.13 $
 & $1.32\,10^{ -3} $
 & $0 $
 & $0 $
 \\
 \hfill $10^{ -10}\times$
 & $17.7 $
 & $13.1 $
 & $8.6 $
 & $4.32 $
 & $1.37 $
 & $0.172 $
 & $1.73\,10^{ -3} $
 & $0 $
 & $0 $
 \\
 & $23.9 $
 & $17.8 $
 & $11.8 $
 & $5.9 $
 & $1.87 $
 & $0.234 $
 & $2.18\,10^{ -3} $
 & $0 $
 & $0 $
 \\
 \hline
 $p_T= 300 $
 & $8.5 $
 & $6.2 $
 & $3.98 $
 & $1.9 $
 & $0.55 $
 & $0.057 $
 & $2.7\,10^{ -4} $
 & $0 $
 & $0 $
 \\
 \hfill $10^{ -10}\times$
 & $11.2 $
 & $8.2 $
 & $5.3 $
 & $2.52 $
 & $0.73 $
 & $0.076 $
 & $3.67\,10^{ -4} $
 & $0 $
 & $0 $
 \\
 & $15.2 $
 & $11.1 $
 & $7.2 $
 & $3.46 $
 & $1.01 $
 & $0.103 $
 & $4.4\,10^{ -4} $
 & $0 $
 & $0 $
 \\
 \hline
\end{tabular}
\end{center}
\end{table}
\begin{table}[htb]
\begin{center}
\caption{As in table \ref{bprod:tab:diff1}, for $p_T$ from 320 to 500.
In the entries with a $*$ the cross section is too small to be computed
reliably.}
\label{bprod:tab:diff4}
\vskip0.2cm
\begin{tabular}{|l|c|c|c|c|c|c|c|}
\hline
$y$ & 0& 1&1.5& 2&2.5& 3
 &3.5\\
\hline
 $p_T= 320 $
 & $5.5 $
 & $3.97 $
 & $2.5 $
 & $1.14 $
 & $0.304 $
 & $0.0256 $
 & $4.36\,10^{ -5} $
 \\
 \hfill $10^{ -10}\times$
 & $7.3 $
 & $5.2 $
 & $3.3 $
 & $1.51 $
 & $0.403 $
 & $0.0334 $
 & $6.5\,10^{ -5} $
 \\
 & $9.9 $
 & $7.2 $
 & $4.52 $
 & $2.08 $
 & $0.55 $
 & $0.0453 $
 & $7.3\,10^{ -5} $
 \\
 \hline
 $p_T= 340 $
 & $3.67 $
 & $2.6 $
 & $1.6 $
 & $0.7 $
 & $0.17 $
 & $0.0114 $
 & $5.1\,10^{ -6} $
 \\
 \hfill $10^{ -10}\times$
 & $4.85 $
 & $3.44 $
 & $2.12 $
 & $0.93 $
 & $0.226 $
 & $0.0149 $
 & $7.2\,10^{ -6} $
 \\
 & $6.6 $
 & $4.69 $
 & $2.89 $
 & $1.27 $
 & $0.309 $
 & $0.0199 $
 & $7.7\,10^{ -6} $
 \\
 \hline
 $p_T= 360 $
 & $2.49 $
 & $1.74 $
 & $1.04 $
 & $0.437 $
 & $0.097 $
 & $5.1\,10^{ -3} $
 & $3.56\,10^{ -7} $
 \\
 \hfill $10^{ -10}\times$
 & $3.29 $
 & $2.31 $
 & $1.38 $
 & $0.58 $
 & $0.128 $
 & $6.6\,10^{ -3} $
 & $4.33\,10^{ -7} $
 \\
 & $4.47 $
 & $3.13 $
 & $1.89 $
 & $0.79 $
 & $0.175 $
 & $8.7\,10^{ -3} $
 & $4.63\,10^{ -7} $
 \\
 \hline
 $p_T= 380 $
 & $17.1 $
 & $11.8 $
 & $6.9 $
 & $2.78 $
 & $0.56 $
 & $0.0223 $
 & $6.4\,10^{ -8} $
 \\
 \hfill $10^{ -11}\times$
 & $22.8 $
 & $15.7 $
 & $9.2 $
 & $3.7 $
 & $0.74 $
 & $0.0288 $
 & $1.24\,10^{ -7} $
 \\
 & $30.9 $
 & $21.4 $
 & $12.6 $
 & $5.1 $
 & $1 $
 & $0.038 $
 & $1.46\,10^{ -7} $
 \\
 \hline
 $p_T= 400 $
 & $12 $
 & $8.2 $
 & $4.69 $
 & $1.79 $
 & $0.326 $
 & $9.7\,10^{ -3} $
 &
 \\
 \hfill $10^{ -11}\times$
 & $16 $
 & $10.8 $
 & $6.2 $
 & $2.38 $
 & $0.43 $
 & $0.0124 $
 & $*$
 \\
 & $21.7 $
 & $14.8 $
 & $8.5 $
 & $3.26 $
 & $0.58 $
 & $0.0163 $
 &
 \\
 \hline
 $p_T= 420 $
 & $8.6 $
 & $5.8 $
 & $3.21 $
 & $1.17 $
 & $0.1 92 $
 & $4.16\,10^{ -3} $
 &
 \\
 \hfill $10^{ -11}\times$
 & $11.3 $
 & $7.6 $
 & $4.25 $
 & $1.55 $
 & $0.251 $
 & $5.2\,10^{ -3} $
 & $*$
 \\
 & $15.5 $
 & $10.4 $
 & $5.8 $
 & $2.13 $
 & $0.342 $
 & $6.8\,10^{ -3} $
 &
 \\
 \hline
 $p_T= 440 $
 & $6.2 $
 & $4.09 $
 & $2.23 $
 & $0.78 $
 & $0.114 $
 & $1.73\,10^{ -3} $
 & $0 $
 \\
 \hfill $10^{ -11}\times$
 & $8.2 $
 & $5.4 $
 & $2.96 $
 & $1.03 $
 & $0.149 $
 & $2.14\,10^{ -3} $
 & $0 $
 \\
 & $11.2 $
 & $7.4 $
 & $4.04 $
 & $1.41 $
 & $0.201 $
 & $2.79\,10^{ -3} $
 & $0 $
 \\
 \hline
 $p_T= 460 $
 & $4.53 $
 & $2.94 $
 & $1.57 $
 & $0.52 $
 & $0.068 $
 & $6.9\,10^{ -4} $
 & $0 $
 \\
 \hfill $10^{ -11}\times$
 & $6 $
 & $3.9 $
 & $2.08 $
 & $0.69 $
 & $0.089 $
 & $8.5\,10^{ -4} $
 & $0 $
 \\
 & $8.2 $
 & $5.3 $
 & $2.84 $
 & $0.94 $
 & $0.119 $
 & $1.09\,10^{ -3} $
 & $0 $
 \\
 \hline
 $p_T= 480 $
 & $3.34 $
 & $2.14 $
 & $1.11 $
 & $0.351 $
 & $0.0407 $
 & $2.61\,10^{ -4} $
 & $0 $
 \\
 \hfill $10^{ -11}\times$
 & $4.43 $
 & $2.84 $
 & $1.48 $
 & $0.465 $
 & $0.053 $
 & $3.16\,10^{ -4} $
 & $0 $
 \\
 & $6 $
 & $3.87 $
 & $2.01 $
 & $0.63 $
 & $0.071 $
 & $4.07\,10^{ -4} $
 & $0 $
 \\
 \hline
 $p_T= 500 $
 & $2.49 $
 & $1.57 $
 & $0.8 $
 & $0.239 $
 & $0.0244 $
 & $9.3\,10^{ -5} $
 & $0 $
 \\
 \hfill $10^{ -11}\times$
 & $3.31 $
 & $2.09 $
 & $1.06 $
 & $0.316 $
 & $0.0319 $
 & $1.1\,10^{ -4} $
 & $0 $
 \\
 & $4.51 $
 & $2.85 $
 & $1.44 $
 & $0.429 $
 & $0.0423 $
 & $1.45\,10^{ -4} $
 & $0 $
 \\
 \hline
\end{tabular}
\end{center}
\end{table}
The central values we obtained are also plotted
in figs.~\ref{bprod:fig:diffpt} and \ref{bprod:fig:diffy}, so that the
wide kinematic range of heavy flavour production can be appreciated by
a glance.
More detailed rapidity distributions at low momenta are shown in
fig.~\ref{bprod:fig:diffy1}.
First of all, we see that the differential cross section spans many
orders of magnitude. At a luminosity of $10^{34}{\rm cm}^{-2}{\rm
sec}^{-1}$ each $\mu b$ of cross section corresponds to $10^4$ events
per second, or (roughly) $10^{11}$ events per year. Thus, at the level
of $10^{-11}$ in the plot there should be one event per year per bin
of $p_T$ and $y$.  The $p_T$ spectrum starts to drop fast for $p_T$
larger than the heavy quark mass, dropping even faster as
the threshold region is approached. The rapidity distributions have the
typical shape of a wide plateau, dropping at the edge of the phase space,
and becoming narrower for larger transverse momenta. At the LHC the gluon
fusion production mechanism is dominant, as can bee seen in
fig.~\ref{bprod:fig:totqqqg}.
There one can see that the quark-antiquark annihilation component
is below the gluon fusion component by more than one order of magnitude
in the $p_T$ range considered, while the quark-gluon term becomes more
important at larger $p_T$. We remind the reader that the cross section
for $qq\to b+X$ is not included in the NLO calculation. One may thus worry
about a loss of accuracy in the result, since the quark-quark luminosity
at the LHC are by far the largest for high transverse momenta $b$ production.
This problem, however, is dealt with appropriately in the resummation
formalism for high $p_T$ heavy flavour production, where a quark-quark
fusion contribution does indeed appear.
\clearpage
\subsubsection{Understanding Tevatron data}
\begin{figure}[htb]
\begin{center}
\includegraphics[width=0.6\textwidth,clip]{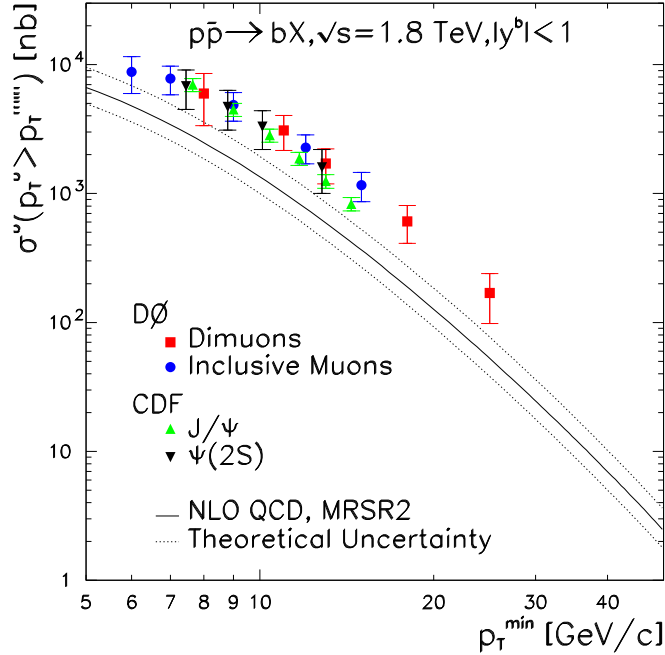}
\caption{The integrated $p_T$ distribution for single $b$ production measured 
at the Tevatron, and the corresponding QCD prediction.}
\label{fig:tevpt}
\end{center}
\end{figure}

It is well known that Tevatron data for the integrated transverse
momentum spectrum in $b$ production are systematically larger than QCD
predictions. This problem has been around for a long time,
although it has become less severe with time.
The present status of this issue is summarized in 
fig.~\ref{fig:tevpt}.
A similar discrepancy is also observed in UA1 data
(see ref.~\cite{bprod:Buras2}
for details).

The theoretical prediction has a considerable uncertainty, which is
mainly due to neglected higher-order terms in the perturbative
expansion. In our opinion, it is not unlikely that we may have to
live with this discrepancy, which is certainly disturbing, but not
strong enough to question the validity of perturbative QCD
calculations. In other words, the QCD ${\cal O}(\as^3)$
corrections for this process are above 100\% of the Born term, and
thus it is not impossible that higher order terms may give
contributions of the same size. Nevertheless, it is useful to look for
higher-order perturbative effects and non-perturbative effects that
may enhance the cross section.

For values of $p_T$ much larger than the $b$ quark mass, large
logarithms of the ratio $p_T/m_b$ arise in the coefficients of the
perturbative expansion.  Techniques are available to resum this class
of logarithms to all orders. In ref.~\cite{bprod:CacciariGrecoNason} the
NLO cross section for the production of a massless parton $i$ (a gluon
or a massless quark) has been folded with the NLO fragmentation
function for the transition $i\to b$ \cite{bprod:MeleNason}. The evolution
of the fragmentation functions resums all terms of order $\as^n
\log^n(p_T/m_b)$ and $\as^{n+1} \log^n (p_T/m_b)$.  All the dependence
on the $b$-quark mass lies in the boundary conditions for the
fragmentation function. The result is then matched with the full NLO
cross section, which contains the exact dependence on $m_b$ up to
order $\as^3$, in a way that avoids double counting.  Corrections to
the result of ref.~\cite{bprod:CacciariGrecoNason} are either of order
$\as^4\log^i(p_T/m_b)$, with $i\le 2$, or of order $\as^4$ times
positive powers of $m_b/\sqrt{p_T^2+m_b^2}$.

\begin{figure}[htb]
\begin{center}
\includegraphics[width=0.6\textwidth,clip]{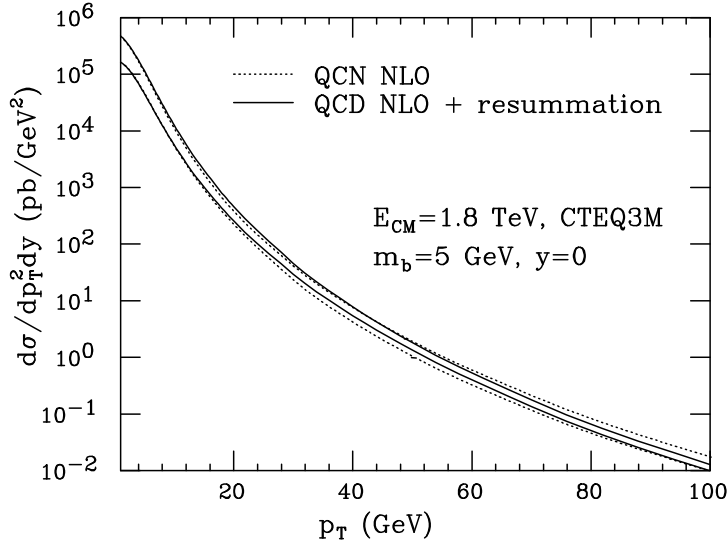}
\caption{Single-inclusive $p_T$ distribution for $b$ production 
at the Tevatron energy: pure QCD and resummed results.}
\label{fig:bandms}
\end{center}
\end{figure}

\begin{figure}[htb]
\begin{center}
\includegraphics[width=0.6\textwidth,clip]{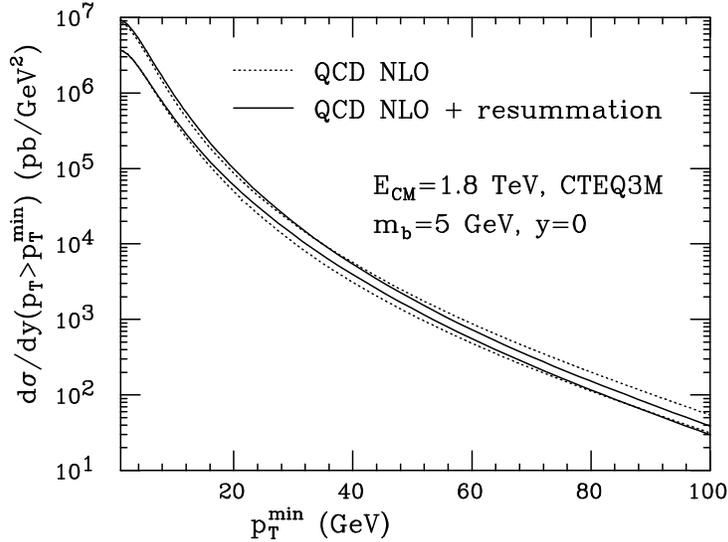}
\caption{Integrated $p_T$ distribution for $b$ production 
at the Tevatron energy: pure QCD and resummed results.}
\label{fig:bandmsint}
\end{center}
\end{figure}

Figures~\ref{fig:bandms}-\ref{fig:bandmsint} show the differential and
integrated $b$-quark $p_T$ distribution obtained in the
fragmentation function approach of ref.~\cite{bprod:CacciariGrecoNason},
compared to the standard fixed-order NLO result. It should be noted
that for high transverse momenta the scale dependence is significantly
reduced with respect to the fixed-order calculation. Furthermore, it can 
be seen from fig.~\ref{fig:bandmsint} that, for $10 \gsim p_T\gsim 30$~GeV,
the result of the fragmentation-function approach lies slightly 
above the fixed-order NLO calculation. 
This has been interpreted in ref.~\cite{bprod:CacciariGrecoNason}
as an evidence for large, positive higher order corrections.
Unfortunately, their effect is not easy to quantify. These higher order
terms are in fact computed in a massless approximation, and thus
fail at low transverse momenta.
In figs.~\ref{fig:bandms}-\ref{fig:bandmsint} these terms are suppressed by
a factor that becomes smaller and smaller at low $\pt$.
A more detailed discussion of this point can be found in the original
reference. Here, we simply conclude that some evidence for large higher
order terms in the intermediate transverse momentum region is present,
although difficult to quantify.

Finally, notice that the overall
effect of the inclusion of higher-order logarithms is a steepening of
the $p_T$ spectrum. This is quite natural, since multiple radiation
is accounted for in the resummation procedure.

It has been argued that an intrinsic transverse momentum
for the incoming partons may explain the discrepancy observed at the Tevatron.
In fact, large values (up to 4 GeV) of the average transverse momentum of the 
incoming partons have been invoked to explain direct photon production data
\cite{bprod:Huston}. Such large values, much larger than typical QCD
scales, are clearly incompatible with the usual application of perturbative
QCD. Thus, evidence for such a large
intrinsic transverse momentum cannot be claimed on the basis of
a single observable. In other words, we would need evidence from
several observables, all leading to a similar value of the intrinsic
$k_T$, before we accept such a flaw in the usual perturbative QCD description.
Nevertheless, in the following we will perform the exercise of applying
very large intrinsic transverse momenta to the heavy flavour production
process.
This procedure will lead to an increase in the $b$ transverse momentum
spectrum. We will also show, however, that other variables, very sensitive
to an intrinsic transverse momentum, that should be strongly affected,
do not show any evidence of that.

There are several possible ways to implement the presence of a
non-zero transverse momentum of the colliding partons, and the choice
is, to a large extent, arbitrary.  We implemented it in the FMNR code
in the following way.  We call $\vec{p}_{\sss T}(Q\overline{Q})$ the
total transverse momentum of the pair.  For each event, in the
longitudinal centre-of-mass frame of the heavy-quark pair, we boost
the $Q\overline{Q}$ system to rest.  We then perform a transverse
boost, which gives the pair a transverse momentum equal to
$\vec{p}_{\sss T}(Q\overline{Q})+ \vec{k}_{\sss T}(1)+\vec{k}_{\sss
T}(2)$; $\vec{k}_{\sss T}(1)$ and $\vec{k}_{\sss T}(2)$ are the
transverse momenta of the incoming partons, which are chosen randomly,
with their moduli distributed according to
\begin{equation}
\frac{1}{N}\frac{dN}{dk_{\sss T}^2}=\frac{1}{\langle k_{\sss T}^2 \rangle}
      \exp(-k_{\sss T}^2/\langle k_{\sss T}^2 \rangle).
\end{equation}
The reader can find more details in ref.~\cite{bprod:Buras2}.

\begin{figure}[htb]
\begin{center}
\includegraphics[width=0.6\textwidth,clip]{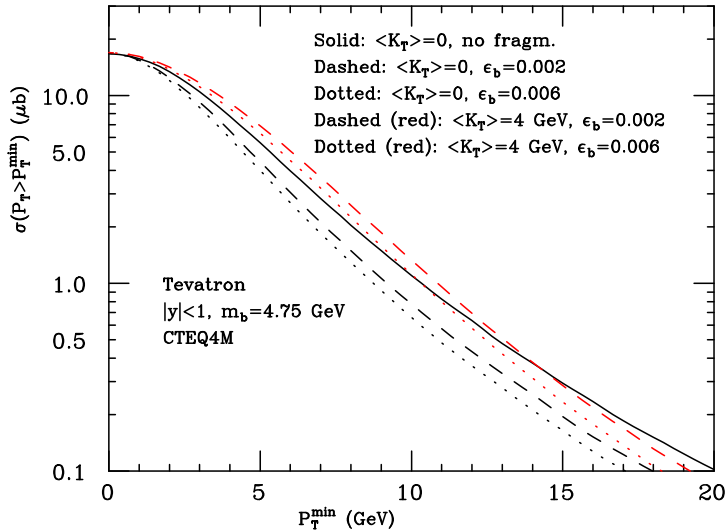}
\caption{
The $b$ cross section at the Tevatron:
the effect of a large intrinsic transverse momentum,
and the sensitivity to the fragmentation parameter $\epsilon_b$.}
\label{fig:bkick}
\end{center}
\end{figure}

In fig.~\ref{fig:bkick} we show the effect
of an intrinsic $k_T$ generated in this way, with the (unphysically large)
choice
$\langle k_{\sss T}\rangle = 4$~ GeV
(in fig.~\ref{fig:bkick}, the sensitivity to the
$\epsilon_b$ parameter in the fragmentation function is also shown;
fragmentation will be discused in more detail in the next subsection.)
We see that, for $p_T^{min}<20$ GeV, the $k_T$ effect is sizeable,
even in the presence of fragmentation, provided we allow for
unphysically large intrinsic $k_T$.  

\begin{figure}[htb]
\begin{center}
\includegraphics[width=0.6\textwidth,clip]{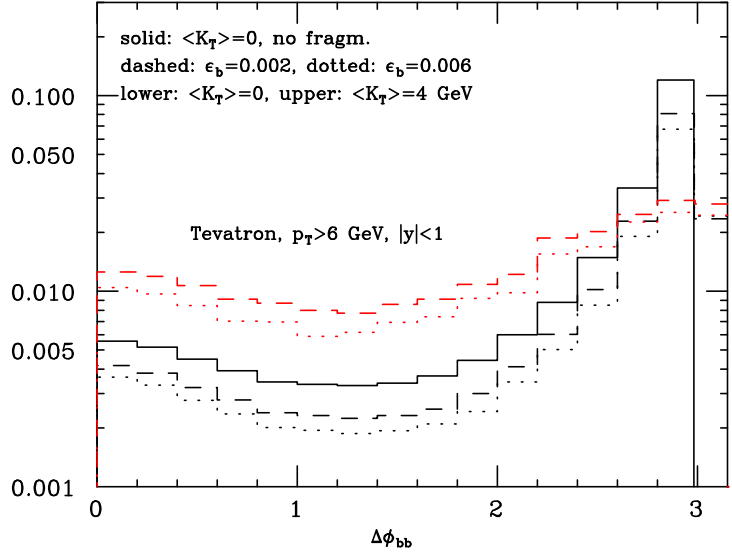}
\caption{The $b$-$\bar{b}$ azimuthal correlation at the Tevatron:
the effect of a large intrinsic transverse momentum,
and the sensitivity to the fragmentation parameter $\epsilon_b$.}
\label{fig:bazi}
\end{center}
\end{figure}

\begin{figure}[htb]
\begin{center}
\includegraphics[width=0.6\textwidth,clip]{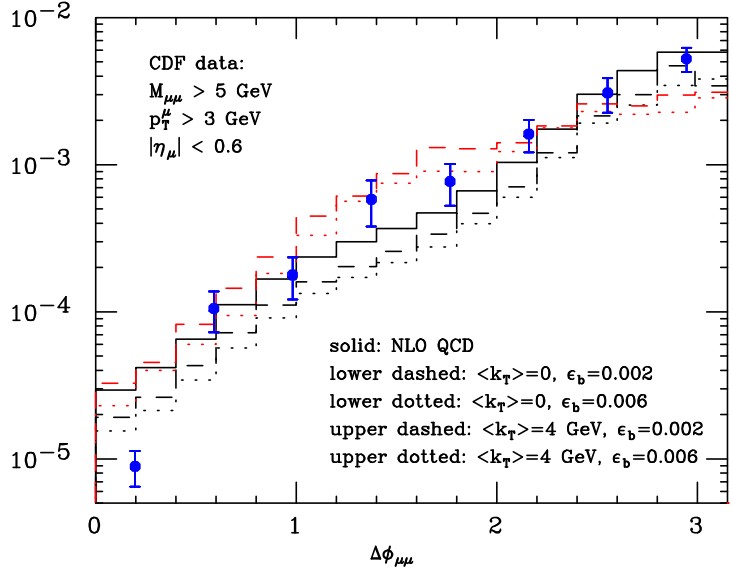}
\caption{CDF results on azimuthal correlations compared with
the perturbative calculation, with and without intrinsic $k_T$.}
\label{fig:cdfazi}
\end{center}
\end{figure}

\begin{figure}[htb]
\begin{center}
\includegraphics[width=0.6\textwidth,clip]{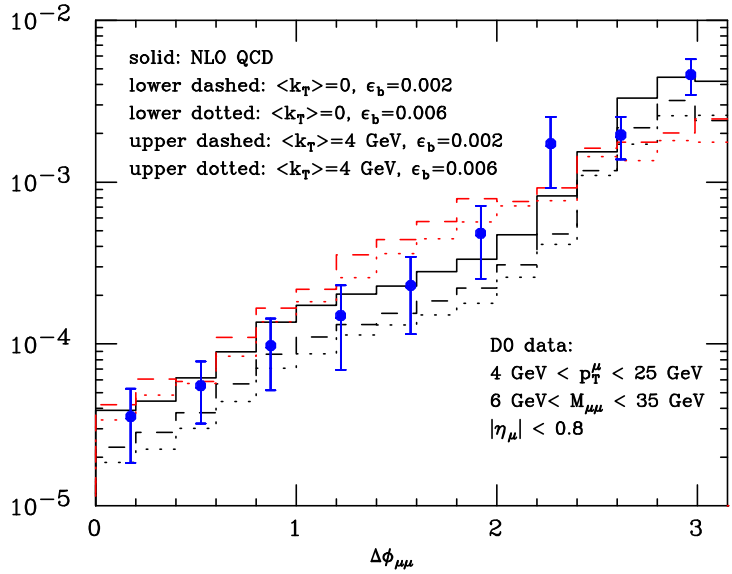}
\caption{D0 results on azimuthal correlations compared with
the perturbative calculation, with and without intrinsic $k_T$}
\label{fig:d0azi}
\end{center}
\end{figure}

It is fair to ask whether such large values are
compatible with other observables. There is a particular class of
observables that are particularly sensitive to the intrinsic
transverse momentum.  One example is the azimuthal distance
$\Delta\phi$ between the directions of the produced $b$ and
$\overline{b}$.  The $\Delta\phi$ distribution is trivial at leading
order: $b$ and $\bar{b}$ are emitted back-to-back, and therefore
\begin{equation}
\frac{d\sigma}{d\Delta\phi}\propto \delta(\phi-\pi).
\end{equation}
An intrinsic $k_T$ of the colliding partons has the effect of
smearing the $\delta$ function. For
$\langle k_T\rangle=4\;$GeV the effect is quite dramatic, as can be seen in
fig.~\ref{fig:bazi}.
Is such an important effect consistent with the observed
azimuth correlations? 
The CDF and D0 collaborations have measured the azimuthal correlation
of muon pairs produced in $b$ decays.
In order to compare
with these data sets, we have implemented in the FMNR code the semileptonic
decay of $b$ quarks. We have assumed that the muon energy is distributed
according to the prediction of the spectator model~\cite{bprod:CCM}
with massless leptons. We have also checked that the muon energy distribution
given by \Py\ leads to similar results.
Our results are shown in figs.~\ref{fig:cdfazi} and \ref{fig:d0azi},
where CDF and D0 data are superimposed to the perturbative QCD
prediction, with and without an intrinsic $k_T$ with
$\langle k_T\rangle=4\;$GeV. Tevatron data do not seem to favour such a large
intrinsic transverse momentum. The measured distributions
are more peaked at $\Delta\phi=\pi$ than the theoretical curve
with $\langle k_T\rangle=4\;$GeV. The effect of Peterson
fragmentation is also shown in both cases.
We thus conclude that the data does not seem to favour large $k_T$ effects.

\subsubsection{Single-incusive distributions and correlations at the LHC}
\begin{figure}[htb]
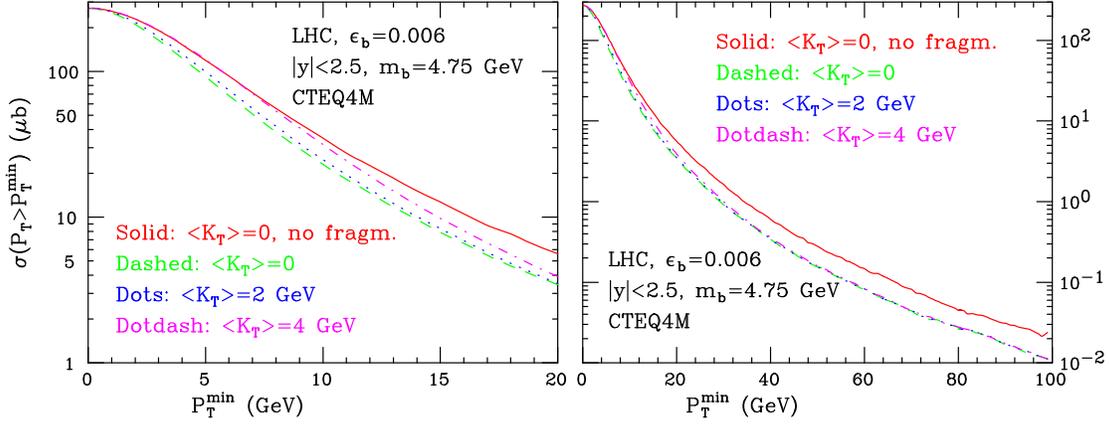

\begin{center}
\includegraphics[height=0.35\textwidth,clip]{ptlow_lhc.eps}
\includegraphics[height=0.35\textwidth,clip]{pthigh_lhc.eps}
\caption{Cross section with a transverse momentum cut at the LHC.}
\label{fig:ptlow_lhc}
\end{center}
\end{figure}

\begin{figure}[htb]
\begin{center}
\includegraphics[width=0.6\textwidth,clip]{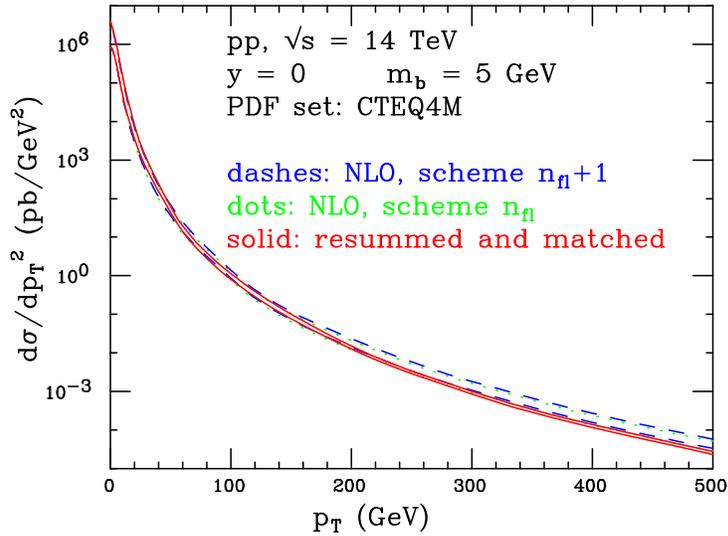}
\caption{Differential cross section for $b$ production at the LHC.
The bands are obtained by varying independently the renormalization and
factorization scales by a factor of 2 above and below the central
value, which is the $b$ transverse  mass.}
\label{fig:cacciari-lhc}
\end{center}
\end{figure}

In this subsection, we will follow the pragmatic assumption
that the discrepancy observed at the Tevatron may either be attributed
to a problem in the overall normalization of the cross section, or to
the presence of effects, either perturbative or not, that distort the
spectrum. We will continue to model these effects as fragmentation
effects and intrinsic transverse momentum effects, and see if the LHC
can distinguish among the two. In fig.~\ref{fig:ptlow_lhc} we plot
the $b$ cross section with a transverse momentum cut.
From the figure it is quite clear that the effects of fragmentation
and the effects of an intrinsic transverse momentum kick manifest
themselves in quite a different way. In particular, at $\pt>20$~GeV
even the effect of a very large transverse momentum kick is small,
while fragmentation has a strong impact. On the other hand, the transverse
momentum kick increases the cross section in the intermediate $\pt$ region,
with a maximum around 7~GeV. The $\pt$ coverage offered by the
combined LHC experiments will allow an effective discrimination
of the two kinds of effects. For completeness, we also show in
fig.~\ref{fig:cacciari-lhc} a comparison of the fixed-order calculation of the
single-inclusive spectrum, using the fixed-order calculation in two
different schemes for the light flavour, and the matched-resummed
result.
As in the Tevatron case, the band obtained with the resummation procedure
is much narrower at large transverse momentum. The corresponding uncertainty
does not include fragmentation function uncertainties, that will be discussed
in more detail further on.

\begin{figure}[htb]
\begin{center}
\includegraphics[width=0.6\textwidth,clip]{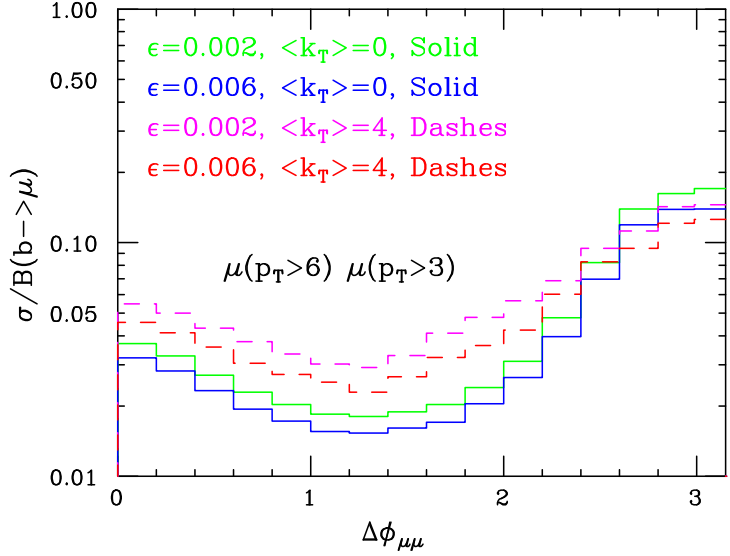}
\caption{Azimuthal correlations in muon pairs at the LHC.}
\label{fig:rebinned_lhcazi1}
\end{center}
\end{figure}

As an example of what could be discriminated at the LHC using correlations,
we present in fig.~\ref{fig:rebinned_lhcazi1} the azimuthal correlation
of the muons coming from semileptonic $B$ decays, using typical
cuts that are implemented in the LHC experiments triggers for $B$
studies.
The curves are obtained with different values of the $\epsilon_b$ parameter
for the fragmentation function, and with or without a very large intrinsic
transverse momentum for the incoming partons. As one can expect,
the $\epsilon_b$ parameter affects only the total rate in this case,
while the primordial transverse momentum has a considerable
effect on the shape of the distribution. This example
shows that, even with very simple experimental setup, at the LHC it will be
possible to test important features of the differential distributions.

\subsection{Fragmentation function formalism}
In analogy with the case of charm production, the agreement
between theory and data improves if one does not include any fragmentation
effects. It is then natural to ask whether the fragmentation
functions commonly used in these calculations are appropriate.
Following the LEP measurements, fragmentation functions have appeared
to be harder then previously thought. It will be interesting to see whether
SLD new data \cite{bprod:SLDfrag} will help in clarifying this issue.

The effect of a non-perturbative fragmentation function on the $p_T$ spectrum
is easily quantified if one assumes a steeply-falling transverse momentum
distribution for the produced $b$ quark
\begin{equation}
\frac{d\sigma}{d p_T} = A p_T^{-M}\,,
\end{equation}
The corresponding distribution for the hadron is
\begin{equation}
\frac{d\sigma_{\rm had}}{d p_T} = A \int {\hat p}_T^{-M}\,
       \delta(p_T-z{\hat p}_T) D(z)\, dz \,d{\hat p}_T
=A  p_T^{-M} \int_0^1 dz z^{M-1} D(z)\,.
\end{equation}
We can see that the hadron spectrum is proportional to the quark
spectrum times the $M^{\rm th}$ moment of the fragmentation function $D(z)$.
Thus, the larger the moment,
the larger the enhancement of the spectrum.

\begin{figure}
\begin{center}
    \includegraphics[width=0.6\textwidth,clip]{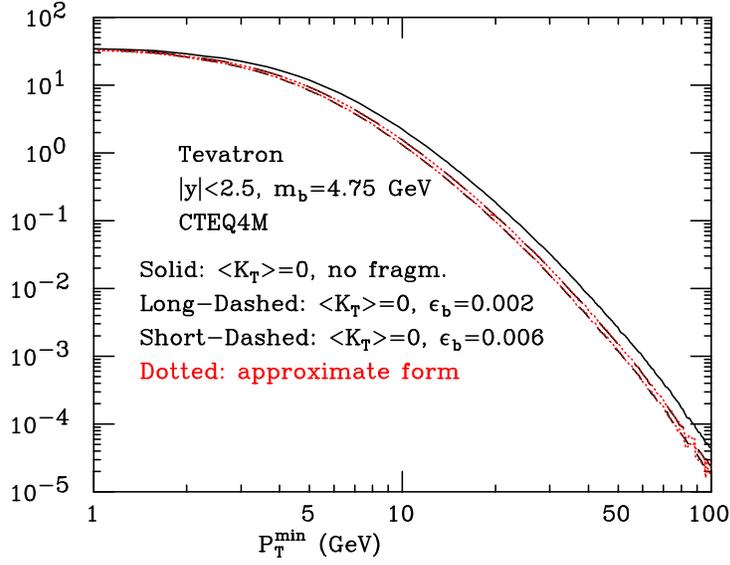}
    \caption{
The effect of a Peterson fragmentation function of the inclusive $b$
cross section. The (red) dotted lines correspond to the approximation
eq.~(\ref{apprfragmeff}), and are almost indistinguishable from
the exact results.}
    \label{fig:frageff}
\end{center}
\end{figure}

In practice, the value of $M$ will be slightly dependent upon $p_T$.
We thus define a $p_T$ dependent $M$ value
\begin{equation}\label{apprfragmeff}
\frac{d\log\sigma(p_T>p_T^{\rm cut})}{d\log p_T^{\rm cut}}=
-M(p_T^{\rm cut})+1
\end{equation}
and
\begin{equation}
\sigma_{\rm had}(p_T>p_T^{\rm cut})
=\sigma(p_T>p_T^{\rm cut})\times
\int_0^1 dz z^{M(p_T^{\rm cut})-1} D(z)\,.
\end{equation}
This gives an excellent approximation to the effect of the
fragmentation function, as can be seen from fig.~\ref{fig:frageff}.

Since the second moment of the fragmentation function is well
constrained by $e^+e^-$ data, it is sensible to ask for what shapes of
the fragmentation function, for fixed $\langle z \rangle$, one gets the
highest value for $\langle z^{M-1} \rangle$. We convinced ourselves that
the maximum is achieved by the functional form
\begin{equation}
D(z) = A\delta(z)+B\delta(1-z)
\end{equation}
which gives
\begin{equation}
\langle z\rangle = \frac{B}{A+B}\,;\quad
\langle z^{M-1}\rangle = \frac{B}{A+B}\,.
\end{equation}
This is however not very realistic: somehow, we expect a
fragmentation function which is concentrated at high values of $z$,
and has a tail at small $z$. We convinced ourselves that, if we impose
the further constraint that $D(z)$ should be monotonically increasing,
one gets instead the functional form
\begin{equation}
\label{maxform}
D(z) = A+B\delta(1-z)\,,
\end{equation}
which gives
\begin{equation}
\langle z\rangle = \frac{A/2 + B}{A+B}\,;\quad
\langle z^{M-1}\rangle = \frac{A/M + B}{A+B}\,.
\end{equation}
We computed numerically the $M^{th}$ moments of the Peterson form,
\begin{equation}
D(z)\propto \frac{1}{z\left(1-\frac{1}{z}-\frac{\epsilon}{1-z}\right)^2}
\end{equation}
of the form
\begin{equation}
\label{powerform}
D(z) \propto z^\alpha (1-z)^\beta
\end{equation}
for $\beta=1$ (Kartvelishvili), for which
\begin{equation}
\langle z^{M-1} \rangle =\frac{\Gamma(\alpha+M)\Gamma(\alpha+\beta+2)}
{\Gamma(\alpha+1)\Gamma(\alpha+\beta+M+1)}\,,
\end{equation}
of the form of Collins and Spiller
\begin{equation}
D(z)\propto\frac{
\left(\frac{1-z}{z}+\frac{(2-z)\epsilon}{1-z}\right)\left(1+z^2\right)}
{\left(1-\frac{1}{z}-\frac{\epsilon}{1-z}\right)^2}
\end{equation}
and of the form in eq.~(\ref{maxform}),
at fixed values of $\langle z \rangle$ corresponding to the
choices $\epsilon_b=0.002$ and $0.006$ in the Peterson form.
We found that the $p_T$ distribution at the Tevatron, for $p_T$ 
in the range $10$ to $100$ GeV, behaves like $p_T^{-M}$, with $M$ around 5.
Therefore, we present in tables~\ref{tabfrag1} and \ref{tabfrag2}
values of the $4^{\rm th}$,
$5^{\rm th}$ and $6^{\rm th}$ moments of the above-mentioned
fragmentation functions.
\begin{table}[htbp]
\begin{center}
\caption{Values of the $4^{\rm th}$, $5^{\rm th}$ and $6^{\rm th}$ moment,
at fixed $\langle z \rangle$ (corresponding to $\epsilon_b=0.002$
in the Peterson form), for different forms of the fragmentation function.}
\label{tabfrag1}
\leavevmode
\begin{tabular}{|l|r|r|r|}
 \hline
 $\langle z \rangle=0.879$ & $M=4$ & $M=5$ & $M=6$ \\
\hline
Peterson  & 0.711  &  0.649 & 0.595 \\
\hline
Kartvelishvili  & 0.694  &  0.622 & 0.562 \\
\hline
Collins-Spiller & 0.729  &  0.677 & 0.633 \\
\hline
Maximal (eq. (\ref{maxform})) & 0.818  &  0.806 & 0.798 \\
\hline
\end{tabular}
\end{center}
\end{table}
\begin{table}[htbp]
 \begin{center}
\caption{Values of the $4^{\rm th}$, $5^{\rm th}$ and $6^{\rm th}$ moment,
at fixed $\langle z \rangle$ (corresponding to $\epsilon_b=0.006$
in the peterson form), for different forms of the fragmentation function.}
\label{tabfrag2}
 \leavevmode
\begin{tabular}{|l|r|r|r|}
 \hline
 $\langle z \rangle=0.828$ & $M=4$ & $M=5$ & $M=6$ \\
\hline
Peterson & 0.611  &  0.535 & 0.474 \\
\hline
Kartvelishvili & 0.594  &  0.513 & 0.447 \\
\hline
Collins-Spiller & 0.626  &  0.559 & 0.505 \\
\hline
Maximal (eq. (\ref{maxform})) & 0.742  &  0.724 & 0.713 \\
\hline
\end{tabular}
\end{center}
\end{table}
We thus find that keeping the second moment fixed the variation of the
hadronic $p_T$ distribution obtained by varying the shape of the
fragmentation function among commonly used models is between 5\% and
13\% for both values of $\epsilon_b$. It thus seems difficult to enhance
the transverse momentum distribution by suitable choices of the form
of the fragmentation function. With the extreme choice of
eq.~(\ref{maxform}), one gets at most a variation of 50\% for the
largest value of $\epsilon_b$ and $M$. It would be interesting to see
if such an extreme choice is compatible with $e^+e^-$ fragmentation function
measurements.

\section{A STUDY OF HEAVY QUARK NON-PERTURBATIVE
     FRAGMENTATION IN \herw\label{sec:HVQfrag}\protect
\footnote{Section coordinators: S.~Frixione and M.L.~Mangano}}
\labelsection{bprod:MLMFrix}
In this Section we present the results of a phenomenological study of
the non-perturbative hadronization of $b$-quarks.
According to the standard QCD picture, distributions for an observable
hadron $H$ can be computed by convoluting the short-distance cross section
$\hat{\sigma}(p)$ with a fragmentation function $D^{(h)}_H(z)$
that describes the way in which the heavy quark $h$ hadronizes into $H$:
\beq
{\rm d}\sigma_H(p)=\int dz\,D^{(h)}_H(z)\,{\rm d}\hat{\sigma}(p/z).
\label{xsecfrag}
\eeq 
The precise definition of $D^{(h)}_H(z)$ depends on how much of
the heavy quark evolution after its production is absorbed into the
perturbative part $\hat{\sigma}(p)$, and how much is assigned to the
non-perturbative component parameterised by $D^{(h)}_H(z)$. Since
perturbation theory (PT) is well defined for a massive quark, the
standard prescription is to absorb into $\hat{\sigma}(p)$ not only the
hard matrix elements, but also the perturbative part of the
fragmentation function, defined by the evolution in $Q^2$ down to a
scale equal to the heavy quark mass $m_h$. $D^{(h)}_H(z)$ will
therefore account for the transition of an ``on-shell'' quark $h$ into
the hadron $H$. The assumptions built into eq.~(\ref{xsecfrag}) are that
$D^{(h)}_H(z)$ depends neither on the type of hard process, nor on the
scale at which $h$ was produced.
Under these assumptions, $D^{(h)}_H(z)$ can be
extracted from data in one given reaction (typically, $e^+e^-$),
and eventually used to predict the cross section in some other 
reaction ($p\bar{p}$, DIS and so on).

QCD factorization theorems indicate that this universality of
$D^{(h)}_H(z)$ holds in the asymptotic limit, and up to corrections of
order $m_h/Q$, $Q$ being the scale of the hard process. The size of
these corrections cannot be calculated, today, in any rigorous way.  A
possible approach to this problem is to turn to the phenomenological
models of hadronization implemented in QCD-based parton-shower Monte
Carlo (PSMC) codes. In PSMC the full final-state kinematical
configuration is available at both the parton and hadron levels.
Therefore, it is possible to ``measure'' $D^{(h)}_H(z)$ using
eq.~(\ref{xsecfrag}), both ${\rm d}\sigma_H$ and ${\rm d}\hat{\sigma}$
being known.  In the present section, we carry out this program
using the PSMC \herw~\cite{bprod:Marchesini:1992ch}. \herw\ evolves quarks
according to perturbative QCD down to small scales. The quarks are
paired up at the end of the evolution into colour singlet clusters,
which are then decayed to the physical hadrons using phenomenological
distributions.  The study of the heavy quark hadronisation process in
\herw\ will allow us to test the universality assumption, and to
measure the size of possible deviations.

We should stress that, at this moment, our conclusions are only
relevant for the hadronization model implemented in \herw; other
PSMC's, which treat the hadronization process differently (for
example, by adopting a string model), may well lead to different
conclusions.

In order to precisely define our procedure for extracting $D^{(h)}_H(z)$,
we need to consider in more details the way in which \herw\
generates events. Regardless of the type of initial-state 
particles, we can distinguish the following steps.

\begin{itemize}

\item {\it Hard subprocess:} at this stage, the PSMC generates the
kinematics for the basic $2\to 2$ hard reaction. We denote the 
momentum of the $b$-quark (or antiquark) as $p_b^{hard}$.

\item {\it Parton shower:} the partons resulting from the hard
subprocess undergo successive branchings, until their
virtuality is smaller than a fixed cutoff value. We denote the 
momentum of the $b$-quark at the end of this phase as $p_b^{ps}$.

\item {\it Gluon splitting and cluster formation:} the gluons present
at the end of the shower are decayed into light-quark pairs. Colour-singlet,
two-body clusters are formed, according to colour parenthood and
closeness in the phase-space. If there exist one or more cluster whose
mass is too large (relative to a given threshold), part of the cluster
rest energy is transformed into new $q\bar{q}$
pairs, and new clusters are defined. 
In this process, energy-momentum is redistributed among the cluster
elements, and the momentum of the $b$-quark can therefore
be modified with respect to $p_b^{ps}$. The momentum of the 
$b$ quark after completion of the clustering process
will be denoted by $p_b^{glsp}$.

\item {\it Cluster decay and hadron formation:} the clusters decay
into observable hadrons, according to the flavour and to tabulated 
mass spectra. We therefore obtain $b$-flavoured hadrons, whose momentum
we denote as $p_B$.

\end{itemize}
The hard subprocess and parton shower stages are based on 
perturbative QCD.
Thus, we identify the predictions given by \herw\ at the end
of the parton shower with the cross section $\hat{\sigma}$
that appears in eq.~(\ref{xsecfrag}). On the other hand,
the gluon splitting and cluster decay stages do not contain
QCD information, as they are performed according to a phenomenological
model. The $g\to q\bar{q}$ splitting and the decay kinematics
are induced by simple phase-space considerations. We thus identify
these stages as the long-distance, non perturbative part of 
the process, which gives rise to $D^{(h)}_H(z)$. 
We therefore determine the fragmentation function by comparing the results
for $p_b^{ps}$ and $p_B$, defining, on an event-by-event basis,
the following variables:
\beq
z_1=\frac{\vec{p}_B\cdot\vec{p}_b^{\,ps}}{\abs{\vec{p}_b^{\,ps}}^2},
\;\;\;\;\;\;
z_2=\frac{E_B+\vec{p}_B\cdot\hat{p}_b^{\,ps}}{E_b^{ps}+\abs{\vec{p}_b^{\,ps}}},
\label{zdef}
\eeq
where $\hat{p}_b^{\,ps}=\vec{p}_b^{\,ps}/\abs{\vec{p}_b^{\,ps}}$. 
Our conclusions will apply to both $z_1$ and $z_2$; thus, we will
collectively denote them by $z$. In hadronic collisions, the
momenta and energies have to be substituted by transverse momenta and
transverse energies respectively. Our results are summarized in
table~\ref{tab:Mmoments}; we considered $e^+e^-$ collisions at
$\sqrt{S}=91.2$~GeV and $p\bar{p}$ collisions at $\sqrt{S}=1.8$~TeV.
In the table, we present four of the (normalized) Mellin moments of
the $z$ distribution, defined as follows:
\beq
\mu_n=\int dz\,z^n\,D^{(h)}_H(z)\,\Big/\,\int dz\,D^{(h)}_H(z).
\label{Mmomdef}
\eeq
Usually, $0\le z\le 1$. In the present case, as we will see, we can also
have $z>1$; thus, in eq.~(\ref{Mmomdef}) the range of integration coincide
with the support of $D^{(h)}_H(z)$.
\begin{table}
\begin{center}
\caption{\label{tab:Mmoments}
Normalized Mellin moments of the $b$-quark non-perturbative fragmentation 
function. Results are given for the case of $e^+e^-$ collisions at
$\sqrt{S}=91.2$~GeV and for $p\bar{p}$ collisions at $\sqrt{S}=1.8$~TeV.
All numbers have a statistical accuracy of $\pm0.01$.}
\begin{tabular}{|l||c|c|c|c||c|c|c|c|} \hline
& \multicolumn{4}{c||}{$e^+e^-$} 
& \multicolumn{4}{c|}{$p\bar{p}$} 
\\ \hline
& $\mu_1$ & $\mu_2$ & $\mu_3$ & $\mu_4$ 
& $\mu_1$ & $\mu_2$ & $\mu_3$ & $\mu_4$ 
\\ \hline\hline
$\phantom{10<}\abs{\vec{p}_b^{\,ps}}<5$~GeV
& 0.87 &  0.78 &  0.71 & 0.66
& 0.95 &  0.94 &  0.94 & 0.96
\\ \hline
$10<\abs{\vec{p}_b^{\,ps}}<15$~GeV
& 0.92 &  0.85 &  0.80 & 0.76
& 0.85 &  0.74 &  0.66 & 0.60
\\ \hline
$20<\abs{\vec{p}_b^{\,ps}}<25$~GeV
& 0.92 &  0.86 &  0.81 & 0.76
& 0.83 &  0.71 &  0.63 & 0.57
\\ \hline
$30<\abs{\vec{p}_b^{\,ps}}<35$~GeV
& 0.92 &  0.85 &  0.80 & 0.75
& 0.82 &  0.70 &  0.62 & 0.56
\\ \hline
\end{tabular} 
\end{center}                                                            
\end{table}
The Mellin moments appearing in table~\ref{tab:Mmoments} have been
evaluated by considering bins in $\abs{\vec{p}_b^{\,ps}}$ (in the case
of hadronic collisions, the momentum is the transverse one). In
$e^+e^-$
collisions larger (smaller) values of $\abs{\vec{p}_b^{\,ps}}$ 
correspond to less (more) energy lost to gluons. 
In hadronic collisions larger (smaller)
values of $\abs{\vec{p}_b^{\,ps}}$ are more likely to correspond to
larger (smaller) values of the hard process momentum before evolution.
In either case, dependence of
$D^{(h)}_H(z)$ on $\abs{\vec{p}_b^{\,ps}}$ signals therefore a
departure from universality. 

By inspection of the table, we see that $D^{(h)}_H(z)$ is scale-%
independent to a very good extent (the situation appears to be slightly
better in the case of $e^+e^-$ collisions), except for the very low
$p_b^{ps}$ region; this is what we should expect, since in that
region the factorization theorem on which eq.~(\ref{xsecfrag}) is
based is bound to fail. On the other hand, there seems to emerge
a clear difference between the fragmentation functions extracted from 
$e^+e^-$ and $p\bar{p}$ ``data'', the latter being substantially softer
than the former. The first moment, which is the average
value of the fragmentation variable, changes by about 10\%. This
variation can change the rate of predicted $b$-hadrons in
hadronic collisions by almost 50\%.

This suggests that transporting to hadronic collisions
the non-perturbative fragmentation functions obtained by fitting 
$e^+e^-$ data may not be correct.
Of course, a much more detailed investigation
on the subject is required before reaching a firm conclusion; however,
this simple exercise of ours shows that universality should not
be taken for granted.

We now concentrate on the separate role played in the fragmentation process 
by the purely perturbative evolution and by the non-perturbative
gluon-splitting phase, before the cluster formation and decay
take place. We shall confine ourselves to the case of 
$e^+e^-$ collisions. The variables relevant to our study are the following:
\begin{enumerate}
\item Energy fraction retained {\em during} the perturbative evolution:
\beq
z_{ps} = \frac{2\abs{\vec{p}_b^{\,ps}}}{\sqrt{S}}
= \frac{\abs{\vec{p}_b^{\,ps}}}{\abs{\vec{p}_b^{\,hard}}},
\eeq
where $\sqrt{S}$ is the $e^+e^-$ CM energy.
\item Energy fraction retained {\em during} the gluon-splitting:
\beq
z_{glsp} = \frac{\abs{\vec{p}_b^{\,glsp}}}{\abs{\vec{p}_b^{\,ps}}}.
\eeq
\item Energy fraction left {\em after} the perturbative evolution
and the gluon-splitting:
\beq
z = z_{ps} \times z_{glsp} = \frac{2\abs{\vec{p}_b^{\,glsp}}}{\sqrt{S}}
= \frac{\abs{\vec{p}_b^{\,glsp}}}{\abs{\vec{p}_b^{\,hard}}},
\eeq
\end{enumerate}

\begin{figure}
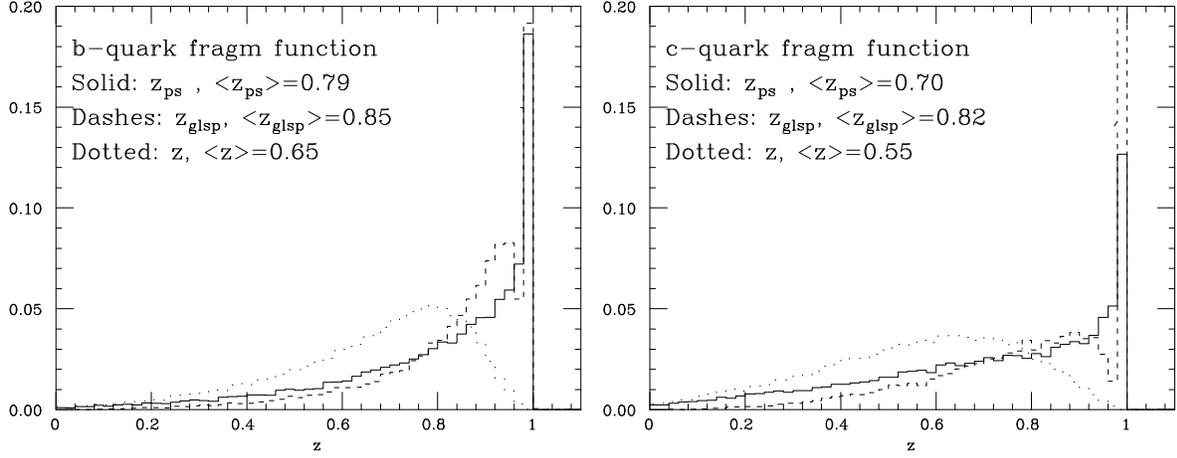

\centerline{
  \includegraphics[width=0.48\textwidth,clip]{fm_bfrag.eps} \hfil
  \includegraphics[width=0.48\textwidth,clip]{fm_cfrag.eps} }
\caption{\label{fig:bfrag} 
Fragmentation functions for $b$ (left) and $c$ (right) quarks,
produced in $e^+e^-$ collisions at $\sqrt{s}=91.2$~GeV.}
\end{figure}

The left panel in fig.~\ref{fig:bfrag} shows the three
distributions for $b$ quarks at $\sqrt{S}=91.2$~GeV. 
The solid histogram represents the distribution of
$z_{ps}$. The distribution has the shape of a Gribov-Lipatov, with no
indication of a Sudakov turn-over at large $z_{ps}$. The dotted line
is the distribution in $z$. A strong deformation of the purely
perturbative curve is clearly seen.  The dashed line corresponds to
the $z_{glsp}$ distribution. This is part of what the MC treats as a
non-perturbative component of the fragmentation process. The peak of
the dashed histogram at $z_{glsp}=1$ corresponds to events where the
cluster containing the heavy quark does not need to be further split,
while the tail corresponds to events where the invariant mass of the
heavy-quark cluster is too large, and additional light-quark
pairs have to be produced by hand. Notice that almost as much energy
is lost during this non-perturbative phase, as is lost during the
perturbative evolution.
                
\begin{figure}
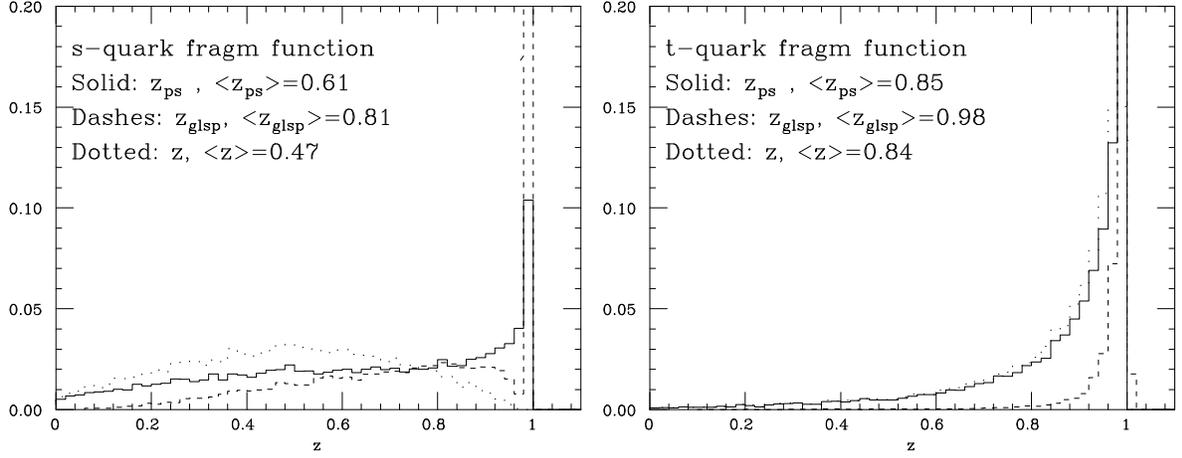

\centerline{
  \includegraphics[width=0.48\textwidth,clip]{fm_sfrag.eps} \hfil
  \includegraphics[width=0.48\textwidth,clip]{fm_tfrag.eps} }
\caption{\label{fig:sfrag} 
Fragmentation functions for $s$ (left) and $t$ (right) quarks,
produced in $e^+e^-$ collisions at 
$\sqrt{s}=91.2$ (left) and 2000 (right)~GeV.}
\end{figure}

For comparison, we show the same set of curves for the evolution of
the charm quark (right panel of fig.~\ref{fig:bfrag}). Notice that while
the effect of the perturbative evolution is to soften the quark
spectrum relative to the $b$-quark case, the amount of energy lost due
to gluon splitting is similar ($\langle z^c_{glsp}\rangle=0.82$, 
as opposed to $\langle z^b_{glsp}\rangle=0.85$). This is bizarre, since 
one expects the non-perturbative part to scale with $1/m_h$. The same 
result is found for the fragmentation of the $s$ quark (left panel of
fig.~\ref{fig:sfrag}). Here $\langle z^s_{glsp}\rangle$ 
is $0.81$. Again, a violation of the expected $1/m_h$ scaling.

Things improve for the top quark, whose distributions for $\sqrt{S}=2$~TeV 
are shown on the right panel of fig.~\ref{fig:sfrag}. The gluon-splitting 
part has only a minor impact on the overall spectrum of the top quark.

We are a bit bothered by the dominant role played by the
gluon-splitting phase.  By comparison, the next step in the evolution,
namely the cluster formation and decay, plays only a minor role, as
will be shown next. We would have anticipated that the cluster
formation and decay should be the place where most of the
non-perturbative physics should show up. This suggests that
the thresholds for the perturbative evolution in the MC shold be lowered,
so that the impact of the non-perturbative gluon splitting phase is
reduced, and purely perturbative Sudakov effects can manifest
themselves.


\begin{figure}
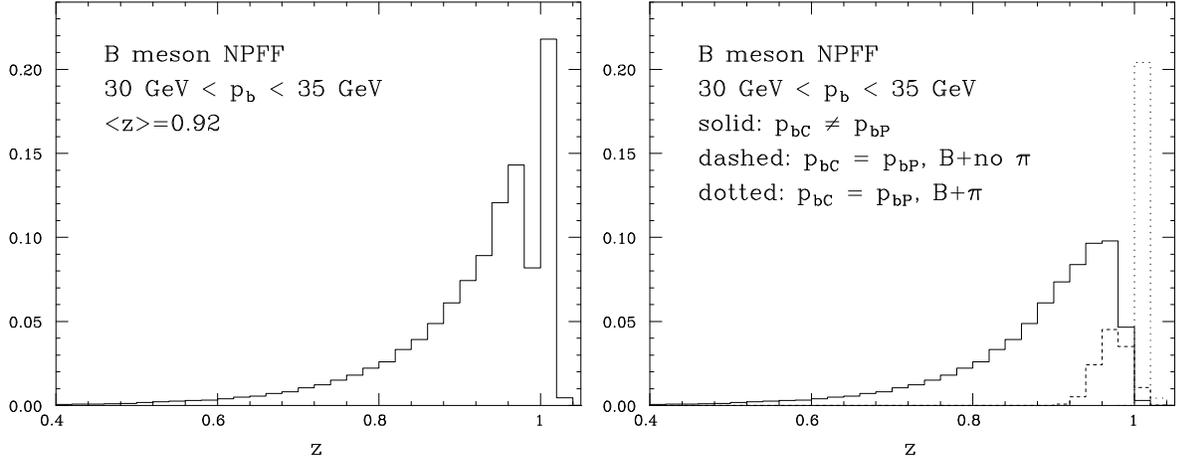

\centerline{
  \includegraphics[width=0.48\textwidth,clip]{fm_bmes1.eps} \hfil
  \includegraphics[width=0.48\textwidth,clip]{fm_bmes2.eps} }
\caption{ \label{fig:z}
$B$-meson non-perturbative fragmentation function in $e^+e^-$ at 
$\sqrt{S}=91.2$~GeV, for 30~GeV$<\abs{\vec{p}_b}<$35~GeV. 
CLDIR=1, CLSMR=0. See the text for details.
}
\end{figure}

We now turn again to the non-perturbative part of the fragmentation
function. The most striking feature, that cannot be inferred from the
simple study of Mellin moments as done in table~\ref{tab:Mmoments}, is
the presence of a double peak in the high-$z$ region (see the left
panel of fig.~\ref{fig:z}). A first peak (which we will call peak \AP)
is seen at $z$ values around 0.97. A second peak (peak \BP) is at
$z=1.01$ (we have a $z$-bin size of $0.02$. We verified that the
events contributing to the peak \BP\ do not have $z=1+\epsilon$,
i.e., the peak is not due to a roundoff error). The latter peak is
higher than the former.

The origin of this double peak can be traced back to the following facts.
First, the momentum of the emitted $B$ meson is very strongly correlated
with the momentum of the $b$ quark which enters the cluster. Therefore,
the $z$ distribution closely reflects the mass spectrum of the light
hadron emitted, together with the $B$ meson, in the cluster decay.
Second, the peak \BP\ is almost entirely due to events where the
cluster decay into \mbox{$B+\pi$}: at this peak, $z>1$ because the
mass of the pion is lighter than the mass of the lightest quarks
in \herw. More in detail, we have observed the following facts.

\begin{itemize}

\item
The structure of the double peak is strongly influenced by the value
taken by the two input parameters {\rm\bf CLDIR} and {\rm\bf CLSMR}.
If the default is used ({\rm\bf CLDIR=1}, {\rm\bf CLSMR=0}), the
double peak is observed (see the plot on the left of fig.\ref{fig:z}). On
the other hand, by setting {\rm\bf CLSMR}{\bf $\ne$0}, the $z$
distributions display a single peak (broader that the previous ones)
at about $z=0.97$. For small {\rm\bf CLSMR} values and large $b$
momenta, a second peak at $z=1.01$ tends to re-appear, although
smaller than observed before.

\item
The double peak disappears also if one chooses {\rm\bf CLDIR=0},
as in older \herw\ versions. In this case, the $z$ distributions
peak at about $z=0.9$, this peak being much broader that those obtained
with {\rm\bf CLDIR=1}, regardless of the value of {\rm\bf CLSMR}.

\item
We then set {\rm\bf CLDIR=1} and {\rm\bf CLSMR=0}. For any given
$B$ meson, we looked for the parent cluster ${\cal C}=\{b_C q\}$,
and for the parent bottom quark, $b_P$ (the parent quark is defined
as in the \herw\ routine {\rm\bf HWCHAD}). We observed what follows.

\begin{itemize}
\item
Plotting the $z$ distributions for the events with
$\vec{p}_{b_C}\ne\vec{p}_{b_P}$ (i.e.\/ events where the original
cluster was split), we see a single peak,
at the same $z$ value as for the peak \AP\ (solid line,
plot on the right of fig.\ref{fig:z}).

\item
The $z$ distributions for events such that $\vec{p}_{b_C}=\vec{p}_{b_P}$
display again a double peak. The two peaks are at the same $z$ values
as peaks \AP~and \BP, the latter one being by far dominant.

\item
Selecting only events with $\vec{p}_{b_C}=\vec{p}_{b_P}$, we found
that the peak at the position of \BP~corresponds to those clusters
decaying into a $B$ meson and a $\pi$, while the peak at the
position of \AP~is relevant for all the other two-body decays
(dotted and dashed lines respectively, plot on the right of 
fig.\ref{fig:z}).

\end{itemize}

\end{itemize}
Overall, notice also that the amount of energy retained after 
the gluon-splitting phase is of the same size as that retained at the
end of the full hadronization process, indicating that cluster
formation and decay have a minor impact on the total amount of energy
lost during the non-perturbative part of the evolution.

We were also able to reproduce the previous findings with a very
simple model. Given a momentum for a quark $b$, we generate randomly
the momentum for a light quark $q$, to be combined with $b$ into
a cluster, which eventually decays into a $B$ meson and a particle
of given mass $m_P$. The momentum of the quark $q$ is allowed to have
a (small) transverse momentum with respect to the direction
of the quark $b$. After evaluating the cluster mass, we performed
the decay in the rest frame of the cluster, either in a isotropic
manner (thus mimicking the choice {\rm\bf CLDIR=0}), or by letting
the momentum of the meson $B$ to be parallel to that of the quark 
$b$ (which corresponds to {\rm\bf CLDIR=1} and {\rm\bf CLSMR=0}).
In the latter case, depending upon the value of $m_P$, we got
a peak for $z<1$ (if $m_P>m_q$) or $z>1$ (if $m_P<m_q$).

In conclusion, the $z$ distributions we find when using \herw\ 
seem not to contain a lot of dynamical information, the most
important features being those implemented in the cluster-decay routine.
If the decay is not smeared out ({\rm\bf CLDIR=0}), we get a 
structure which is very difficult to reconcile with the idea
of fragmentation we have from QCD. After smearing, the distribution
still has a $z>1$ tail which will be extremely difficult to
fit with a function vanishing for $z\to 1$. This problem is
related to the fact that the mass of the lightest quarks in the MC is 
320 MeV, that is much larger than the pion mass.
We performed a test by reducing the light quark masses to 20~MeV, and
increasing the shower cutoff {\rm\bf VQCUT} in such a way as to maintain the
default value of the effective infrared threshold. The double peak
structure, as expected, disappeared. It remains to be seen, however,
whether such a small value of the quark masses is, more generally,
acceptable.

\section{A STUDY OF THE $\bb$ PRODUCTION MECHANISM IN PHYTIA\protect
\footnote{Section coordinators: S.~Gennai, A.~Starodumov, F.~Palla, 
R.~Dell'Orso }}
\labelsection{bprod:gennai}
\subsection{Introduction} 
In this section, we present a study on $\bb$ production performed
within the CMS collaboration using the Monte Carlo package \Py\ 5.75
as an event generator.  In particular, we investigate the influence of
the cut-off on the hard interaction transverse momentum $\pthat$ on
the production of $\bb$ events.
 
In Monte Carlo programs, $\bb$ pairs in hadron collisions are produced
by the mechanisms of gluon fusion, gluon splitting and flavour
excitation. All of them give contributions of the same order to the
total cross section, but they give rise to different kinematical
configurations of the final state.
 
There are two ways to generate $\bb$ events in \Py: 
\begin{itemize} 
\item Using a steering card MSEL=5, a gluon fusion mechanism 
($gg \rightarrow \bb$) is mainly simulated.  Each event contains at 
least one $\bb$ pair. 
\item Using MSEL=1, all QCD $2 \rightarrow 2$  
processes are simulated. In this case, all production mechanisms 
contribute to the $\bb$ production, but the probability to find a $\bb$ 
pair in the event is less than 1$\%$. 
\end{itemize} 
About one million events have been simulated in CMS with MSEL=1, in 
order to have a sample with all $\bb$ production mechanisms and 
default \Py cut-off, not to introduce any bias in the kinematics. 
The selection efficiency of triggered events out of this sample is 
quite low. In order to have higher signal statistics, in some cases 
one can use kinematical cuts which are different from the \Py 
default.~\cite{bprod:cms:trig}. 
 
\subsection{$\bb$  production} 
\begin{figure}[htb] 
\begin{center} 
\resizebox{12cm}{12cm}{\includegraphics{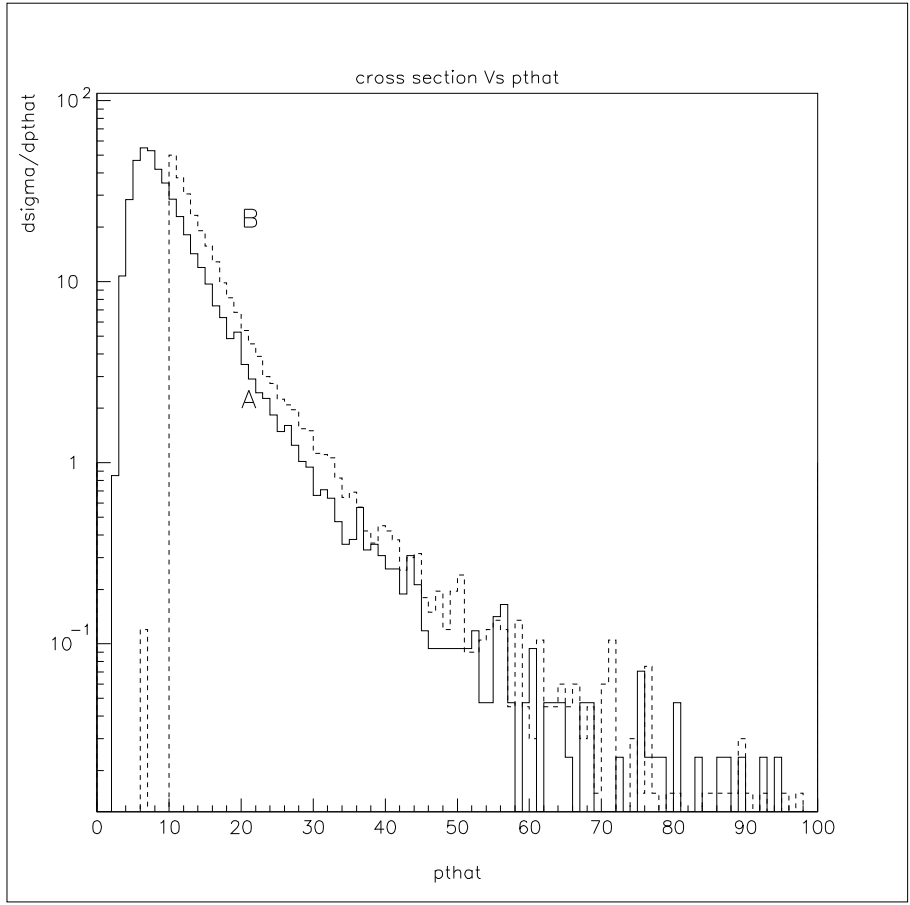}} 
\caption{Cross section with the two different cut-off on $\hat{P_{t}}$. 
The solid line is with the default cut-off (1 GeV). 
The picture is obtained with \Py\ 5.75.} 
\label{fig:cross} 
\end{center} 
\end{figure}

Two samples have been prepared to investigate the influence of the 
$\pthat$ cut on the production of $\bb$ events.  Both of them have 
been generated using MSEL=1 and contain events with only one $\bb$
pair.
Only events with $\pthat\ge 10$~GeV have been selected in the samples. 
The first sample (SAMPLE~A) has been generated 
with the default $\pthat$ cut of 1~GeV and only 
events with $\pthat \ge 10$~GeV were selected. The second sample 
(SAMPLE B) has been generated with $\pthat \ge 10$~ GeV. In both 
samples the following processes contribute: 
\begin{eqnarray} 
\label{1} 
&&gg \rightarrow q\overline{q} \\ 
\label{2} 
&&gq_{i} \rightarrow  gq_{i}  \\ 
\label{3} 
&& gg  \rightarrow  gg\;. 
\end{eqnarray} 
$\bb$ pair is produced by gluon splitting $g \rightarrow \bb$ 
in initial or final state shower evolution (processes (\ref{1}) to
(\ref{3})) or 
in the hard interaction (process (\ref{1})).  
For both samples A and B, we have computed  
the $\bb$ production cross section  
\begin{equation} 
\sigma_{ \bb }^{tot}  =  \frac{N_{b\overline{b}}}{N^{tot}}\sigma^{tot}\;, 
\end{equation} 
where $N_{b\overline{b}}$ is the number of $\bb$ events with $\hat{P_{t}} 
\ge 10$~GeV, $N^{tot}$ is the total number of generated events,
and $\sigma^{tot}$
is the total cross section (given by \Py).
We find that 
\begin{itemize} 
\item  for sample A, $\sigma_{\bb}^{tot}=150$~$\mu$b. 
The gluon fusion contribution 
is about 20~$\mu$b, while the gluon splitting contributions are 
$\sim30$~$\mu$b and $\sim 100$~$\mu$b 
for processes (\ref{2}) and (\ref{3}) respectively; 
\item  for sample B, $\sigma_{\bb}^{tot}$=257 $\mu$b.  Gluon fusion and 
gluon splitting contributions are at the same level as in sample A. 
In this case, however, there are also contributions from the processes 
$bg \rightarrow bg$ and $bq \rightarrow bq$ of about 110 $\mu$b. 
In the following we will 
call these contributions flavour excitation. 
\end{itemize} 
Figure~\ref{fig:cross} illustrates the difference in the $\bb$ 
production cross sections due to the additional contribution of the 
flavour excitation mechanism in sample~B. 
The effect has the following explanation. When the default 
$\hat{P_{t}}$ cut-off is used, \Py\ generates processes in the low 
energy approximation, i.e. there are no heavy quarks inside the parton 
distribution. This approach changes if one uses a different 
$\hat{P_{t}}$ cut-off: the parton distributions in this case include 
also $b$ and $c$ quarks. As a consequence, samples A and B are 
different in two respects: values of the cross sections, and set of 
production mechanisms.  The difference in the cross section is not 
very important, because the results are usually normalized to the 
total $\bb$ cross section of 500~$\mu$b. On the other hand, the 
different production mechanisms could be more dangerous, as they can lead 
to different kinematical distributions, and therefore affect the 
efficiencies of physical selection. 
 
\subsection{Kinematics} 
\begin{figure}[htb] 
\begin{center} 
\resizebox{8cm}{5cm}{\includegraphics{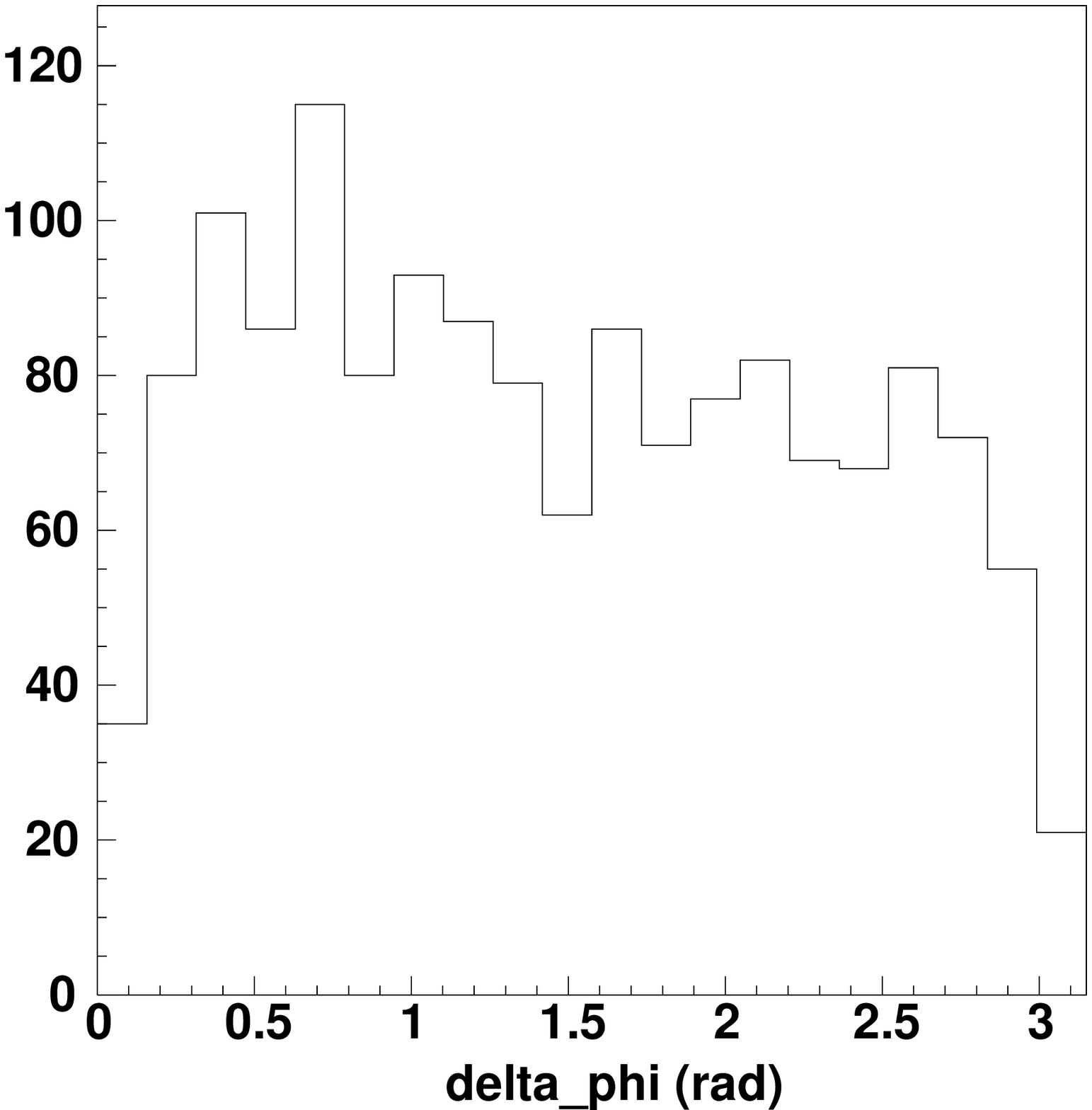}} 
\resizebox{8cm}{5cm}{\includegraphics{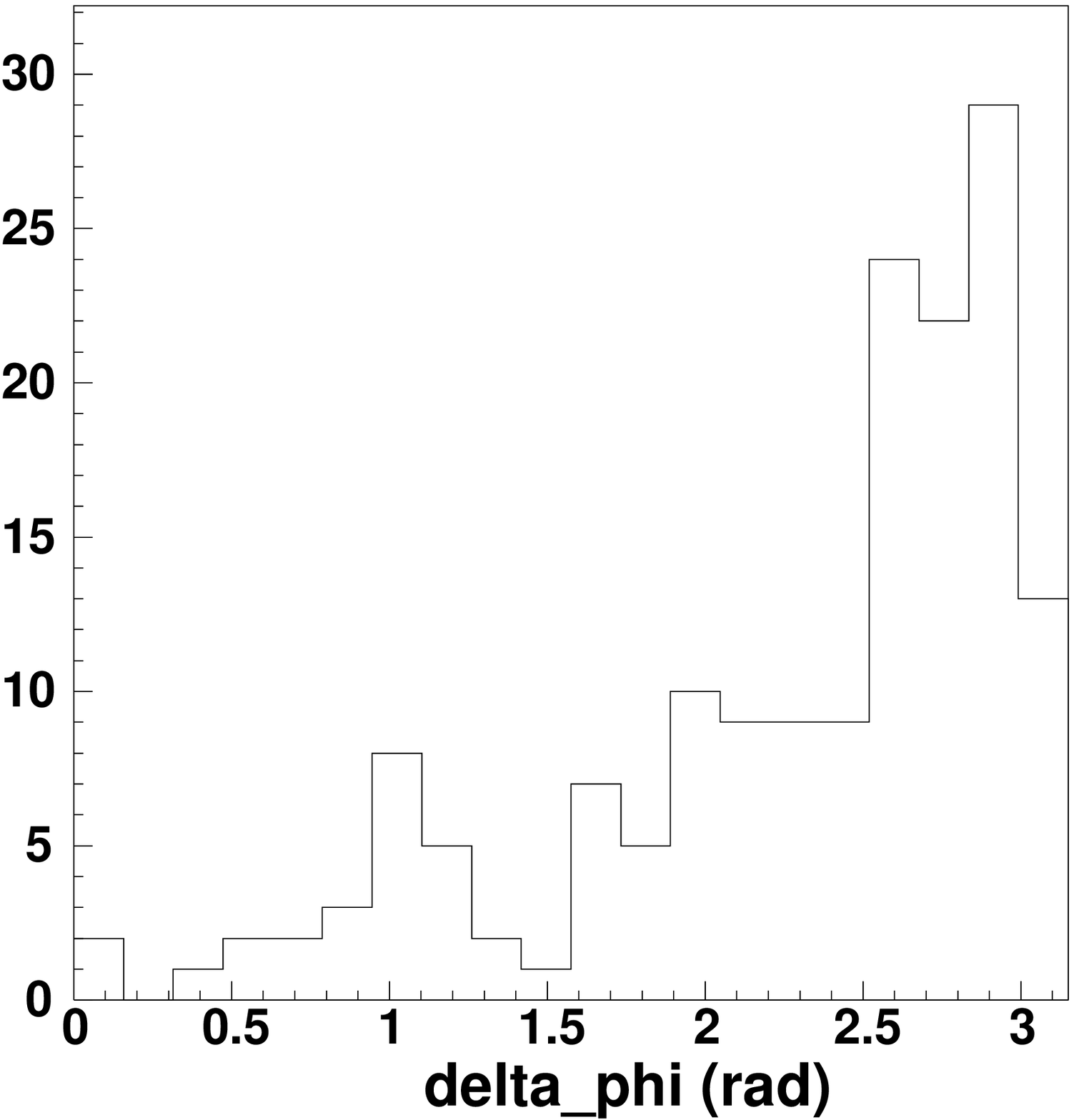}} 
\resizebox{8cm}{5cm}{\includegraphics{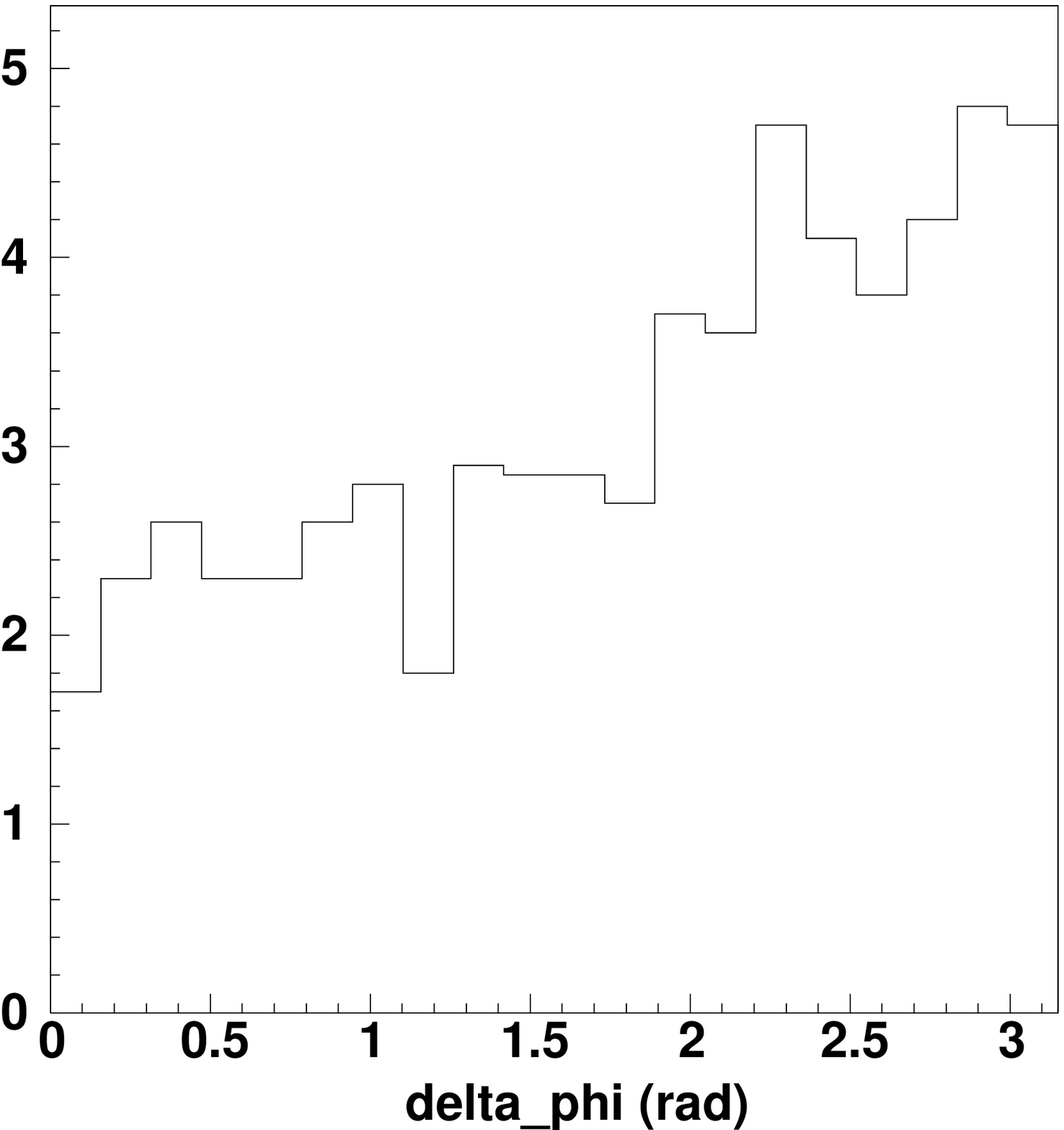}} 
\caption 
{Angle between the two $b$ quarks in the transverse plane: the upper one 
is gluon splitting, the middle is gluon fusion, the last is related to 
flavour excitation. In the plots the values are not normalized.} 
\label{fig:distr} 
\end{center} 
\end{figure}

The main kinematical parameters which define the signature of $\bb$ 
event are the transverse momenta and pseudorapidities of the $b$ 
quarks, and the angular distance $\Delta\phi$ between their directions in the 
transverse plane. The first two parameters have similar distributions 
in both samples. The $\Delta\phi$ distribution is shown in 
fig.~\ref{fig:distr} for the three different mechanisms. 
For what concerns gluon splitting, the distribution is slightly peaked 
at small $\Delta\phi$. The angle between the two $b$-quarks produced 
by the gluon-fusion mechanism has a peak at $\Delta \phi \sim \pi$, as 
expected, since in the process $gg \rightarrow \bb$ the b-quarks are 
produced back-to-back in the transverse plane.  The last distribution corresponds to the flavour excitation production mechanism, for which the back-to-back topology is preferred. We can conclude that the total $\Delta \phi$ distributions of sample A and sample B are slightly different. Some care should be taken about this, as it could affect the estimated efficiency of selection cuts.
 
\subsection{\Py\ 6.125} 
We have studied the same problem using the new 6.125 version  
\Py. We have generated two new samples A and B with \Py\ 6.125, 
and we have found the following results: 
\begin{itemize} 
\item  Sample A:  
the $\bb$ production cross section is $\sigma_{\bb}^{tot}$=220 $\mu$b. 
Gluon fusion contributes $\sim$ 47 $\mu$b and total gluon splitting gives 
173~$\mu$b. 
\item  Sample B: $\bb$ production 
cross section is $\sigma_{\bb}^{tot}$=465~$\mu$b. In this case, gluon fusion is $\sim$ 51 $\mu$b and 
total gluon splitting is $\sim$ 193 $\mu$b. The contribution from the flavour 
excitation is about 221 $\mu$b. 
\end{itemize} 
Also the way \Py\ 6.125 generates $\bb$ 
pairs depends on the $\pthat$ cut-off. The difference with \Py\ 5.75 values, are due to different total
cross section in the two versions.
 
\subsection{Interpretation}
Many of the features of \Py\ illustrated in this section are easily
explained\footnote{T.~Sj\"ostrand and E.~Norrbin, private communication.}.
It turns out that \Py\ treats differently processses with a low and high
$\pthatmin$. The limit is related to the scale of multiple
interactions, which is fixed to 2~GeV in the older versions, and was
made energy dependent in \Py~6, being 3.2~GeV at the LHC energy.
When $\pthatmin$ is above this scale, the hard process is selected
according to conventional matrix elements. Below this scale,
the hardest interaction is instead taken from the naive jet cross section
multiplied by a ``Sudakov style'' form factor, that represents the probability
that higher $p_T$ interactions did not take place in the rest of the
event. Since this procedure implies the computation of all parton-parton
scattering processes, the choice was made to exclude from it the incoming
$b$ and $c$ components, to save time in the computation.
This feature is no longer considered useful in modern times, the computers
being much faster. Thus, in \Py\ 6.138, also the $b$ and $c$
processes will be implemented in the low $p_T$ mode.

The difference in the total cross section in \Py~5.7 and 6.1
have a physical origin, since 6.1 uses newer parton distributions
that, according to HERA data, are more singular in the small~$x$ region.

The authors of \Py\ recommend the following procedure for the
generation of $b$ events. Parton fusion and flavour excitation
can be generated separately; the relevant massive
matrix elements are used for parton
fusion, and one can go to the limit $p_T \to 0$ with this process.
Gluon splitting cannot be generated separately: all hard processes
must be generated, excluding parton fusion and flavour excitation,
and one should look for the heavy flavour. Multiple interactions
are there switched off, in order to avoid a double-counting of the jet
cross section. This is adequate for the study of the $b$ production
properties, but clearly does not fully represent the structure of the
underlying event. In future \Py\ versions, when flavour excitation
is included in the minimum bias machinery with multiple interaction,
this latter should offer an almost equivalent alternative, but still without
the correct mass treatment of the parton fusion process near threshold.
Other limitations still remain from complex problems related to the
treatment of beam remnants; therefore, flavour excitation is only enabled
for the hardest interaction in the multiple-interaction scenario.

A sample of commented code is included below.
By using different flags ({\tt MEKIND}=0,1,2) three samples will be generated:
parton fusion, flavour excitation and gluon splitting.
\begin{verbatim}
      INTEGER KFINTMP(-40:40)
C... Multiple interactions switched off
      MSTP(81)=0
      PARP(81)=0.D0
      PARP(82)=0.D0
C... Maximum virtuality in ISR is PARP(67)*Q**2
C... This is the default in Pythia 6.137
      PARP(67)=1.D0
C... Choose heavy quark (bottom=5, charm=4)
      MASSIVE=5
C... Helper variable
      HQMASS=PMAS(MASSIVE,1)
C... Choose the kind of heavy quark production:
C... MEKIND is a local variable set to 0, 1 or 2
      IF (MEKIND==0) THEN  ! Massive matrix elements
        MSEL=MASSIVE
      ELSE IF (MEKIND==1) THEN  ! Flavour excitation
        MSEL=1
        CKIN(3)=HQMASS
        CKIN(5)=CKIN(3)
      ELSE IF (MEKIND==2) THEN  ! Gluon splitting (ISR, FSR)
        MSEL=1
        CKIN(3)=HQMASS
        CKIN(5)=CKIN(3)
      END IF
C... More restrictive cuts can be put here.
C... Example, 100 events in total.
      NEVENTS=100
C *** EVENT LOOP ***
      IF (MEKIND==1) NEVENTS=NEVENTS/2
C.... Loop over incoming partons
      DO ISIDE=1,2  
        IF (MEKIND/=1.AND.ISIDE==1) THEN
          GOTO 100
        ELSE IF (MEKIND==1) THEN
C... Only for flavour excitation:
C... Make backup copy of KFIN array
          DO IKF=-40,40
            KFINTMP(IKF)=KFIN(ISIDE,IKF)
C... Remove all incoming partons:
            KFIN(ISIDE,IKF)=0
          END DO
C... Select only b/bbar as incoming partons:
          KFIN(ISIDE, MASSIVE)=1
          KFIN(ISIDE,-MASSIVE)=1
        END IF
        DO IEV=1,NEVENTS
C... Generate an event
          CALL PYEVNT
C... For gluon splitting, remove events with HQ in the hard interaction
C... to avoid double counting:
          IF (MEKIND==2) THEN
            DO I=5,8
              IF (ABS(K(I,2))==MASSIVE) GOTO 50
            END DO
          END IF
C... Analysis...
50      END DO
C... Print statistics
      CALL PYSTAT(1)
C... Restore KFIN matrix:
      IF (MEKIND==1.AND.ISIDE==1) THEN
        DO IKF=-40,40
          KFIN(ISIDE,IKF)=KFINTMP(IKF)
        END DO
      END IF
100   END DO
\end{verbatim}


\section{ASYMMETRIES\protect\footnote{Section coordinators: E. Norrbin
and R.~Vogt}}
\labelsection{bprod:asymmetry}
\subsection{Introduction}
Sizeable leading particle asymmetries between e.g. $\D^-$ and $\D^+$
have been observed in several fixed target experiments
\cite{bprod:Asymobs}. It is of interest to investigate to what extent these
phenomena translate to bottom production and higher energies. No
previous experiment has observed asymmetries for bottom hadrons due to
limited statistics or other experimental obstacles.  Bottom
asymmetries are in general expected to be smaller than for charm
because of the larger bottom mass, but there is no reason why they
should be absent. In the fixed target experiment HERA-B, bottom
asymmetries could very well be large \cite{bprod:Norrbin} even at central
rapidities, but the conclusion of the present study is that
asymmetries at the LHC are likely to be small. In the following we
study possible asymmetries between $\mathrm{B}$ and
$\overline{\mathrm{B}}$ hadrons at the LHC within the Lund string
fragmentation model \cite{bprod:Norrbin:PLB442} and the intrinsic heavy
quark model~\cite{bprod:Brodsky:PLB93}.

In the string fragmentation model \cite{bprod:AGIS:PRP97}, the
perturbatively produced heavy quarks are colour connected to the beam
remnants. This gives rise to beam-drag effects where the heavy hadron
can be produced at larger rapidities than the heavy quark.  The
extreme case in this direction is the collapse of a small string,
containing a heavy quark and a light beam remnant valence quark of the
proton, into a single hadron. This gives rise to flavour correlations
which are observed as asymmetries. Thus, in the string model, there
can be coalescence between a perturbatively produced bottom quark and
a light quark in the beam remnant producing a leading bottom hadron.

There is also the possibility to have coalescence between the light
valence quarks and bottom quarks already present in the proton,
because the wave function of the proton can fluctuate into Fock
configurations containing a $\b \bbar$ pair, such as
$|\u\u\d\b\bbar\rangle$.  In these states, two or more gluons are
attached to the bottom quarks, reducing the amplitude by ${\cal
O}(\alphas^2)$ relative to parton fusion \cite{bprod:Vogt:NPB438}.  The
longest-lived fluctuations in states with invariant mass $M$ have a
lifetime of ${\cal O}( 2 P_{\rm lab}/M^2)$ in the target rest frame,
where $P_{\rm lab}$ is the projectile momenta. Since the comoving
bottom and valence quarks have the same rapidity in these states, the
heavy quarks carry a large fraction of the projectile momentum and can
thus readily combine to produce bottom hadrons with large longitudinal
momenta. Such a mechanism can then dominate the hadroproduction rate
at large $\xF$. This is the underlying assumption of the intrinsic
heavy quark model \cite{bprod:Brodsky:PLB93}, in which the wave function
fluctuations are initially far off shell. However, they materialize as
heavy hadrons when light spectator quarks in the projectile Fock state
interact with the target \cite{bprod:Brodsky:NPB369}.

In both models the coalescence probability is largest at small
relative rapidity and rather low transverse momentum where the
invariant mass of the $\Qbar\q$ system is small, enhancing the binding
amplitude. One exception is at very large $\pperp$, where the collapse
of a scattered valence quark with a $\bbar$ quark from the parton
shower is also possible, giving a further (small) source of leading
particle asymmetries in the string model.

\subsection{Lund String Fragmentation}
\label{Asym:Lundstring}
Before describing the Lund string fragmentation model, some words on
the perturbative heavy quark production mechanisms included in the
Monte Carlo event generator \Py \cite{bprod:Pythia:ref} used in this study
is in order. We study $\p\p$ events with one hard interaction because
events with no hard interaction are not expected to produce heavy
flavours and events with more than one hard interaction --- multiple
interactions --- are beyond the scope of this initial study and
presumably would not influence the asymmetries. After the hard
interaction is generated, parton showers are added, both to the
initial (ISR) and final (FSR) state.  The branchings in the shower are
taken to be of lower virtualities than the hard interaction
introducing a virtuality (or time) ordering in the event.  This
approach gives rise to several heavy quark production mechanisms,
which we will call \textit{pair creation}, \textit{flavour excitation}
and \textit{gluon splitting}.  The names may be somewhat misleading
since all three classes create pairs at $\g \to \Q\Qbar$ vertices, but
it is in line with the colloquial nomenclature.  The three classes are
characterized as follows.
\begin{description}
\item [Pair creation] The hard subprocess is one of the two LO
parton fusion processes $\g\g \to \Q\Qbar$ or $\q\qbar \to \Q\Qbar$.
Parton showers do not modify the production cross sections, but only shift kinematics. 
For instance, in the LO description, the $\Q$ and $\Qbar$ have to emerge 
back-to-back in azimuth in order to conserve momentum, while the parton shower 
allows a net recoil to be taken by one or several further partons. 
\item [Flavour excitation] A heavy flavour from the parton distribution
of one beam particle is put on mass shell by scattering against a parton of 
the other beam, i.e. $\Q\q \to \Q\q$ or $\Q\g \to \Q\g$. When the $\Q$ is not a
valence flavour, it must come from a branching $\g \to \Q\Qbar$ of the
parton-distribution evolution. In most current sets of parton-distribution
functions, heavy-flavour distributions are assumed to vanish for
virtuality scales $Q^2 < m_{\Q}^2$. The hard scattering must therefore 
have a virtuality above $m_{\Q}^2$. When the initial-state shower is 
reconstructed backwards \cite{bprod:Sjostrand:PLB157}, the $\g \to \Q\Qbar$ branching
will be encountered, provided that $Q_0$, the lower cutoff of the shower,
obeys $Q_0^2 < m_{\Q}^2$. Effectively the processes therefore become
at least $\g\q \to \Q\Qbar\q$ or $\g\g \to \Q\Qbar\g$, with the possibility 
of further emissions. In principle, such final states could also be obtained 
in the above pair-creation case, but the requirement that the hard scattering
must be more virtual than the showers avoids double counting.
\item [Gluon splitting] A $\g \to \Q\Qbar$ branching occurs in the 
initial- or final-state shower but no heavy flavours are produced in the hard 
scattering. Here the dominant $\Q\Qbar$ source is gluons in 
the final-state showers since time-like gluons emitted in the initial state 
are restricted to a smaller maximum virtuality. Except at high energies, 
most initial state gluon splittings instead result in flavour excitation, already covered above.
An ambiguity of terminology exists with initial-state evolution chains where
a gluon first branches to $\Q\Qbar$ and the $\Q$ later emits another gluon
that enters the hard scattering. From 
an ideological point of view, this is flavour excitation, since it is related 
to the evolution of the heavy-flavour parton distribution. From a practical 
point of view, however, we choose to classify it as gluon splitting, 
since the hard scattering does not contain any heavy flavours. 
\end{description}

In summary, the three classes above are then characterized by having
2, 1 or 0, respectively, heavy flavours in the final state of the LO
hard subprocess.  Another way to proceed is to add next-to-leading
order (NLO) perturbative processes, i.e the $\mathcal{O}(\alphas^3)$
corrections to the parton fusion \cite{bprod:NDE2}
\cite{bprod:Beenakker2}.  However,
with our currently available set of calculational tools, the NLO
approach is not so well suited for exclusive Monte Carlo studies where
hadronization is added to the partonic picture.

Flavour excitation and gluon splitting give significant contributions
to the total b cross section at LHC energies and thus must be
considered when this is of interest, see the following. However, NLO
calculations probably do a better job on the total b cross section
itself (while, for the lighter $c$ quark, production in parton showers
is so large that the NLO cross sections are more questionable).  The
shapes of single heavy quark spectra are not altered as much as the
correlations between $\Q$ and $\Qbar$ when extra production channels
are added.  Similar observations have been made when comparing NLO to
LO calculations \cite{bprod:NDE2} \cite{bprod:MNR}.
Likewise, asymmetries between
single heavy quarks are also not changed much by adding further
production channels, so for simplicity we consider only the pair
creation process here.

After an event has been generated at the parton level we add
fragmentation to obtain a hadronic final state. We use the Lund string
fragmentation model. Its effects on charm production were described
in \cite{bprod:Norrbin:PLB442}. Here we only summarize the main points.

In the string model, confinement is implemented by spanning strings
between the outgoing partons. These strings correspond to a
Lorentz-invariant description of a linear confinement potential with
string tension $\kappa \approx 1$~GeV/fm. Each string piece has a
colour charge at one end and its anticolour at the other. The double
colour charge of the gluon corresponds to it being attached to two
string pieces, while a quark is only attached to one. A diquark is
considered as being in a colour antitriplet representation, and thus
behaves (in this respect) like an antiquark. Then each string contains
a colour triplet endpoint, a number (possibly zero) of intermediate
gluons and a colour antitriplet end. An event will normally contain
several separate strings, especially at high energies where $\g \to
\q\qbar$ splittings occur frequently in the parton shower.

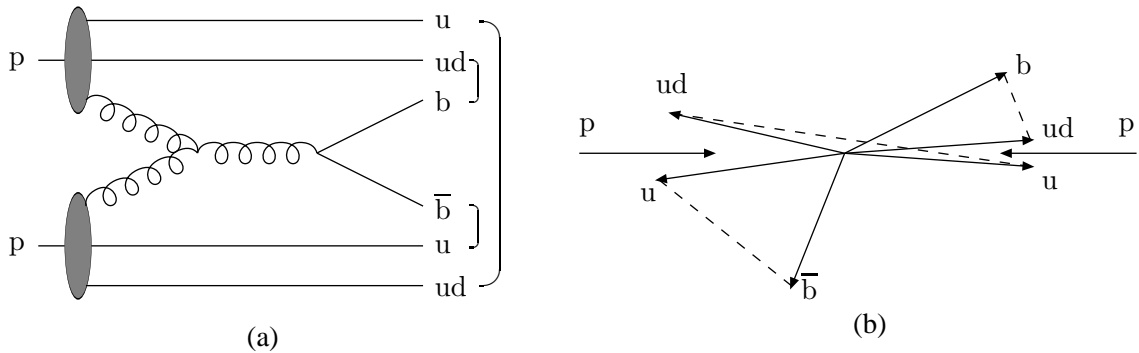
\begin{figure}[htb]
\begin{picture}(210,130)(-10,-10)
\Text(100,-5)[]{(a)}
\Text(10,30)[r]{$\p$}
\Text(10,100)[r]{$\p$}
\Line(15,30)(25,30)
\Line(15,100)(25,100)
\GOval(30,30)(20,5)(0){0.5}
\GOval(30,100)(20,5)(0){0.5}
\Gluon(33,45)(75,65){4}{4}
\Gluon(33,85)(75,65){4}{4}
\Gluon(75,65)(120,65){4}{4}
\Line(120,65)(160,45)
\Line(120,65)(160,85)
\Text(165,45)[l]{$\bbar$}
\Text(165,85)[l]{$\b$}
\Line(33,15)(160,15)
\Line(33,115)(160,115)
\Text(165,15)[l]{$\u\d$}
\Text(165,115)[l]{$\u$}
\Line(35,30)(160,30)
\Line(35,100)(160,100)
\Text(165,30)[l]{$\u$}
\Text(165,100)[l]{$\u\d$}
\put(178,37.5){\oval(7,16)[r]}
\put(182,65){\oval(15,100)[r]}
\put(178,92.5){\oval(7,16)[r]}
\end{picture}
\hspace{5mm}
\begin{picture}(210,130)(0,-15)
\Text(110,-5)[]{(b)}
\LongArrow(0,60)(50,60)
\Text(0,70)[l]{$\p$}
\LongArrow(210,60)(160,60)
\Text(210,70)[r]{$\p$}
\LongArrow(100,60)(170,65)
\Text(175,70)[l]{$\u\d$}
\LongArrow(100,60)(160,90)
\Text(165,94)[l]{$\b$}
\LongArrow(100,60)(30,50)
\Text(26,47)[t]{$\u$}
\LongArrow(100,60)(80,10)
\Text(90,10)[r]{$\bbar$}
\LongArrow(100,60)(35,75)
\Text(35,87)[]{$\u\d$}
\LongArrow(100,60)(170,55)
\Text(175,49)[l]{$\u$}
\DashLine(170,65)(160,90){4}
\DashLine(30,50)(80,10){4}
\DashLine(35,75)(170,55){4}
\end{picture}\\[2mm]
\caption{Example of a string configuration in a $\p\p$ collision.
(a) Graph of the process, with brackets denoting the final colour singlet
subsystems. (b) Corresponding momentum space picture, with dashed lines
denoting the strings.
\label{asym:colourflow}}
\end{figure}

The string topology can be derived from the colour flow of the hard
process with some ambiguity arising from colour-suppressed terms.
Consider e.g. the LO process $\g\g \to \b\bbar$ where two distinct
colour topologies are possible. Representing the proton remnant by a
$\u$ quark and a $\u\d$ diquark (alternatively $\d$ plus $\u\u$), one
possibility is to have the three strings $\b$--$\u\d$, $\bbar$--$\u$
and $\u$--$\u\d$, fig.~\ref{asym:colourflow}, and the other is
identical except the $\b$ is instead connected to the $\u\d$ diquark
of the other proton because the initial state is symmetric.

Once the string topology has been determined, the Lund string
fragmentation model \cite{bprod:AGIS:PRP97} can be applied to describe the
nonperturbative hadronization. To first approximation, we assume that
the hadronization of each colour singlet subsystem, i.e. string, can
be considered separately from that of all the other
subsystems. Presupposing that the fragmentation mechanism is
universal, i.e. process-independent, the good description of
$\e^+\e^-$ annihilation data should carry over.  The main difference
between $\e^+\e^-$ and hadron--hadron events is that the latter
contain beam remnants which are colour-connected with the
hard-scattering partons.

Depending on the invariant mass of a string, practical considerations
lead us to distinguish the following three hadronization
prescriptions:
\begin{description}
\item [Normal string fragmentation] 
In the ideal situation, each string has a large invariant mass. Then
the standard iterative fragmentation scheme, for which the assumption
of a continuum of phase-space states is essential, works well. The
average multiplicity of hadrons produced from a string increases
linearly with the string `length', which means logarithmically with
the string mass. In practice, this approach can be used for all
strings above some cutoff mass of a few GeV.
\item [Cluster decay]
If a string is produced with a small invariant mass, perhaps only a
single two-body final state is kinematically accessible. In this case
the standard iterative Lund scheme is not applicable. We call such a
low-mass string a cluster and consider its decay separately. When
kinematically possible, a $\Q$--$\qbar$ cluster will decay into one
heavy and one light hadron by the production of a light $\q\qbar$ pair
in the colour force field between the two cluster endpoints with the
new quark flavour selected according to the same rules as in normal
string fragmentation. The $\qbar$ cluster end or the new $\q\qbar$
pair may also denote a diquark. In the latest version of \Py,
anisotropic decay of a cluster has been introduced, where the mass
dependence of the anisotropy has been matched to string fragmentation.
\item [Cluster collapse]
This is the extreme case of cluster decay, where the string mass is so
small that the cluster cannot decay into two hadrons.  It is then
assumed to collapse directly into a single hadron which inherits the
flavour contents of the string endpoints. The original continuum of
string/cluster masses is replaced by a discrete set of hadron masses,
mainly $\mathrm{B}$ and $\mathrm{B}^*$ (or the corresponding baryon
states). This mechanism plays a special r\^ole since it allows flavour
asymmetries favouring hadron species that can inherit some of the
beam-remnant flavour contents. Energy and momentum is not conserved in
the collapse so that some energy-momentum has to be taken from, or
transferred to, the rest of the event. In the new version, a scheme
has been introduced where energy and momentum are shuffled locally in
an event.
\end{description}

We assume that the nonperturbative hadronization process does not
change the perturbatively calculated total rate of bottom
production. By local duality arguments \cite{bprod:Bloom:PRD4}, we further
presume that the rate of cluster collapse can be obtained from the
calculated rate of low-mass strings. In the process $\e^+\e^- \to
\c\cbar$ local duality suggests that the sum of the $\mathrm{J}/\psi$
and $\psi'$ cross sections approximately equal the perturbative
$\c\cbar$ production cross section in the mass interval below the
$\D\Dbar$-threshold. Similar arguments have also been proposed for
$\tau$ decay to hadrons \cite{bprod:BNP:NPB373} and shown to be accurate.
In the current case, the presence of other strings in the 
event also allows soft-gluon exchanges to modify
parton momenta as required to obtain the correct hadron masses.
Traditional factorization of short- and long-distance physics would
then also preserve the total bottom cross section. Local duality and
factorization, however, do not specify \textit{how} to conserve the
overall energy and momentum of an event when a continuum of $\bbar\d$
masses is to be replaced by a discrete $\mathrm{B}^0$. In practice,
however, the different possible hadronization mechanisms do not affect
asymmetries much. The fraction of the string-mass distribution below
the two particle threshold effectively determines the total rate of
cluster collapse and therefore the asymmetry.

The cluster collapse rate depends on several model parameters. The
most important ones are listed here with the \Py~parameter values that
we have used.  The \Py~parameters are included in the new default
parameter set in \Py~\texttt{6.135} and later versions.
\begin{itemize}
\item {\bf Quark masses}
The quark masses affect the threshold of the
string-mass distribution. Changing the quark mass shifts the string-mass
threshold relative to the fixed mass of the lightest two-body hadronic
final state of the cluster. Smaller quark
masses imply larger below-threshold production and an increased asymmetry.
The new default masses are \texttt{PMAS(1)}$=m_\u=$ \texttt{PMAS(2)}$=m_\d=$ 0.33D0,
\texttt{PMAS(3)}$=m_\s=$ 0.5D0, \texttt{PMAS(4)}$=m_\c=$ 1.5D0
and \texttt{PMAS(5)}$=m_\b=$ 4.8D0.
\item {\bf Width of the primordial $\kperp$ distribution.}
If the incoming partons are given small $\pperp$ kicks in the initial state,
asymmetries can appear at larger $\pperp$ since the beam remnants are given compensating
$\pperp$ kicks, thus allowing collapses at larger $\pperp$.
The new parameters are \texttt{PARP(91)=1.D0} and \texttt{PARP(93)=5.D0}.
\item {\bf Beam remnant distribution functions (BRDF).}
When a gluon is picked out of the proton, the rest of the proton forms
a beam remnant consisting, to first approximation, of a quark and a
diquark. How the remaining energy and momentum should be split between
these two is not known from first principles. We therefore use
different parameterizations of the splitting function and check the
resulting variations. We find significant differences only at large
rapidities where an uneven energy-momentum splitting tend to shift
bottom quarks connected to a beam remnant diquark more in the
direction of the beam remnant, hence giving rise to asymmetries at
very large rapidities. We use an intermediate scenario in this study,
given by \texttt{MSTP(92)=3}.
\item {\bf Threshold behaviour between cluster decay and collapse.} Consider
a $\b\dbar$ cluster with an invariant mass at, or slightly above, the
two particle threshold. Should this cluster decay to two hadrons or
collapse into one?  In one extreme point of view, a $\mathrm{B}\pi$
pair should always be formed when above this threshold, and never a
single $\mathrm{B}$. In another extreme, the two-body fraction would
gradually increase at a succession of thresholds: $\mathrm{B}\pi$,
$\mathrm{B}^*\pi$, $\mathrm{B}\rho$, $\mathrm{B}^*\rho$, etc., where
the relative probability for each channel is given by the standard
flavour and spin mixture in string fragmentation.  In our current
default model, we have chosen to steer a middle course by allowing two
attempts (\texttt{MSTJ(17)=2}) to find a possible pair of hadrons.
Thus a fraction of events may collapse to a single resonance also
above the $\mathrm{B}\pi$ threshold, but $\mathrm{B}\pi$ is
effectively weighted up. If a large number of attempts had been
allowed (this can be varied using the free parameter
\texttt{MSTJ(17)}), collapse would only become possible for cluster masses below the
$\mathrm{B}\pi$ threshold.
\end{itemize}

\begin{figure}[htb]
\begin{center}
\mbox{\epsfig{file=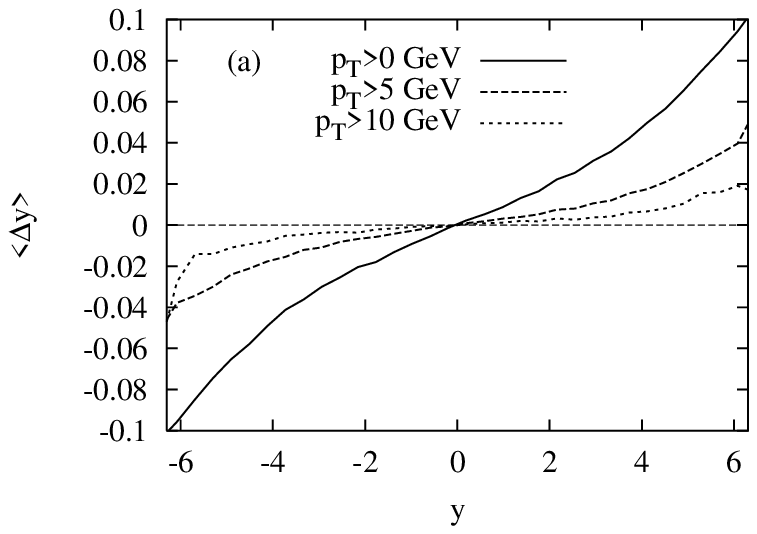}}
\mbox{\epsfig{file=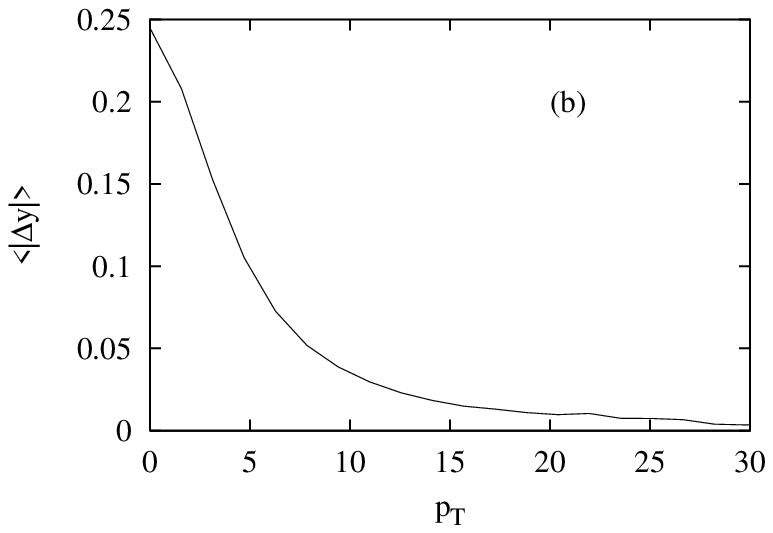}}
\end{center}
\caption{(a) Average rapidity shift $\Delta y = \langle y_\mathrm{B} - y_\mathrm{b} \rangle$
as a function of $y$ for some different $\pperp$ cuts. (b) Average rapidity shift
$\langle |\Delta y| \rangle$ in the direction of ``the other end of the string''
that the bottom quark is connected to, i.e. ignoring the sign of the shift.}
\label{asym:drageffect}
\end{figure}

The colour connection between the produced heavy quarks and the beam
remnants in the string model gives rise to an effect called beam
remnant drag. In an independent fragmentation scenario the light cone
energy momentum of the quark is simply scaled by some factor picked
from a fragmentation function.  Thus, on average the rapidity is
conserved in the fragmentation process.  This is not necessarily so in
string fragmentation, where both string ends contribute to the
four-momentum of the produced heavy hadron. If the other end of the
string is a beam remnant, the hadron will be shifted in rapidity in
the direction of the beam remnant resulting in an increase in
$|y|$. This beam-drag is shown qualitatively in
fig.~\ref{asym:drageffect}, where the rapidity shift is shown as a
function of rapidity and transverse momentum. This shift is not
directly accessible experimentally, only indirectly as a discrepancy
between the shape of perturbatively calculated quark distributions and
the data.

\subsection{Intrinsic Heavy Quarks}
\label{Asym:Intrinsic}
 
The wavefunction of a hadron in QCD can be represented as a
superposition of Fock state fluctuations, e.g. $\vert n_\V
\rangle$, $\vert n_\V \g \rangle$, $\vert n_\V \Q \Qbar \rangle$,
\ldots components where $n_\V \equiv \u\u\d$ for a proton.
When the projectile scatters in the target, the
coherence of the Fock components is broken and the fluctuations can
hadronize either by uncorrelated fragmentation as for leading twist
production or coalescence with
spectator quarks in the wavefunction \cite{bprod:Brodsky:PLB93}
\cite{bprod:Brodsky:NPB369}.
The intrinsic heavy quark Fock components are generated by virtual
interactions such as $\g \g \rightarrow \Q \Qbar$ where the
gluons couple to two or more projectile valence quarks. Intrinsic
$\Q\Qbar$ Fock states are dominated by configurations with
equal rapidity constituents so that, unlike sea quarks generated
from a single parton, the intrinsic heavy quarks carry a large
fraction of the parent momentum \cite{bprod:Brodsky:PLB93}.
 
The frame-independent probability distribution of an $n$--particle $\b
\bbar$ Fock state is
\begin{eqnarray}
\frac{dP^n_{\rm ib}}{dx_i \cdots dx_n} = N_n 
\frac{\delta(1-\sum_{i=1}^n x_i)}{(m_h^2 - \sum_{i=1}^n
(\widehat{m}_i^2/x_i) )^2} \, \, ,
\label{icdenom}
\end{eqnarray}
where $\widehat{m}_i^2 =k^2_{\perp,i}+m^2_i$ is the effective transverse mass
of the $i^{\rm th}$ particle and $x_i$ is the light-cone momentum
fraction.   The probability, $P^n_{\rm ib}$, is normalized by $N_n$
and $n=5$ for baryon production from the $|n_\V \b\bbar \rangle$ 
configuration.  The delta function
conserves longitudinal momentum.  The dominant Fock configurations
are closest to the light-cone energy shell and therefore the invariant
mass, $M^2 = \sum_i \widehat{m}_i^2/ x_i$, is minimized. 
Assuming $\langle \vec k_{\perp, i}^2 \rangle$ is
proportional to the square of the constituent quark mass, we choose
$\widehat{m}_\q = 0.45$ GeV, $\widehat{m}_\s =
0.71$ GeV, and $\widehat{m}_\b = 5$ GeV \cite{bprod:VBH:NPB360}
\cite{bprod:VBH:NPB383}.
 
The $\xF$ distribution for a single bottom hadron produced from an
$n$-particle intrinsic bottom state can be related to $P^n_{\rm ib}$
and the inelastic $\p\p$ cross section by
\begin{eqnarray}
\frac{\sigma^H_{\rm ib}(\p\p)}{d\xF} = \frac{dP_H}{d\xF} \sigma_{\p\p}^{\rm in}
\frac{\mu^2}{4 \widehat{m}_\b^2}  \alpha_s^4(M_{\b\bbar}) \, \, .
\label{icsign}
\end{eqnarray}
The probability distribution is the sum of all contributions from the $|n_\V \b
\bbar \rangle$ and the $|n_\V \b
\bbar \q \qbar \rangle$ configurations with $\q = \u$, $\d$, and $\s$
and includes uncorrelated fragmentation and coalescence, as described below, 
when appropriate \cite{bprod:Vogt:LBNL43095}.
The factor of $\mu^2/4 \widehat{m}_\b^2$ arises from the soft
interaction which breaks the coherence of the Fock state.  We take
$\mu^2 \sim 0.1$ GeV$^2$ \cite{bprod:GV:B539}. The intrinsic charm
probability, $P^5_{\rm ic} = 0.31$\%, was determined from analyses of
the EMC charm structure function data \cite{bprod:EMC:PLB110}. The intrinsic
bottom probability is scaled from the intrinsic charm probability by
the square of the transverse masses, $P_{\rm ib} = P_{\rm ic}
(\widehat{m}_\c/\widehat{m}_\b)^2$.  The intrinsic bottom cross
section is reduced relative to the intrinsic charm cross section by a
factor of $\alphas^4(M_{\b\bbar})/\alphas^4(M_{\c\cbar})$
\cite{bprod:VB:NPB438}.  Taking these factors into account, we obtain
$\sigma^5_{\rm ib}(p N) \approx 7$ nb at 14 TeV.

There are two ways of producing bottom hadrons from intrinsic
$\b\bbar$ states.  The first is by uncorrelated fragmentation.  If we
assume that the $\b$ quark fragments into a $\mathrm{B}$ meson, the
$\mathrm{B}$ distribution is
\begin{eqnarray}
\frac{d P^{nF}_{\rm ib}}{dx_{\mathrm{B}}} = \int dz \prod_{i=1}^n dx_i
\frac{dP^n_{\rm ib}}{dx_1 \ldots dx_n} \frac{D_{B/b}(z)}{z} \delta(x_{\mathrm{B}} - z
x_\b) \, \, ,
\label{icfrag}
\end{eqnarray}
These distributions are assumed for all intrinsic bottom production by uncorrelated
fragmentation with $D_{H/b}(z) = \delta(z-1)$.  At low $\pperp$, this
approximation should not be too bad, as seen in fixed target production
\cite{bprod:VBH:NPB383}.

If the projectile has the corresponding valence quarks, the bottom
quark can also hadronize by coalescence with the valence
spectators. The coalescence distributions are specific for the
individual bottom hadrons. It is reasonable to assume that the
intrinsic bottom Fock states are fragile and can easily materialize
into bottom hadrons in high-energy, low momentum transfer reactions
through coalescence.  The coalescence contribution to bottom hadron
production is
\begin{eqnarray}
\frac{d P^{nC}_{\rm ib}}{dx_H} = \int \prod_{i=1}^n dx_i
\frac{dP^n_{\rm ib}}{dx_1 \ldots dx_n} \delta(x_H - x_{H_1}-\cdots - 
x_{H_{n_\V}}) \, \, . 
\label{iccoalD}
\end{eqnarray}
where the coalescence function is simply a delta function combining
the momentum fractions of the quarks in the Fock state configuration
that make up the valence quarks of the final-state hadron.

Not all bottom hadrons can be produced from the minimal intrinsic
bottom Fock state configuration, $|n_\V \b\bbar \rangle$.  However,
coalescence can also occur within higher fluctuations of the intrinsic
bottom Fock state.  For example, in the proton, the $\mathrm{B}^-$ and
$\Xi_\b^0$ can be produced by coalescence from $|n_\V \b
\bbar\u\ubar \rangle$ and $|n_\V \b\bbar\s\sbar \rangle$ configurations.
These higher Fock state probabilities can be obtained using earlier results
on $\psi \psi$ pair production \cite{bprod:VB:NPB478} \cite{bprod:VB:PLB349}. 
If all the measured $\psi \psi$ pairs \cite{bprod:Badpsipsi} arise
from $|n_\V \c \cbar\c\cbar \rangle$ 
configurations, $P_{\rm icc} \approx 4.4\%\ P_{\rm
ic}$ \cite{bprod:VB:PLB349} \cite{bprod:RV:NPB446}.
It was found that the probability of a
$|n_\V \c\cbar\q\qbar \rangle$
state was then $P_{\rm icq} =
(\hat m_\c/\hat m_\q)^2 P_{\rm icc}$ \cite{bprod:VB:NPB478}.  
If we then assume $P_{\rm ibq} = (\hat m_\c/\hat m_\b)^2
P_{\rm icq}$, we find that 
\begin{eqnarray}
P_{\rm ibq} \approx \left( \frac{\widehat{m}_\c}{\widehat{m}_\b}
\right)^2 \left( \frac{\widehat{m}_\c}{\widehat{m}_\q}
\right)^2 P_{\rm icc} \, \, ,
\label{icrat}
\end{eqnarray}
leading to $P_{\rm ibu} = P_{\rm ibd} \approx 70.4\%\ P_{\rm ib}$
and $P_{\rm ibs} \approx 28.5\%\ P_{\rm ib}$.  To go to still higher
configurations, one can make similar assumptions. However, as more
partons are included in the Fock state, the coalescence
distributions soften and approach the fragmentation distributions,
eventually producing bottom hadrons with less momentum
than uncorrelated fragmentation from the minimal $\b\bbar$
state if a sufficient number of $\q\qbar$ pairs are included.
There is then no longer any advantage to introducing more light
quark pairs into the configuration---the relative probability will
decrease while the potential gain in momentum is not significant.
Therefore, we consider production by fragmentation and coalescence
from the minimal state and the next higher states with $\u\ubar$,
$\d\dbar$ and $\s\sbar$ pairs.
 
The probability distributions entering Eq.~(\ref{icsign}) for
$\mathrm{B}^0$ and $\Bbar^0$ are
\begin{eqnarray} 
\frac{dP_{\mathrm{B}^0}}{d\xF} & = & \frac{1}{2} \left( \frac{1}{10} \frac{dP_{\rm
ib}^{5F}}{d\xF} + \frac{1}{4} \frac{dP_{\rm ib}^{5C}}{d\xF} \right) + 
 \frac{1}{2} \left( \frac{1}{10} \frac{dP_{\rm
ibu}^{7F}}{d\xF} + \frac{1}{5} \frac{dP_{\rm ibu}^{7C}}{d\xF} \right) \nonumber
\\ &   & + \, \frac{1}{2} \left( \frac{1}{10} \frac{dP_{\rm
ibd}^{7F}}{d\xF} + \frac{2}{5} \frac{dP_{\rm ibd}^{7C}}{d\xF} \right) +
 \frac{1}{2} \left( \frac{1}{10} \frac{dP_{\rm
ibs}^{7F}}{d\xF} + \frac{1}{5} \frac{dP_{\rm ibs}^{7C}}{d\xF} \right) \\
 \frac{dP_{\Bbar^0}}{d\xF} & = & 
\frac{1}{10} \frac{dP_{\rm ib}^{5F}}{d\xF} +
 \frac{1}{10} \frac{dP_{\rm ibu}^{7F}}{d\xF} + \frac{1}{2} \left( \frac{1}{10}
\frac{dP_{\rm ibd}^{7F}}{d\xF} + \frac{1}{8} \frac{dP_{\rm ibd}^{7C}}{d\xF} 
\right) + \frac{1}{10} \frac{dP_{\rm ibs}^{7F}}{d\xF} \, \, .
\end{eqnarray}
See Ref.~\cite{bprod:Vogt:LBNL43095} for more details and the
probability distributions of other bottom hadrons.

\subsection{Model predictions}
\begin{figure}[htb]
\begin{center}
\mbox{\epsfig{file=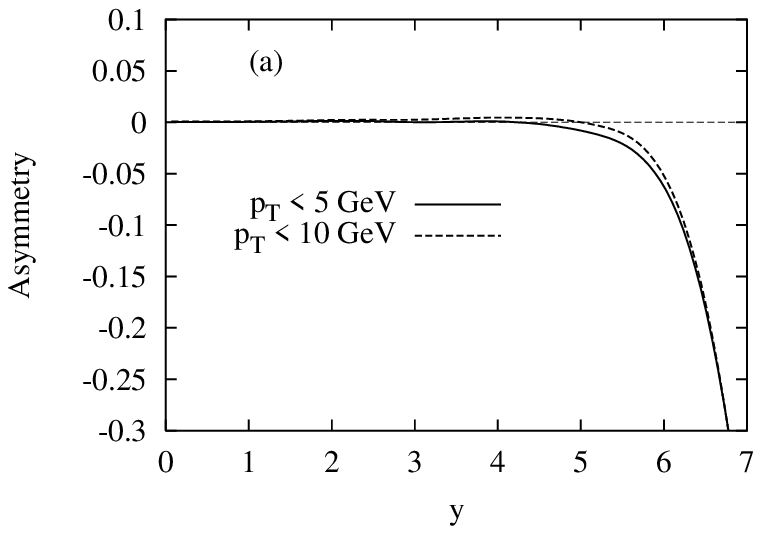}}
\mbox{\epsfig{file=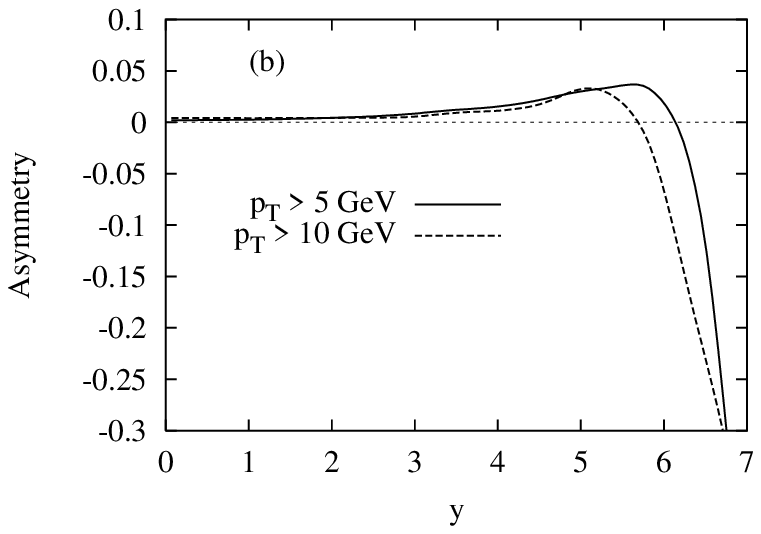}}
\end{center}
\caption{The asymmetry, $A=\frac{\sigma(\mathrm{B}^0) - \sigma(\Bbar^0)}
{\sigma(\mathrm{B}^0) + \sigma(\Bbar^0)}$, as a function of rapidity for
different $\pperp$ cuts: (a) $\pperp < 5,10$ GeV and (b) $\pperp > 5, 10$ GeV using
parameter set 1 as described in the text.}
\label{asym:string_asym}
\end{figure}

In this section we present some results from both
models. Figure~\ref{asym:string_asym} shows the asymmetry between
$\mathrm{B}^0$ and $\Bbar^0$ as a function of $y$ for several $\pperp$
cuts in the string model. The asymmetry is essentially zero for
central rapidities and increases slowly with rapidity.  When the
kinematical limit is approached, the asymmetry changes sign for small
$\pperp$ because of the drag-effect since $\b$-quarks are often
connected to diquarks from the proton beam remnant,
fig.~\ref{asym:colourflow}, thus producing $\Bbar^0$ hadrons which are
shifted more in rapidity than $\mathrm{B}^0$. Cluster collapse, on the
other hand, tend to enhance the production of leading particles (in
this case $\mathrm{B}^0$) so the two mechanisms give rise to
asymmetries with different signs. Collapse is the main effect at small
rapidities while eventually at very large $y$, the drag effect
dominates.

In Table~\ref{asym:parameterdep} we study the parameter dependence of
the asymmetry by looking at the integrated asymmetry for different
kinematical regions using three different parameter sets:
\begin{itemize}
\item {\bf Set 1} is the new default as presented in
section~\ref{Asym:Lundstring}.
\item {\bf Set 2} The same as Set 1 except it uses simple counting rules in
the beam remnant splitting, i.e. each quark get on average one third
of the beam remnant energy-momentum.
\item {\bf Set 3} The old parameter set, before fitting to fixed-target
data, is included as a reference. This set is characterized by current
algebra masses, lower intrinsic $\kperp$, and an uneven sharing of
beam remnant energy-momentum.
\end{itemize}

We see that in the central region the asymmetry is generally very
small whereas for forward (but not extremely forward) rapidities and
moderate $\pperp$ the asymmetry is around 1--2\%. In the very forward
region at small $\pperp$, drag asymmetry dominates which can be seen
from the change in sign of the asymmetry. The asymmetry is fairly
stable under moderate variations in the parameters even though the
difference between the old and new parameter sets (Set 1 and 3) are
large in the central region.  Set 1 typically gives rise to smaller
asymmetries.

\begin{table}
\begin{center}
\caption{Parameter dependence of the asymmetry
in the string model. The statistical
error in the last digit is shown in parenthesis (95\% confidence).}
\label{asym:parameterdep}
\begin{tabular}{|l|c|c|c|} \hline
Parameters	& $|y|<2.5$, $\pperp>5$ GeV& $3<|y|<5$, $\pperp>5$ GeV& $|y|>3$, $\pperp<5$ GeV \\ \hline
Set 1		& 0.003(1)		& 0.015(2)		& $-$0.008(1) \\ \hline
Set 2		& $-$0.000(2)		& 0.009(3)		& $-$0.005(2) \\ \hline
Set 3		& 0.013(2)		& 0.020(3)		& $-$0.018(2) \\ \hline
\end{tabular}
\end{center}
\end{table}

\begin{figure}[htb]
\setlength{\epsfxsize=0.95\textwidth}
\setlength{\epsfysize=0.5\textheight}
\centerline{\epsffile{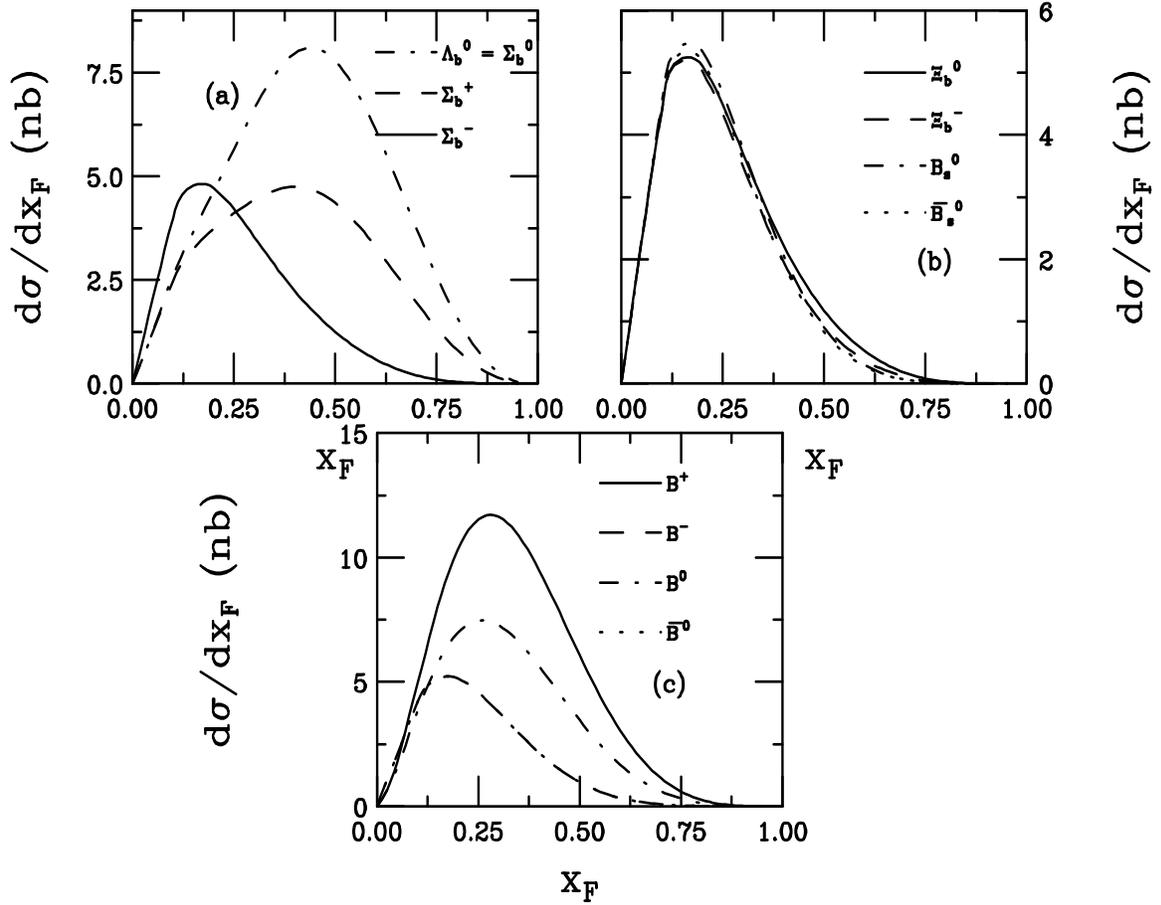}}
\caption[]{ Predictions for bottom hadron production are given for $\p\p$
collisions at 14 TeV.  The bottom baryon distributions are given in (a) for
$\Lambda_\b^0=\Sigma_\b^0$ (dot-dashed), $\Sigma_\b^+$ (dashed), and $\Sigma_\b^-$
(solid).  The bottom-strange distributions are shown in (b) for $\Xi_\b^0$
(solid), $\Xi_\b^-$ (dashed), $\mathrm{B}_\s^0$ (dot-dashed), and $\Bbar_s^0$
(dotted).  In (c), the $\mathrm{B}$ meson distributions are given: $\mathrm{B}^+$ (solid),
$\mathrm{B}^-$ (dashed), $\mathrm{B}^0$ (dot-dashed), and $\Bbar^0$ (dotted).
The $\mathrm{B}^-$ and $\Bbar^0$ distributions are virtually identical.}
\label{pro800}
\end{figure}

The cross sections for all intrinsic bottom hadrons are given as a
function of $\xF$ in fig.~\ref{pro800}. The bottom baryon
distributions are shown in fig.~\ref{pro800}(a).  The $\Lambda_\b^0$
($\Sigma_\b^0$) distributions are the largest and most forward peaked
of all the distributions. The $\Sigma_\b^-$ is the smallest and the
softest, similar to that of the bottom-strange mesons and baryons
shown in fig.~\ref{pro800}(b).  The different coalescence
probabilities assumed for hadrons from the
$|\u\u\d\b\bbar\s\sbar\rangle$ configuration have little real effect
on the shape of the cross section, dominated by independent
fragmentation. Of the $\mathrm{B}$ mesons shown in
fig.~\ref{pro800}(c), the $\mathrm{B}^+$ and $\mathrm{B}^0$ cross
sections are the largest since both can be produced from the 5
particle configuration. The $\mathrm{B}^-$ and $\Bbar^0$ distributions
are virtually identical. We note that the $\xF$ distributions of other
bottom hadrons not included in the figure would be similar to the
bottom-strange hadrons since they would be produced by fragmentation
only.

\begin{figure}[htb]
\setlength{\epsfxsize=0.95\textwidth}
\setlength{\epsfysize=0.25\textheight}
\centerline{\epsffile{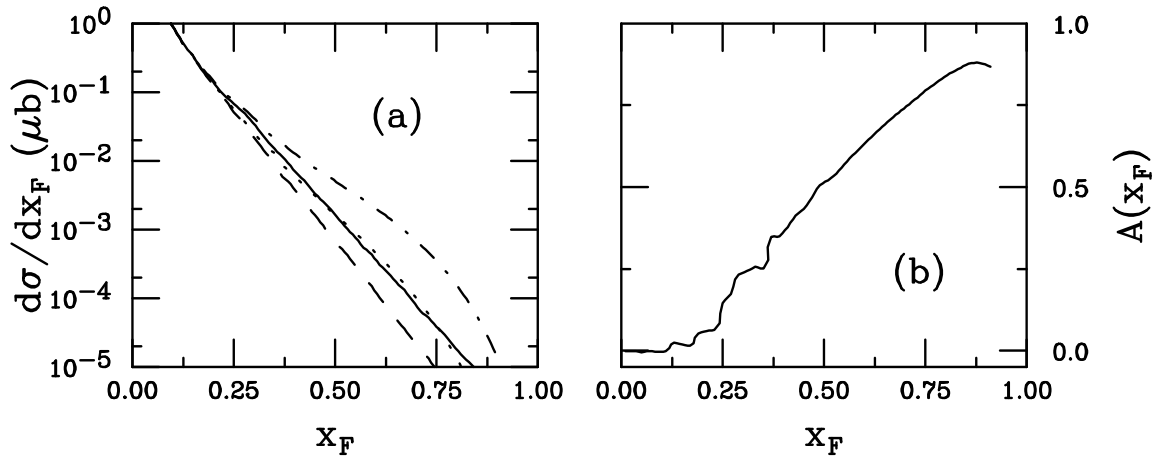}}
\caption[]
{(a) Leading-twist predictions for $\mathrm{B}^0$ (solid) and $\Bbar^0$
(dashed) using independent fragmentation.  Model predictions for
$\mathrm{B}^0$ (dot-dashed) and $\Bbar^0$ (dotted) distributions from
Eq.~(\ref{tcmodel}). (b) The asymmetry between $\mathrm{B}^0$ and
$\Bbar^0$, the dot-dashed and dotted curves in (a), is also given.}
\label{asymlhc}
\end{figure}

The $\xF$ distribution for final-state hadron $H$ is the sum
of the leading-twist fusion and intrinsic bottom components,
\begin{eqnarray}
\frac{d\sigma^H_{hN}}{d\xF} = \frac{d\sigma^H_{\rm lt}}{d\xF} +
\frac{d\sigma^H_{\rm ib}}{d\xF} \, \, .
\label{tcmodel}
\end{eqnarray}
The intrinsic bottom cross sections from Section~\ref{Asym:Intrinsic}
are combined with a leading twist calculation using independent
fragmentation where drag effects are not included.
The leading twist results have been smoothed and
extrapolated to large $\xF$ \footnotetext{Thanks
to J. Klay at UC Davis for extending the curves to large $\xF$.}
to facilitate a comparison with the intrinsic
bottom calculation.  The resulting total $\mathrm{B}^0$ and $\Bbar^0$
distributions are shown in fig.~\ref{asymlhc}, along with the corresponding
asymmetry. Note that since the intrinsic heavy quark $\pperp$ distributions are
more steeply falling than the leading twist, we only consider $\pperp < 5$ GeV.
The distributions are drawn to emphasize the high $\xF$ region where the
distributions differ.  The asymmetry is $\sim 0.1$ at $\xF \sim 0.25$,
corresponding to $y\sim 6.5$.  Therefore, intrinsic bottom should not be a
significant source of asymmetries.

\subsection{Summary}
To summarize, we have studied possible production asymmetries between
$\b$ and $\bbar$ hadrons, especially $\mathrm{B}^0$ and $\Bbar^0$, as
predicted by the Lund string fragmentation model and the intrinsic
heavy quark model.  We find negligible asymmetries for central
rapidities and large $\pperp$ (in general, less than 1\%). For some
especially favoured kinematical ranges such as $y > 3$ and $5 < \pperp
< 10$ GeV the collapse asymmetry could be as high as 1--2\%. Intrinsic
bottom becomes important only for $\xF>0.25$ and $\pperp<5$ GeV,
corresponding to $y>6.5$.

\section{QUARKONIUM PRODUCTION\protect\footnote{Section coordinators:
M.~Kr\"amer, F.~Maltoni, M.A.~Sanchis-Lozano}}
\labelsection{bprod:quarkonium}
The production of charmonium and bottomonium states at high-energy
colliders has been the subject of considerable interest during the
past few years. New experimental results from $p\bar{p}$, $ep$ and
$e^+e^-$ colliders have become available, some of which revealed
dramatic shortcomings of earlier quarkonium production models. In
theory, progress has been made on the factorization between the short
distance physics of heavy-quark creation and the long-distance physics
of bound state formation. The colour-singlet model
\cite{bprod:Berger:1981ni} \cite{bprod:Baier:1981uk} has been
superseded by a consistent
and rigorous framework, based on the use of non-relativistic QCD
(NRQCD) \cite{bprod:Bodwin:1995jh}, an effective field theory that includes
the so-called colour-octet mechanisms.  On the other hand, the colour
evaporation model \cite{bprod:Fritzsch:1977ay} \cite{bprod:Halzen:1977rs}
\cite{bprod:Gluck:1978zm} of
the early days of quarkonium physics has been revived
\cite{bprod:Gavai:1995in} \cite{bprod:Schuler:1996ku}
\cite{bprod:Amundson:1997qr} \cite{bprod:Edin:1997zb}. However,
despite the recent theoretical and experimental developments the range
of applicability of the different approaches is still subject to
debate, as is the quantitative verification of factorization. Because
the quarkonium mass is still not very large with respect to the QCD
scale, in particular for the charmonium system, non-factorizable
corrections \cite{bprod:Brodsky:1997tv} \cite{bprod:Hoyer:1999ha}
\cite{bprod:Hoyer:1999dr} may not be
suppressed enough, if the quarkonium is not part of an isolated jet,
and the expansions in NRQCD may not converge very well. In this
situation a global analysis of various processes is
mandatory in order to assess the importance of different quarkonium
production mechanisms, as well as the limitations of a particular
theoretical framework (for reviews on different quarkonium
production processes see e.g. \cite{bprod:Braaten:1996pv}
\cite{bprod:Beneke:1997av} \cite{bprod:Kramer:1997ws}.)
By the time the LHC starts operating, new
experimental data from the Tevatron and HERA as well as theoretical
progress, e.g.\ in the calculation of higher-order corrections, will
have significantly improved the present picture and will allow more
precise predictions than what is possible at present. In the
following, we will therefore focus on the general phenomenological
implications of the NRQCD approach for quarkonium production at the
LHC, rather than aiming at a detailed and comprehensive numerical
analysis. Based on the information provided by the present Tevatron
data we will derive predictions for observables crucial for future LHC
analyses, such as differential cross sections and quarkonium
polarization.

In the NRQCD approach, the cross section for producing a quarkonium
state $H$ at a hadron collider can be expressed as a sum of terms,
each of which factors into a short-distance coefficient and a
long-distance matrix element: \begin{equation}\label{eq_fac}
\mbox{d}\sigma(pp/p\bar{p} \to H + X) = \sum_n
\mbox{d}\hat{\sigma}(pp/p\bar{p} \to Q\overline{Q}\, [n] + x)\,\langle
{\cal{O}}^{H}\,[n]\rangle , \end{equation} where $n$ denotes the
colour, spin and angular momentum state of an intermediate
$Q\overline{Q}$ pair.  The short-distance cross section
$\mbox{d}\hat{\sigma}$ can be calculated perturbatively in the strong
coupling $\alpha_s$. The NRQCD matrix elements $\langle {\cal O}^H[n]
\rangle$ (see \cite{bprod:Bodwin:1995jh} for their definition) are related
to the non-perturbative transition probabilities from the
$Q\overline{Q}$ state $n$ into the quarkonium $H$. They scale with a
definite power of the intrinsic heavy-quark velocity $v$
\cite{bprod:Lepage:1992tx}. ($v^2 \sim 0.3$ for charmonium and $v^2 \sim
0.1$ for bottomonium.) The general expression (\ref{eq_fac}) is thus a
double expansion in powers of $\alpha_s$ and $v$.

The NRQCD formalism implies that so-called colour-octet processes
associated with higher Fock state components of the quarkonium wave
function must contribute to the cross section. Heavy quark pairs that
are produced at short distances in a colour-octet state can evolve
into a physical quarkonium through radiation of soft gluons at late
times in the production process, when the quark pair has already
expanded to the quarkonium size. Such a possibility is ignored in the
colour-singlet model, where only those heavy quark pairs that are
produced in the dominant Fock state (i.e.\ in a colour-singlet state
and with the spin and angular momentum quantum numbers of the meson)
are assumed to form a physical quarkonium. The most profound
theoretical evidence that the colour-singlet model is incomplete comes
from the presence of infrared divergences in the production cross
sections and decay rates of $P$-wave states.  Within the NRQCD
approach, this problem finds its natural solution since the infrared
singularities are factored into a colour-octet operator matrix element
\cite{bprod:Bodwin:1992ye}. While colour-octet contributions are needed for a
consistent description of $P$-wave quarkonia, they are
phenomenologically even more important for $S$-wave states like
$J/\psi$ or $\Upsilon$. According to the velocity scaling rules,
colour-octet matrix elements for the production of $S$-wave quarkonia
are suppressed by a factor $v^4$ compared to the leading
colour-singlet contributions.  However, as discussed in some detail
below, colour-octet processes can become significant if the
short-distance cross section for producing $Q\overline{Q}$ in a
colour-octet state is enhanced.

\begin{figure}[htb]
\begin{center}
\includegraphics[width=0.8\textwidth,clip]{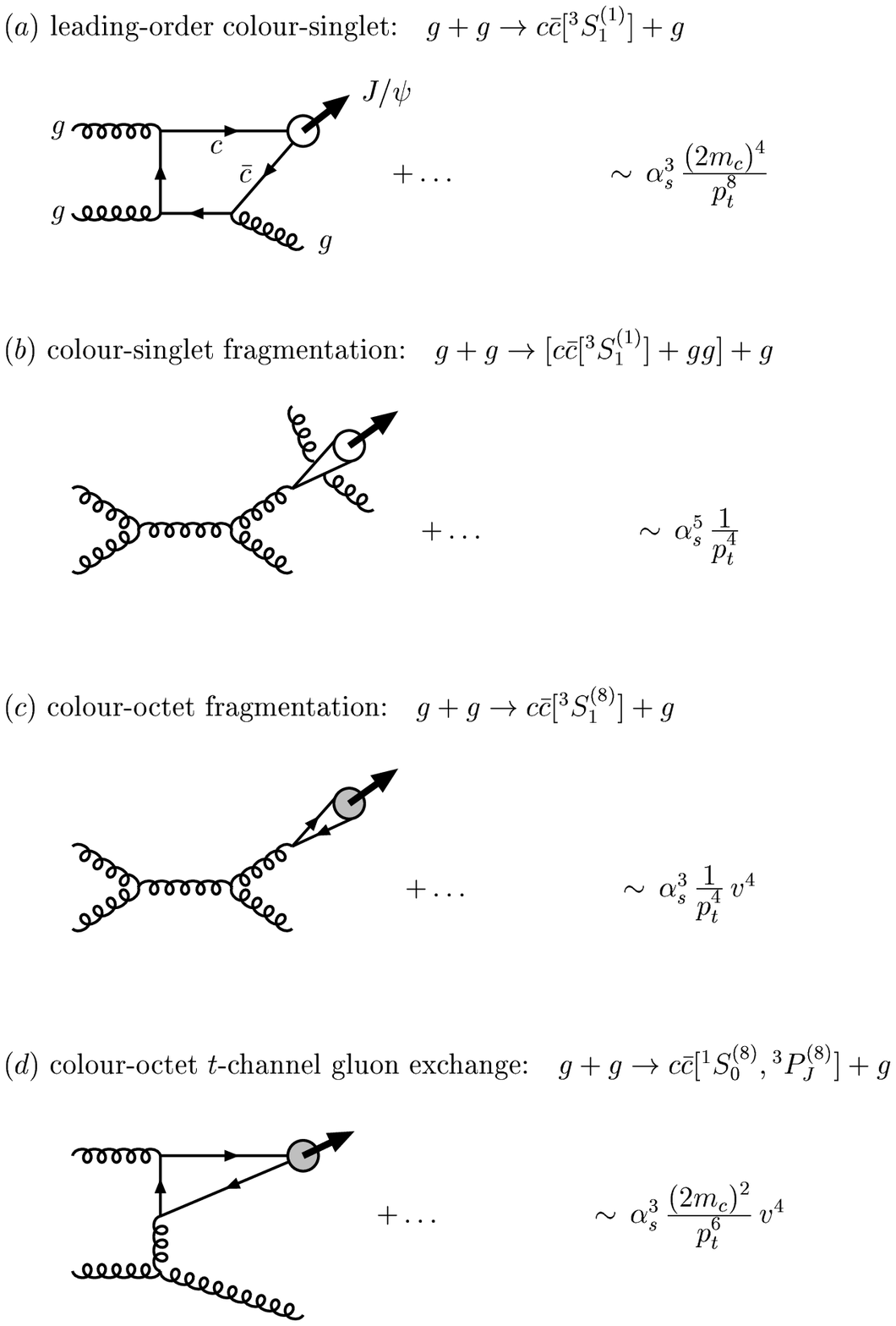}
\caption[]{Generic diagrams for $J/\psi$ 
production in hadron-hadron collisions via colour-singlet and
colour-octet channels.}
\label{fig:quarkonium1}
\end{center}
\end{figure}

The production of $S$-wave charmonium in $p\bar{p}$ collisions at the
Tevatron has attracted considerable attention and has stimulated much
of the recent theoretical development in quarkonium physics.  The CDF
collaboration has measured cross sections for the production of
$J/\psi$ and $\psi(2S)$ states not coming from $B$ or radiative $\chi$
decays, for a wide range of transverse momenta $5\,\mbox{GeV}\;
\simlt\; p_t(\psi) \;\simlt\;20\,\mbox{GeV}$
\cite{bprod:Abe:1992ww} \cite{bprod:Abe:1997jz}.
Surprisingly, the experimental cross sections were found to be orders
of magnitudes larger than the theoretical expectation based on the
leading-order colour-singlet model
\cite{bprod:Baier:1983va} \cite{bprod:Gastmans:1987be}. This result is particularly
striking because the data extends out to large transverse momenta
where the theoretical analysis is rather clean. The shortcoming of the
colour-singlet model can be understood by examining a typical Feynman
diagram contributing to the leading-order parton cross section,
fig.~\ref{fig:quarkonium1}(a). 
At large transverse momentum, the two
internal quark propagators are off-shell by $\sim p_t^2$ so that the
parton differential cross section scales like $\mbox{d}\sigma/\mbox{d}
p_t^2 \sim 1/p_t^8$, as indicated in the figure. On the other hand,
when $p_t \gg 2m_c$ the quarkonium mass can be considered small and
the inclusive charmonium cross section is expected to scale like any
other single-particle inclusive cross section $\sim 1/p_t^4$. The
dominant production mechanism for charmonium at sufficiently large
$p_t$ must thus be via fragmentation
\cite{bprod:Braaten:1993rw}, the production of a parton with large $p_t$
which subsequently decays into charmonium and other partons.  A
typical fragmentation contribution to colour-singlet $J/\psi$
production is shown in fig.~\ref{fig:quarkonium1}(b). While the
fragmentation contributions are of higher order in $\alpha_s$ compared
to the fusion process fig.~\ref{fig:quarkonium1}(a), they are enhanced
by a power $p_t^4/(2m_c)^4$ at large $p_t$ and can thus overtake the
fusion contribution at $p_t \gg 2m_c$. When colour-singlet
fragmentation is included, the $p_t$ dependence of the theoretical
prediction is in agreement with the Tevatron data but the
normalization is still underestimated by about an order of magnitude
\cite{bprod:Cacciari:1994dr} \cite{bprod:Braaten:1994xb}
\cite{bprod:Roy:1994ie}, indicating that an
additional fragmentation contribution is still missing. It is now 
generally believed that gluon fragmentation into colour-octet
${}^3S_1$ charm quark pairs \cite{bprod:Braaten:1995vv}
\cite{bprod:Cacciari:1995yt}, as
shown in fig.~\ref{fig:quarkonium1}(c), is the dominant source of
$J/\psi$ and $\psi(2S)$ at large $p_t$ at the Tevatron. The probability
of forming a $J/\psi$ particle from a pointlike $c\bar{c}$ pair in a
colour-octet ${}^3S_1$ state is given by the NRQCD matrix element
$\langle {\cal O}^{J/\psi}[{}^3S_1^{(8)}] \rangle$ which is suppressed
by $v^4$ relative to the non-perturbative factor of the leading
colour-singlet term. However, this suppression is overcompensated for
by the gain in two powers of $\alpha_s/\pi$ in the short-distance
cross section for producing colour-octet ${}^3S_1$ charm quark pairs
as compared to colour-singlet fragmentation. At ${\cal{O}}(v^4)$ in
the velocity expansion, two additional colour-octet channels have to
be included, fig.~\ref{fig:quarkonium1}(d), which do not have a
fragmentation interpretation at order $\alpha_s^3$ but which become
significant at moderate $p_t\sim 2m_c$
\cite{bprod:Cho:1996vh} \cite{bprod:Cho:1996ce}.
The importance of the ${}^1S_0^{(8)}$
and ${}^3P_J^{(8)}$ contributions cannot be estimated from naive power
counting in $\alpha_s$ and $v$ alone, but rather follows from the
dominance of $t$-channel gluon exchange, forbidden in the
leading-order colour-singlet cross section.

\begin{figure}[htb]
\begin{center}
\includegraphics[width=0.8\textwidth,clip]{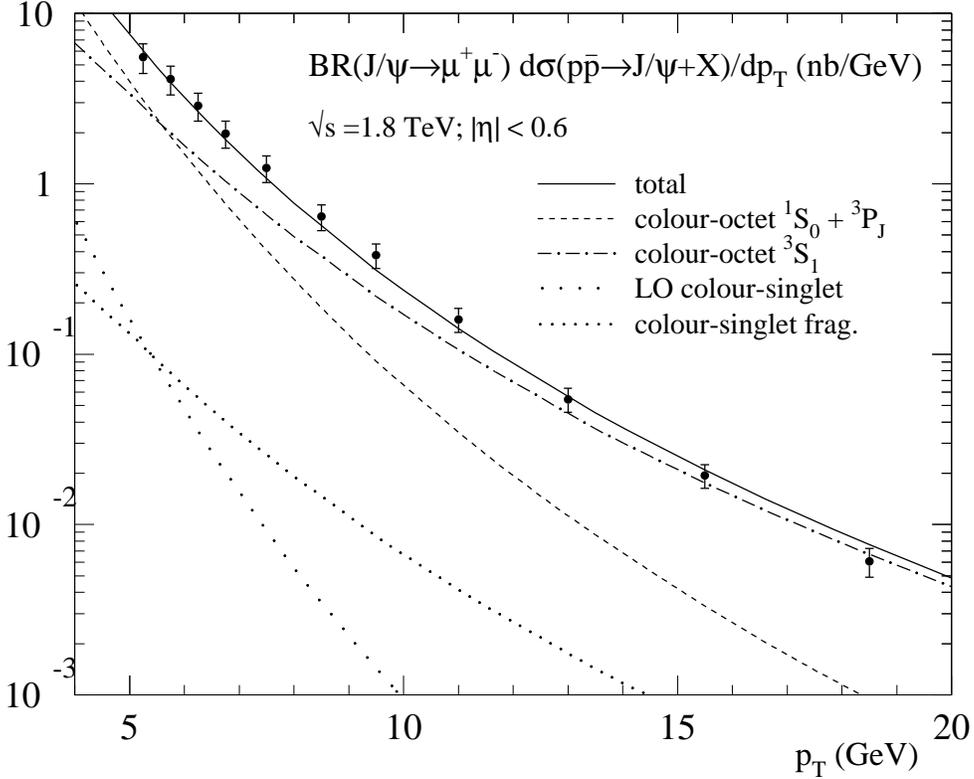}
\caption[]{Colour-singlet and colour-octet
 contributions to direct $J/\psi$ production in $p\bar{p} \to
 J/\psi+X$ at the Tevatron ($\sqrt{s}=1.8$~TeV, pseudorapidity cut
 $|\eta|<0.6$)) together with experimental data from CDF
 \cite{bprod:Abe:1997jz}. Parameters: CTEQ5L parton distribution functions
 \cite{bprod:Lai:1999wy}; factorization and renormalization scale $\mu =
 \sqrt{p_t^2+4m_c^2}$; $m_c = 1.5$~GeV. The leading logarithms
 $(\alpha_s\ln p_t^2/(2m_c)^2)^n$ have been summed by solving the
 Altarelli-Parisi evolution equations for the gluon fragmentation
 function. NRQCD matrix elements as specified in
 Table~\ref{tab:quarkonium1}.}
\label{fig:quarkonium2} 
\end{center}
\end{figure}

The different contributions to the $J/\psi$ transverse momentum
distribution are compared to the CDF data \cite{bprod:Abe:1997jz} in
fig.~\ref{fig:quarkonium2}.
\begin{table}
\begin{center}
\caption{NRQCD matrix elements for charmonium production.
The colour-singlet matrix elements are taken from the potential
model calculation of \cite{bprod:Buchmuller:1981su}
\cite{bprod:Eichten:1995ch}. The
colour-octet matrix elements have been extracted from the CDF data
\cite{bprod:Abe:1997jz}, where $M_k^H(^1\!S_0^{(8)},^3\!P_0^{(8)})\equiv
\langle {\cal O}_8^{H}[{}^1S_0]\rangle + k\,\langle {\cal O}_8^{H}
[{}^3P_0]\rangle/m_c^2$. The errors quoted are statistical
only. Parameters: CTEQ5L parton distribution functions
\cite{bprod:Lai:1999wy}, renormalization and factorization scale
$\mu=(p_t^2+4 m_c^2)^{1/2}$ and $m_c=1.5\,$GeV. The Altarelli-Parisi
evolution has been included for the ${}^3S_1^{(8)}$ fragmentation
contribution. See \cite{bprod:Beneke:1997yw} for further details.}
\label{tab:quarkonium1}
\vskip0.2cm
\renewcommand{\arraystretch}{1.5}
$$
\begin{array}{cccc}
\hline\hline
 H & \langle {\cal{O}}_1^{H} \rangle  & \langle
 {\cal{O}}_8^{H}[{}^3S_1] \rangle  &
 M_{3.5}^{H}({}^1S_0^{(8)},{}^3P_0^{(8)})\\ \hline
 J/\psi   & 1.16~{\rm GeV^3} & (1.19 \pm 0.14)\cdot 10^{-2}~{\rm GeV}^3 &  
 (4.54 \pm 1.11)\cdot 10^{-2}~{\rm GeV}^3 \\[-1mm] 
 \psi(2S) & 0.76~{\rm GeV^3} & (0.50 \pm 0.06)\cdot 10^{-2}~{\rm GeV}^3 & 
 (1.89 \pm 0.46)\cdot 10^{-2}~{\rm GeV}^3  \\[-1mm]
 \chi_{0} & 0.11~{\rm GeV^5} & (0.31 \pm 0.04)\cdot 10^{-2}~{\rm GeV}^3 & 
 \\ \hline \hline
\end{array}
$$
\renewcommand{\arraystretch}{1.0}
\end{center}
\end{table}
As mentioned above, the colour-singlet
model at lowest order in $\alpha_s$ fails dramatically when confronted
with the experimental results. When colour-singlet fragmentation is
included, the prediction increases by more than an order of magnitude
at large $p_t$, but it still falls below the data by a factor of $\sim
30$.  The CDF results on charmonium production can be explained by
including the leading colour-octet contributions and adjusting the
unknown non-perturbative parameters to fit the data. Numerically one
finds the non-perturbative matrix elements to be of
${\cal{O}}(10^{-2}~\mbox{GeV}^3)$, see Table~\ref{tab:quarkonium1},
perfectly consistent with the $v^4$ suppression expected from the
velocity scaling rules. Similar conclusions can be drawn for $\psi(2S)$
production at the Tevatron.

\begin{figure}[htb]
\begin{center}
\includegraphics[width=0.8\textwidth,clip]{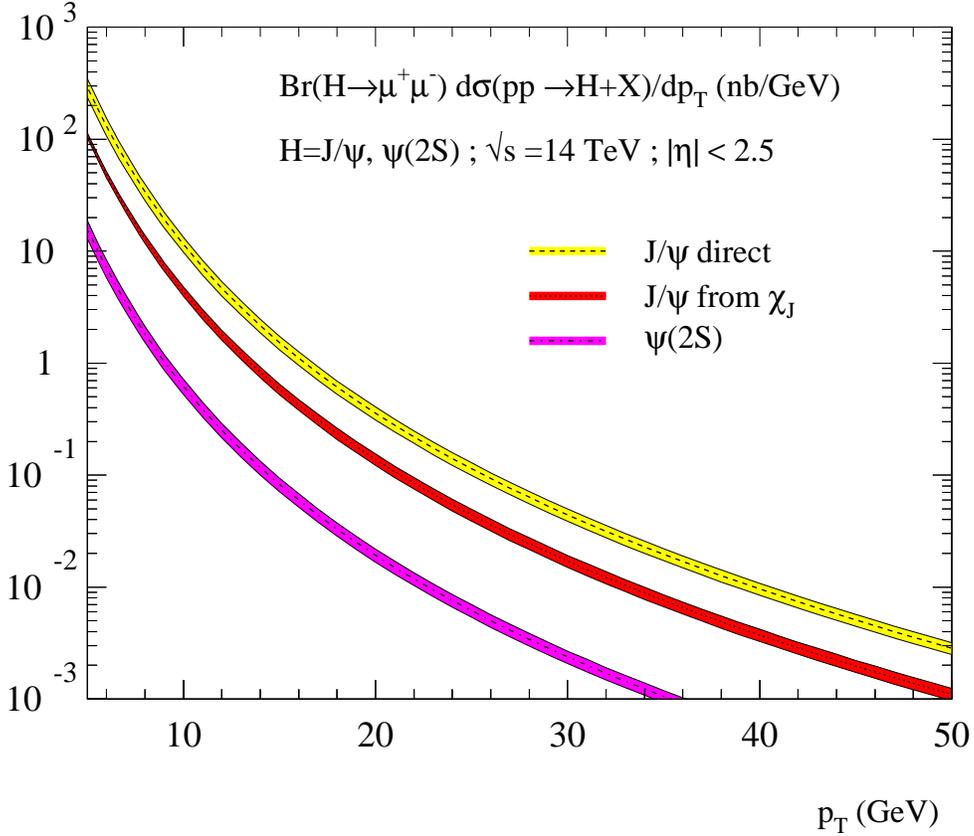}
\caption[]{Cross sections for $J/\psi$ and $\psi(2S)$
production in $pp \to \psi+X$ at the LHC ($\sqrt{s}=14$~TeV,
pseudorapidity cut $|\eta|<2.5$). Parameter specifications as in
fig.~\ref{fig:quarkonium2}. The leading logarithms $(\alpha_s\ln
p_t^2/(2m_c)^2)^n$ have been summed by solving the Altarelli-Parisi
evolution equations for the gluon fragmentation function.
The error bands include the statistical errors in the extraction 
of the NRQCD matrix elements [Table~\ref{tab:quarkonium1}] only.}
\label{fig:quarkonium3} 
\end{center}
\end{figure}

The analysis of the CDF data alone, although very encouraging, does
not strictly prove the phenomenological relevance of colour-octet
contributions because free parameters have to be introduced to fit the
data.  However, if factorization holds the non-perturbative matrix
elements, Table~\ref{tab:quarkonium1}, are universal and can be used
to make predictions for various processes and observables.  Besides a
global analysis of different reactions, the measurement of quarkonium
cross sections at the LHC will be crucial to assess the importance of
the individual production mechanisms and to test factorization. In
fig.~\ref{fig:quarkonium3} we have collected the cross section
predictions for direct $J/\psi$ and $\psi(2S)$ production as well as
the production of $J/\psi$ from radiative $\chi$ decays at the
LHC.
The theoretical curves include the statistical errors in the
extraction of the NRQCD matrix elements
[Table~\ref{tab:quarkonium1}]. There are, however, additional
theoretical uncertainties which might affect the prediction, but which
have not yet been fully quantified. In particular the determination of
the $\langle {\cal O}_8^{\psi}[{}^1S_0]\rangle$ and $\langle {\cal
O}_8^{\psi}[{}^3P_0]\rangle$ matrix elements ($\psi$ denoting $J/\psi$
or $\psi(2S)$) from the Tevatron data is very sensitive to effects
that modify the shape of the charmonium $p_t$ distribution at
relatively small $p_t\;\simlt\; 8$~GeV. Those effects include the
small-$x$ behaviour of the gluon distribution \cite{bprod:Beneke:1997yw},
the evolution of the strong coupling \cite{bprod:Beneke:1997yw}, as well as
systematic effects inherent in NRQCD, such as the inaccurate treatment
of the energy conservation in the hadronization of the colour-octet
$c\bar{c}$ pairs \cite{bprod:Beneke:1997qw}.  Moreover, higher-order QCD
corrections are expected to play an important role, as discussed in
more detail below. The cross sections collected in
fig.~\ref{fig:quarkonium3} should thus not be viewed as firm NRQCD
predictions but will be refined as more experimental and theoretical
information on charmonium production becomes available over the next
few years.

The inclusion of higher-order QCD corrections is required to reduce
the theoretical uncertainty and to allow a more precise prediction of
the LHC cross sections. Next-to-leading order (NLO) calculations for
quarkonium production at hadron colliders are presently available only
for total cross sections \cite{bprod:Kuhn:1993qw}
\cite{bprod:Petrelli:1998ge}. Significant higher-order corrections to
differential distributions are expected from the strong
renormalization and factorization scale dependence of the
leading-order results \cite{bprod:Beneke:1997qw}. Moreover, the NLO
colour-singlet cross section includes processes like $g + g \to
Q\overline{Q}[{}^3S_1^{(1)}] + g + g $ which are dominated by
$t$-channel gluon exchange and scale as $\sim \alpha_s^4
(2m_Q)^2/p_t^6$. At $p_t\gg 2m_Q$ their contribution is enhanced with
respect to the the leading-order cross section,
fig.~\ref{fig:quarkonium1}(a), which scales as $\sim \alpha_s^3
(2m_Q)^4/p_t^8$. This is born out by preliminary studies
\cite{bprod:Petrelli:1999rh} which include part of the NLO hadroproduction 
cross section and by the complete calculation of NLO corrections to
the related process of quarkonium photoproduction
\cite{bprod:Kramer:1996nb}. The NLO colour-singlet cross section may be
comparable in size to the colour-octet ${}^1S_0 $ and ${}^3P_J$
processes, which scale as $ \sim \alpha_s^3 v^4 (2m_Q)^2/p_t^6 $ (see
fig.~\ref{fig:quarkonium1}(d)), and affect the determination of the
corresponding NRQCD matrix elements from the Tevatron data.  A full
NLO analysis is however needed before quantitative conclusions can be
drawn.
 
Another source of potentially large higher-order corrections is the
multiple emission of soft or almost collinear gluons from the initial
state partons. These corrections, as well as effects related to
intrinsic transverse momentum, are expected to modify the shape of the
transverse momentum distribution predominantly at relatively low
values of $p_t\;\simlt\; 2m_Q$. Initial state radiation can be
partially summed in perturbation theory \cite{bprod:Collins:1985kg}, but
so far only total cross sections have been considered in the
literature \cite{bprod:Cacciari:1999sy}.  An estimate of the effect on the
transverse momentum distribution should be provided by
phenomenological models where a Gaussian $k_t$-smearing is added to
the initial state partons. The result of these calculations not only
depends on the average $\langle k_t \rangle$, which enters as a free
parameter, but also on the details of how the smearing is
implemented. Moreover, a lower cut-off has to be provided which
regulates the divergences at $p_t = 0$. Using the NLO calculation for
the total cross section \cite{bprod:Petrelli:1998ge}, one can obtain the
rough estimate that perturbative Sudakov effects should be confined
below $p_t \sim 1-2$ GeV for both charmonium and bottomonium
production at Tevatron energies. Qualitatively, the inclusion of
$k_t$-smearing leads to an enhancement of the short distance cross
section at small $p_t$, which results in smaller values for the fits
of the $\langle {\cal O}_8^{\psi}[{}^1S_0]\rangle$ and $\langle {\cal
O}_8^{\psi}[{}^3P_0] \rangle$ NRQCD matrix elements 
\cite{bprod:Sridhar:1998rt} \cite{bprod:Petrelli:1999rh}.
The actual size of the effect,
however, turns out to be very different for the two models studied in
the literature.

\begin{figure}[htb]
\begin{center}
\includegraphics[width=0.9\textwidth,clip]{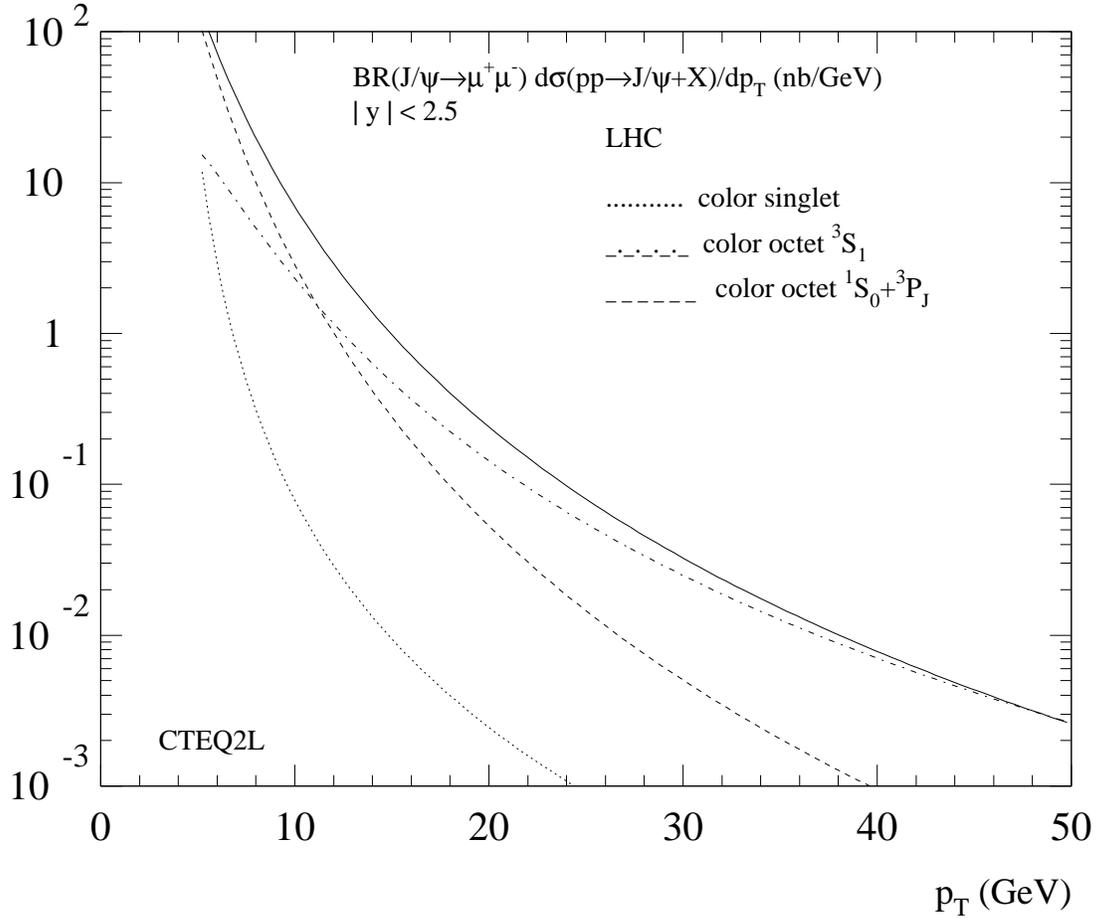}
\caption[]{Cross sections for $J/\psi$
production in $pp \to J/\psi+X$ at the LHC ($\sqrt{s}=14$~TeV,
rapidity cut $|y|<2.5$) obtained from a Monte Carlo event generator
\cite{bprod:Sanchis-Lozano:1999um}. CTEQ2L parton distribution functions
\cite{bprod:Tung:1994ua}; (i) dotted line: colour-singlet, 
(ii) dashed line: colour-octet $^1S_0 + {}^3P_J$, (iii) dot-dashed
line: $^3S_1^{(8)}$, (iv) solid line: all contributions.  NRQCD matrix
elements as specified in \cite{bprod:Sanchis-Lozano:1999um}.}
\label{fig:quarkonium4} 
\end{center}
\end{figure}

An alternative approach to treat the effect of initial state radiation
is by means of Monte Carlo event generators which include multiple
gluon emission in the parton shower approximation. Comprehensive
phenomenological analyses have been carried out for charmonium
production at the Tevatron and at the LHC
\cite{bprod:Cano-Coloma:1997rn} \cite{bprod:Sanchis-Lozano:1999um}
\cite{bprod:Kniehl:1999qy}
using the event generator \Py\ \cite{bprod:Pythia:ref} supplemented
by the leading colour-octet processes \cite{bprod:Cano-Coloma:1997rn}. The
inclusion of initial state radiation as implemented in \Py\ leads to
an enhancement of the short-distance cross section. The size of the
effect is significantly larger than for the Gaussian $k_t$-smearing
mentioned above, and it extents out to large $p_t$. Consequently, the
$\langle {\cal O}_8^{\psi}[{}^1S_0]\rangle$ and $\langle {\cal
O}_8^{\psi}[{}^3P_0] \rangle$ NRQCD matrix elements estimated from the
Monte Carlo analysis of the Tevatron cross sections are significantly
lower than the ones listed in Table~\ref{tab:quarkonium1} (see
\cite{bprod:Cano-Coloma:1997rn} \cite{bprod:Sanchis-Lozano:1999um} for
details).\footnote{Support is added to the Monte Carlo extraction of
the NRQCD matrix elements by analyses of $J/\psi$ production in
inelastic $\gamma p$-scattering \cite{bprod:Cacciari:1996dg}
\cite{bprod:Ko:1996xw} and
$B$ decays \cite{bprod:Beneke:1999ks}, which seem to prefer small values of
$\langle {\cal O}_8^{\psi}[{}^1S_0]\rangle$ and $\langle {\cal
O}_8^{\psi}[{}^3P_0] \rangle$.} Figure~\ref{fig:quarkonium4} shows the
individual contributions to the direct $J/\psi$ cross section at the
LHC as estimated with the \Py\ Monte Carlo
\cite{bprod:Sanchis-Lozano:1999um}.
Note that for consistency the curves are
based on the NRQCD matrix elements extracted from the Monte Carlo
analysis of the Tevatron data rather than on the values of the
leading-order fit listed in Table~\ref{tab:quarkonium1}. One observes
that the final prediction is consistent with the result presented in
fig.~\ref{fig:quarkonium3} within errors. The extrapolation of the
Tevatron fits to LHC energies seems rather insensitive to the details
of the underlying theoretical description, and different approaches
yield similar predictions for the LHC cross sections as long as the
appropriate NRQCD matrix elements are used. The Monte Carlo
implementation \cite{bprod:Cano-Coloma:1997rn} should therefore represent a
convenient and reliable tool for the experimental simulation of
quarkonium production processes at the LHC.

\begin{figure}[htb]
\begin{center}
\includegraphics[width=0.8\textwidth,clip]{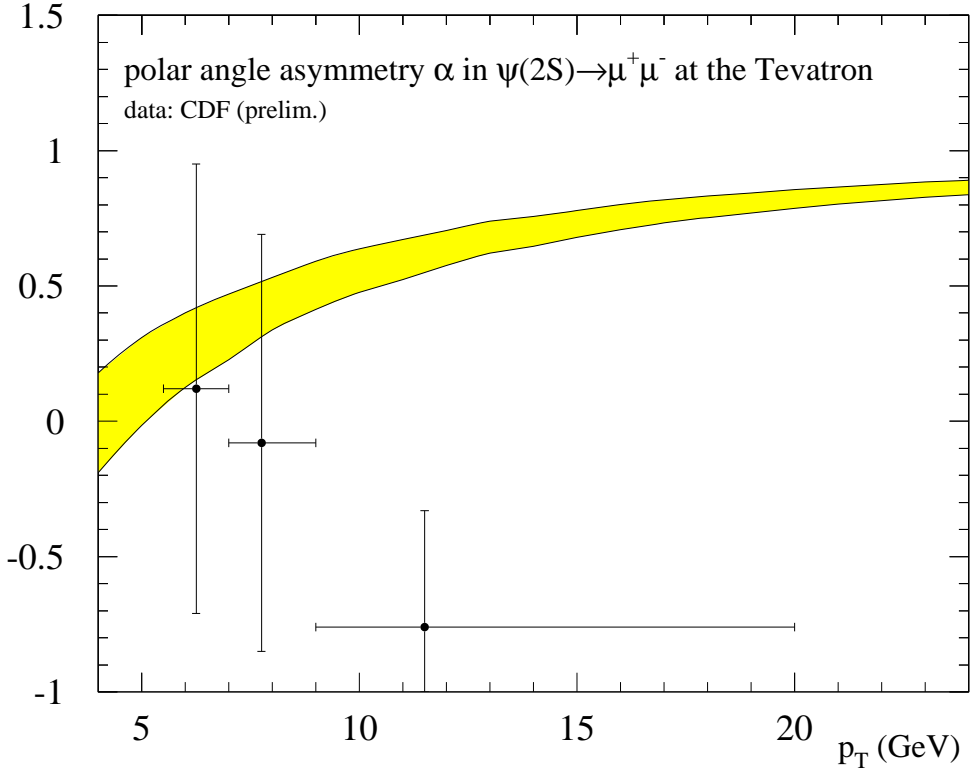}
\caption[]{Polar angle asymmetry $\alpha$ for $\psi(2S)$ production 
in $p\bar{p} \to \psi(2S)(\to \mu^+\mu^-)+X$ at the Tevatron as a
function of $p_t$ compared to preliminary data from CDF
\cite{bprod:CDF-psi}. Parameter specifications as in
fig.~\ref{fig:quarkonium2}.  NLO corrections to the fragmentation
contribution \cite{bprod:Beneke:1996yb} \cite{bprod:Beneke:1997yw}
have not been
included. The error band is obtained as a combination of the
uncertainty (statistical only) in the extraction of the NRQCD matrix
elements [Table~\ref{tab:quarkonium1}] and the limiting cases that
either $\langle {\cal O}_8^{\psi}({}^1S_0)\rangle$ or $\langle {\cal
O}_8^{\psi}({}^3P_0)\rangle$ is set to zero in the linear combination
extracted from the data.}
\label{fig:quarkonium5} 
\end{center}
\end{figure}

A crucial test of the NRQCD approach to charmonium production at
hadron colliders is the analysis of $J/\psi$ and $\psi(2S)$
polarization at large transverse momentum. Recall that at large $p_t$,
$\psi$ production should be dominated by gluon fragmentation into a
colour-octet ${}^3S_1$ charm quark pair, fig.~\ref{fig:quarkonium1}(c).
When $p_t \gg 2m_c$ the fragmenting gluon is effectively on-shell and
transverse.  The intermediate $c\bar{c}$ pair in the colour-octet
${}^3S_1$ state inherits the gluon's transverse polarization and so
does the quarkonium, because the emission of soft gluons during
hadronization does not flip the heavy quark spin at leading order in
the velocity expansion. Consequently, at large transverse momentum one
should observe transversely polarized $J/\psi$ and $\psi(2S)$
\cite{bprod:Cho:1995ih}.  The polarization can be measured through the angular
distribution in the decay $\psi\to l^+l^-$, given by $ \mbox{d}\Gamma
/ \mbox{d}\!\cos\theta \propto 1+\alpha\,\cos^2\theta$, where $\theta$
denotes the angle between the lepton three-momentum in the $\psi$ rest
frame and the $\psi$ three-momentum in the lab frame. Pure transverse
polarization implies $\alpha=1$. Corrections to this asymptotic limit
due to spin-symmetry breaking and higher order fragmentation
contributions have been estimated to be small \cite{bprod:Beneke:1996yb}.
The dominant source of depolarization comes from the colour-octet
fusion diagrams, fig.~\ref{fig:quarkonium1}(d), which are important at
moderate $p_t$. Still, at ${\cal{O}}(v^4)$ in the velocity expansion,
the polar angle asymmetry $\alpha$ can be unambiguously calculated
within NRQCD \cite{bprod:Beneke:1997yw}
\cite{bprod:Leibovich:1997pa} in terms of the
three non-perturbative matrix elements [Table~\ref{tab:quarkonium1}]
that have been determined from the unpolarized cross section. In
fig.~\ref{fig:quarkonium5} we display the theoretical prediction for
$\alpha$ in $\psi(2S)$ production at the Tevatron as function of the
$\psi(2S)$ transverse momentum.
No transverse polarization is expected
at $p_t\sim 5\,$ GeV, but the angular distribution is predicted to
change drastically as $p_t$ increases. A preliminary measurement from
CDF \cite{bprod:CDF-psi} does not support this prediction, but the
experimental errors are too large to draw definite conclusions. A
similar picture emerges from the analysis of $J/\psi$ polarization
\cite{bprod:CDF-jpsi}, where, however, the theoretical analysis is
complicated by the fact that the data sample still includes $J/\psi$
that have not been produced directly but come from decays of higher
excited states \cite{bprod:Braaten:1999qk}.

Polarization measurements are crucial to discriminate the NRQCD
approach from the colour evaporation model, where the cross section
for a specific charmonium state is given as a universal fraction of
the inclusive $c\overline{c}$ production cross section integrated up
to the open charm threshold. In general, the assumption of a single
universal long-distance factor is too restrictive. It implies a
universal $\sigma(\chi_c)/\sigma(J/\psi)$ ratio, which is not
supported by the comparison of charmonium production in 
hadron-hadron and photon-hadron collisions. Still, since the colour
evaporation model allows colour-octet charm quark pairs from gluon
fragmentation to hadronize into charmonium, it can describe the $p_t$
distribution of the Tevatron data
\cite{bprod:Gavai:1995in} \cite{bprod:Schuler:1996ku}
\cite{bprod:Amundson:1997qr} \cite{bprod:Edin:1997zb}. 
In contrast to the NRQCD approach, however, the colour evaporation
model predicts charmonium to be produced unpolarized. The model
assumes unsuppressed gluon emission from the $c\bar{c}$ pair during
hadronization which randomizes spin and colour. This assumption is
clearly wrong in the heavy quark limit where spin symmetry is at work
and soft gluon emission does not flip the heavy quark spin.
Nonetheless, since the charm quark mass is not very large with respect
to the QCD scale, the applicability of heavy quark spin symmetry to
charmonium physics has to be tested by confronting the NRQCD
polarization signature with experimental data.

\begin{figure}[htb]
\begin{center}
\includegraphics[width=0.8\textwidth,clip]{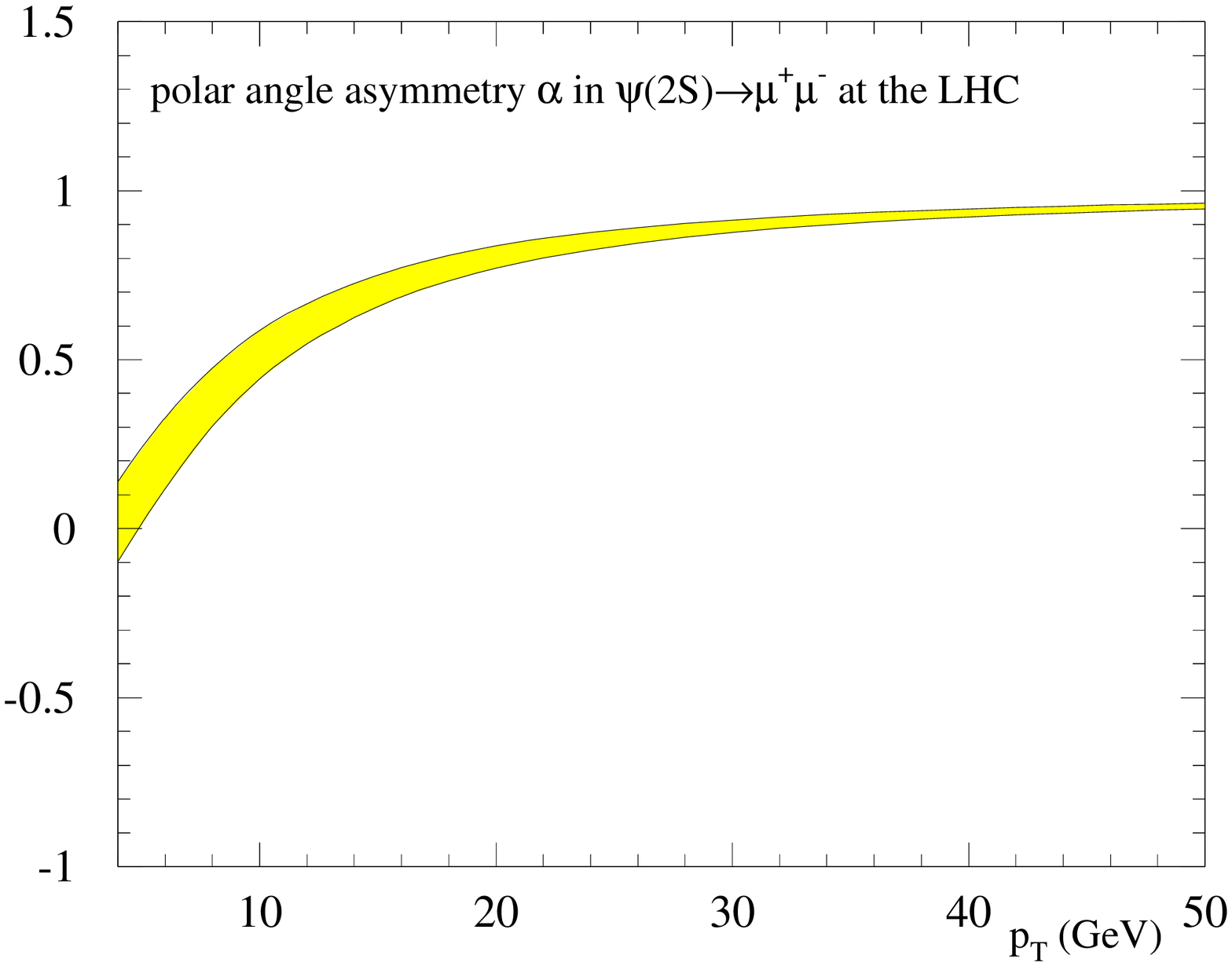}
\caption[]{Polar angle asymmetry $\alpha$ for $\psi(2S)$ production 
in $pp \to \psi(2S)(\to \mu^+\mu^-)+X$ at the LHC as a
function of $p_t$. Parameter specifications as in
fig.~\ref{fig:quarkonium5}.}
\label{fig:quarkonium6} 
\end{center}
\end{figure}

To definitely resolve the issue of quarkonium polarization, a
high-statistics measurement extending out to large transverse momentum
will be necessary. Such a measurement can be carried out at the LHC,
where one expects a polarization pattern similar to that predicted for
the Tevatron, see fig.~\ref{fig:quarkonium6}.
The absence of a
substantial fraction of transverse polarization in $\psi$ production
at large $p_t$ would represent a serious problem for the application
of the NRQCD factorization approach to the charmonium system and might
indicate that the charm quark mass is not large enough for a
nonrelativistic approach to work in all circumstances.

\begin{figure}[htb]
\begin{center}
\includegraphics[width=0.8\textwidth,clip]{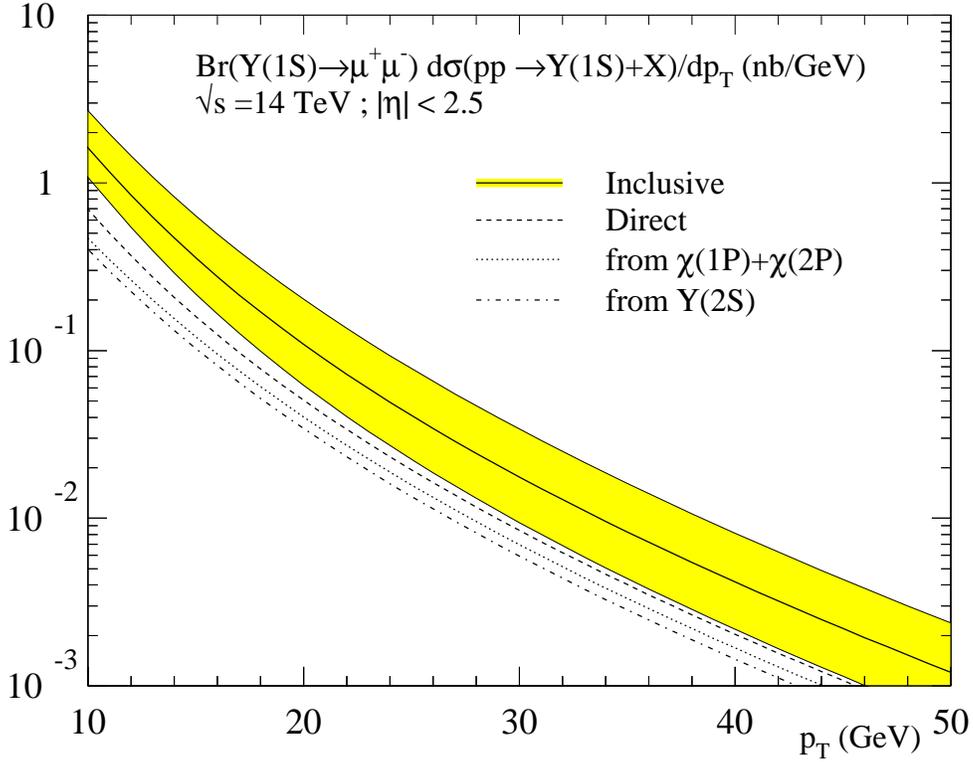}
\caption[]{Cross sections for $\Upsilon(1S)$
production in $pp \to \Upsilon(1S)+X$ at the LHC ($\sqrt{s}=14$~TeV,
pseudorapidity cut $|\eta|<2.5$). Parameters: CTEQ5L parton
distribution functions \cite{bprod:Lai:1999wy}, factorization and
renormalization scale $\mu = \sqrt{p_t^2+4m_b^2}$, $m_b =
4.88$~GeV. The leading logarithms $(\alpha_s\ln p_t^2/(2m_b)^2)^n$
have been summed by solving the Altarelli-Parisi evolution equations
for the gluon fragmentation function. NRQCD matrix elements as
specified in Table~\ref{tab:quarkonium2}. The error band is obtained
by varying the colour-octet matrix elements between half and twice
their central value for illustration.}
\label{fig:quarkonium7} 
\end{center}
\end{figure}

\begin{figure}[htb]
\begin{center}
\includegraphics[width=0.8\textwidth,clip]{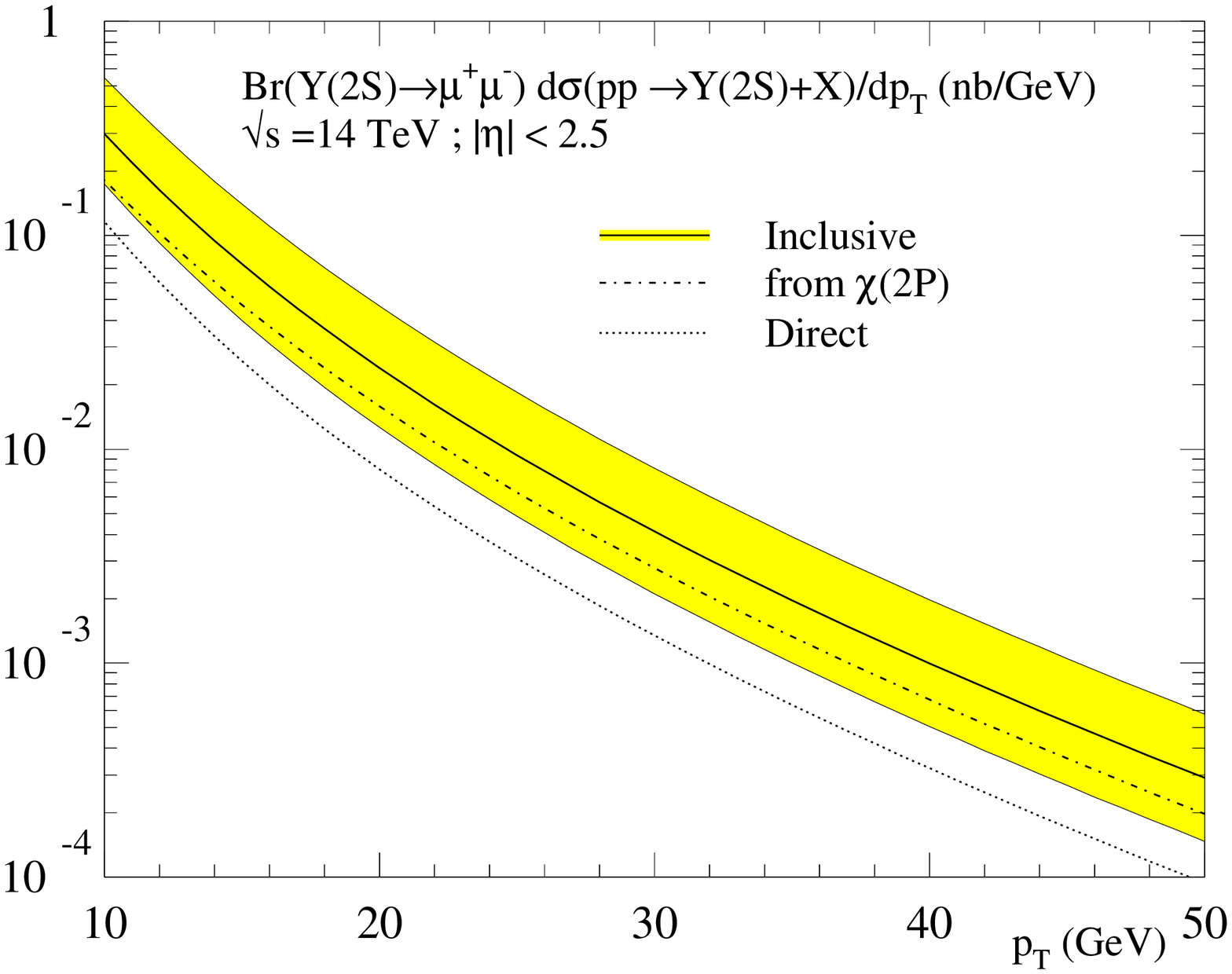}
\caption[]{Same as fig.~\ref{fig:quarkonium7} for $\Upsilon(2S)$
production.}
\label{fig:quarkonium8} 
\end{center}
\end{figure}

The application of NRQCD should be on safer grounds for the
bottomonium system. As $v^2 \sim 0.1$ for bottomonium, higher-order
terms in the velocity expansion (in particular colour-octet
contributions) are expected to be less relevant than in the case of
charmonium. Cross sections for the production of $\Upsilon$ states
have been measured at the Tevatron in the region $p_t
\;\simlt\;20\,\mbox{GeV}$ \cite{bprod:Abe:1995an}
\cite{bprod:Affolder:1999wm} \cite{bprod:CDF-ups}.
The leading-order colour-singlet model predictions underestimate the
data, the discrepancy being, however, much less significant than in
the case of charmonium. Given the large theoretical uncertainties in
the cross section calculation, in particular at small
$p_t\;\simlt\;M_{\Upsilon}$, the need for colour-octet contribution is
not yet as firmly established as for charmonium production. The
inclusion of both next-to-leading order corrections and the
summation of soft gluon radiation is required to obtain a realistic
description of the $\Upsilon$ cross section in the $p_t$-range probed
by present data. Such calculations have not yet been performed, and we
have therefore not attempted a systematic fit
\cite{bprod:Cho:1996vh} \cite{bprod:Cho:1996ce}
of the bottomonium NRQCD matrix
elements. Our predictions for the $\Upsilon$ cross section at the LHC,
figs.~\ref{fig:quarkonium7},\ref{fig:quarkonium8}, are based on a simple
choice of the non-perturbative input parameters
[Table~\ref{tab:quarkonium2}] which is consistent with the present
experimental information from the Tevatron.
\begin{table}
\begin{center}
\caption{ NRQCD matrix elements for bottomonium production.
The colour-singlet matrix elements are taken from the potential model
calculation of \cite{bprod:Buchmuller:1981su}
\cite{bprod:Eichten:1995ch}. The
colour-octet matrix elements have been determined from the CDF data
for $p_t > 8~{\rm GeV} $ \cite{bprod:Affolder:1999wm}, where $\langle {\cal
O}_8^{H}[{}^1S_0]\rangle = \langle {\cal
O}_8^{H}[{}^3P_0]\rangle/m_b^2$ has been assumed for
simplicity. Parameters: CTEQ5L parton distribution functions
\cite{bprod:Lai:1999wy}, renormalization and factorization scale
$\mu=(p_t^2+4 m_b^2)^{1/2}$ and $m_b=4.88\,$GeV.}
\label{tab:quarkonium2}
\vskip0.2cm
\renewcommand{\arraystretch}{1.5}
$$
\begin{array}{cccc}
\hline\hline
 H & \langle {\cal{O}}_1^{H} \rangle  & \langle
 {\cal{O}}_8^{H}[^3S_1] \rangle  
 &\langle {\cal{O}}_8^{H}[^1S_0] \rangle\\ \hline
 \Upsilon(1S)   & 9.28  ~{\rm GeV^3} & 15 \cdot 10^{-2}~{\rm GeV}^3 &
 2.0 \cdot 10^{-2}~{\rm GeV}^3 \\[-1mm] 
 \Upsilon(2S)   & 4.63  ~{\rm GeV^3} & 4.5 \cdot 10^{-2}~{\rm GeV}^3 &
 0.6 \cdot 10^{-2}~{\rm GeV}^3 \\[-1mm]
 \Upsilon(3S)   & 3.54  ~{\rm GeV^3} & 7.5 \cdot 10^{-2}~{\rm GeV}^3 &
 1.0 \cdot 10^{-2}~{\rm GeV}^3 \\[-1mm]
 \chi_{0}(1P)   & 2.03  ~{\rm GeV^5} & 4.0 \cdot 10^{-2}~{\rm GeV}^3 
 \\[-1mm]
 \chi_{0}(2P)   & 2.57  ~{\rm GeV^5} & 6.5 \cdot 10^{-2}~{\rm GeV}^3
 \\ 
\hline \hline
\end{array}
$$
\renewcommand{\arraystretch}{1.0}
\end{center}
\end{table}
The cross sections should
thus not be regarded as firm predictions of NRQCD but rather as
order-of-magnitude estimates. The expected theoretical progress and
more experimental information will allow a more precise prediction in
the near future.

\begin{figure}[htb]
\begin{center}
\includegraphics[width=0.9\textwidth,clip]{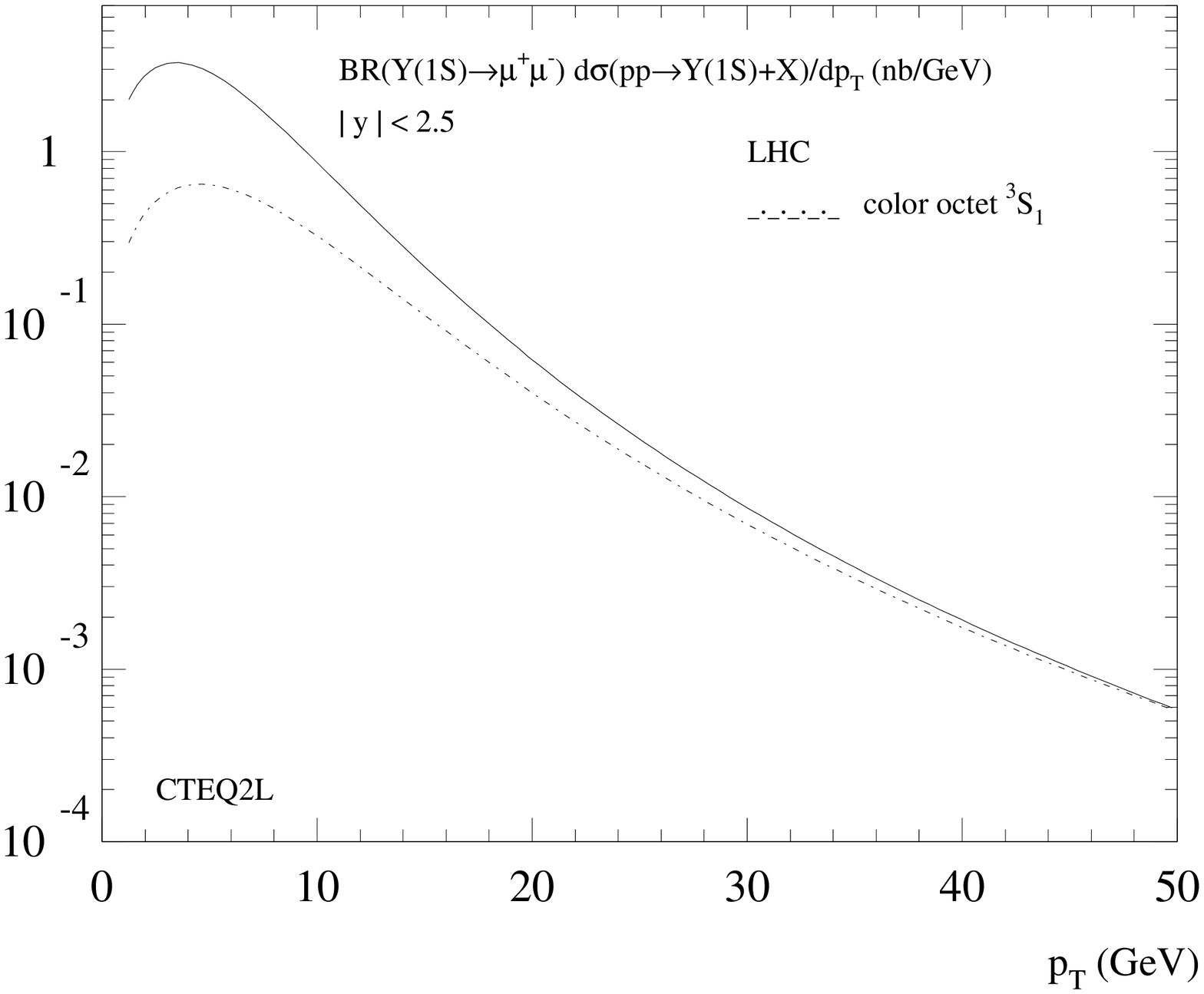}
\caption[]{Cross sections for $\Upsilon(1S)$ production in $pp \to 
\Upsilon(1S)+X$ at the LHC ($\sqrt{s}=14$~TeV, rapidity cut $|y|<2.5$)
obtained from a Monte Carlo event generator \cite{bprod:Domenech:1999qg}.
CTEQ2L parton distribution functions \cite{bprod:Tung:1994ua}; (i) dot-dashed
line: $^3S_1^{(8)}$, (ii) solid line: all contributions.  NRQCD matrix
elements as specified in \cite{bprod:Domenech:1999qg}.}
\label{fig:quarkonium9}
\end{center}
\end{figure}

The impact of initial state gluon radiation on the $\Upsilon$ cross
sections at the Tevatron has been estimated by adding a Gaussian
$k_t$-smearing as discussed previously in the context of charmonium
production. An average $\langle k_t \rangle \sim 3$~GeV and a
$K$-factor $\sim 3$ are found to bring the leading-order
colour-singlet cross section in line with the experimental
$\Upsilon(1S,2S)$ data at $p_t\;\simlt\; M_\Upsilon$
\cite{bprod:Mangano:1995yd}. Similar results have been obtained within a Monte
Carlo analysis \cite{bprod:Domenech:1999qg}, leading to significantly lower
fit values for the colour-octet NRQCD matrix elements than those
determined from a leading-order calculation
[Table~\ref{tab:quarkonium2}]. Moreover, the Monte Carlo results imply
that no feeddown from $\chi$ states produced through colour-octet
${}^3S_1$ $b\bar{b}$ states is needed to describe the inclusive
$\Upsilon$ cross section, in contrast to what is found at
leading-order. The calculation of next-to-leading order corrections
and a systematic treatment of soft gluon radiation within perturbation
theory are required to resolve these
issues. Figure~\ref{fig:quarkonium9} shows the inclusive $\Upsilon(1S)$
cross section at the LHC as obtained from the Monte Carlo calculation
\cite{bprod:Domenech:1999qg}.
The curves are based on the NRQCD matrix
elements extracted from the Monte Carlo analysis of the Tevatron data
\cite{bprod:Domenech:1999qg}. As in the case of charmonium production, one
observes that the final LHC prediction is consistent with the
leading-order result presented in fig.~\ref{fig:quarkonium7} within
errors.

\begin{figure}[htb]
\begin{center}
\includegraphics[width=0.8\textwidth,clip]{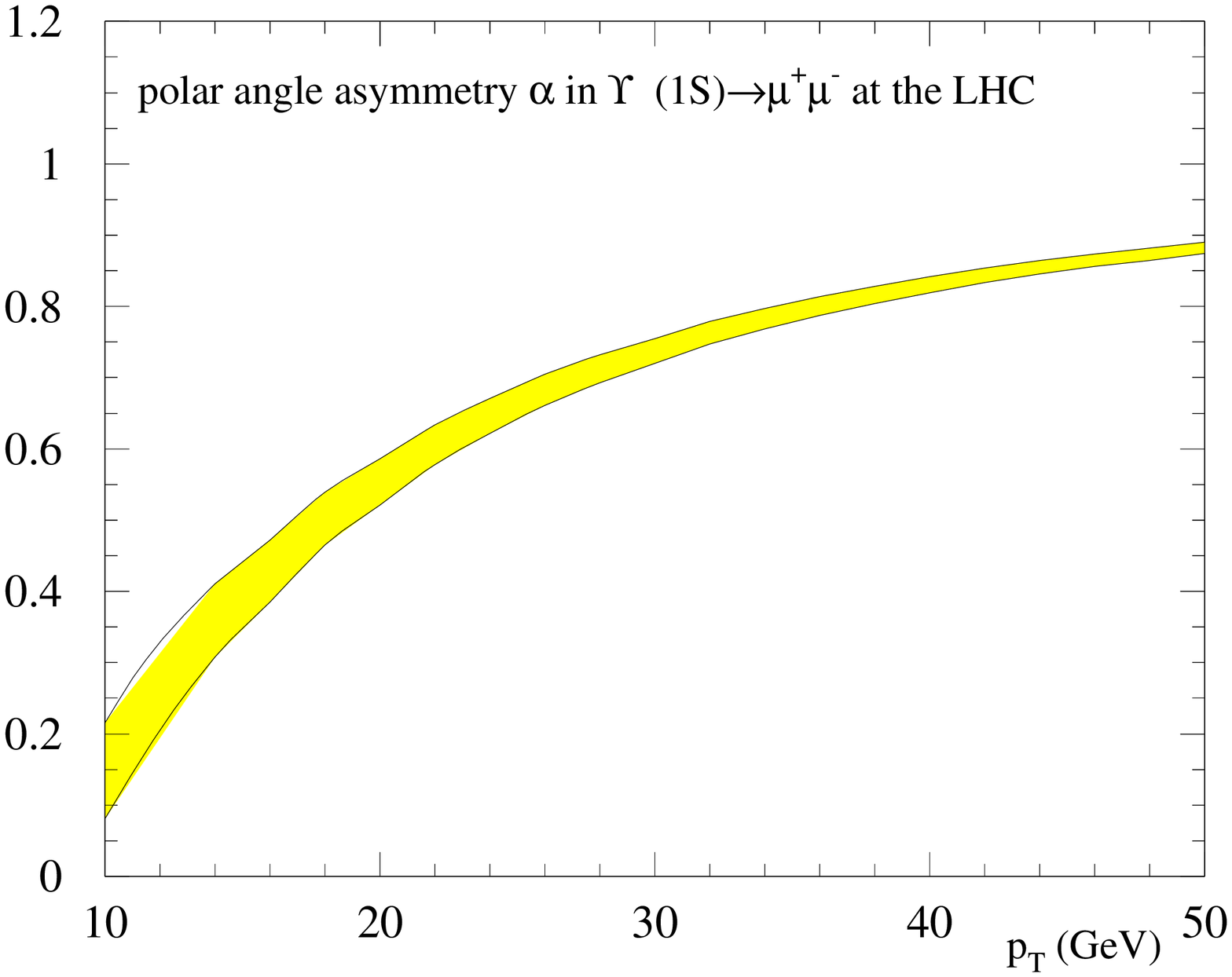}
\caption[]{Polar angle asymmetry $\alpha$ for direct $\Upsilon(1S)$
production in $pp \to \Upsilon(1S)(\to \mu^+\mu^-)+X$ at the LHC as a
function of $p_t$. NRQCD matrix elements as specified in
Table~\ref{tab:quarkonium2} other parameters as in
Figure~\ref{fig:quarkonium7}. The error band reflects the limiting
cases that either $\langle {\cal O}_8^{\psi}({}^1S_0)\rangle$ or
$\langle {\cal O}_8^{\psi}({}^3P_0)\rangle$ is set to zero in the
linear combination extracted from the data.}
\label{fig:quarkonium10}
\end{center}
\end{figure}

Let us finally present the polarization pattern predicted for direct
$\Upsilon(1S)$ production at the LHC, fig.~\ref{fig:quarkonium10},
based on the NRQCD matrix elements of Table~\ref{tab:quarkonium2}.
Higher-order corrections to the gluon fragmentation function
\cite{bprod:Beneke:1996yb} \cite{bprod:Beneke:1997yw}
will lead to a small reduction of
the transverse polarization at large $p_t$ and should be included once
precise data become available. If the charmonium mass is indeed not
large enough for a nonrelativistic expansion to be reliable, the onset
of transverse $\Upsilon$ polarization at $p_t \gg M_{\Upsilon}$ may
become the single most crucial test of the NRQCD factorization
approach.

In summary, we have discussed some of the phenomenological
implications of the NRQCD approach for quarkonium production at the
LHC and presented 'state-of-the-art' predictions for $\psi$ and
$\Upsilon$ differential cross sections and polarization.\footnote{Other
processes that have been studied in the literature include quarkonium
production in association with photons
\cite{bprod:Kim:1997bb} \cite{bprod:Mathews:1999ye} or electroweak bosons
\cite{bprod:Braaten:1999th}, as well as $\eta$ and $P$-wave quarkonium
production \cite{bprod:Mathews:1998nk} \cite{bprod:Sridhar:1996vd}.}
Among the
theoretical issues that need to be addressed in the future are the
calculation of higher-order QCD corrections, the summation of
higher-order terms in the velocity expansion and quantitative insights
in the effect of higher-twist contributions.  Besides a global
analysis of different production processes and observables at various
colliders, quarkonium physics at the LHC will play a crucial role to
assess the importance of colour-octet processes and to conclusively
test the applicability of non-relativistic QCD and heavy-quark spin
symmetry to the charmonium and bottomonium systems.

\noindent

\section{PROSPECTS FOR $b$ PRODUCTION
MEASUREMENTS AT THE LHC\protect\footnote{Section
    coordinator: M. Smizanska and P. Vikas}}
\labelsection{bprod:smizanska}
Of the existing and currently proposed accelerator facilities, the LHC
will yield the largest rate of $b$ quarks.  A well defined program
for $b$ production investigations, and the development of dedicated
detection strategies optimised for ATLAS, CMS and LHCb, are required
for the succesful exploitation of the rich LHC potential.  After an
introduction summarising the main physics motivations, we review the
detector and trigger features relevant for $b$ production in the
LHC experiments. The kinematic ranges accessible to the three
experiments are then described.  Theoretical motivations and possible
measurement methods are presented for single $b$ quark properties,
correlations in $b$ production, multiple heavy flavour production,
polarization, and charge asymmetry effects in $B$-hadron production in
$pp$ interactions.  Based on earlier performance studies, the potential
for these measurements is estimated and some preliminary results are
presented. We conclude with a summary of the present status of the
preparations for $b$-production studies.

\subsection{Introduction}
While many LHC studies have been devoted to $B$-decays, $b$
production has not yet been directly addressed. Even though $b$
decay investigations will provide some information on the production,
at the discussions of this workshop it became clear that they are not
sufficient to cover all aspects of production.

Heavy quark production in high energy hadronic collisions is important
for the study of Quantum Chromodynamics (QCD). Nowadays, QCD is
recognized as a well established and solid theory.  If disagreements
between the theoretical predictions and the experimental data are
found, they will suggest the lack of understanding only of a
particular production mechanism. In many cases these disagreements may
be attributed to a too slow convergence of the perturbation series. In
other cases, there may be important contributions from nonperturbative
effects.  Strictly speaking, the production measurements are not going
to test the principles of QCD, but rather to outline the boundaries,
where the predictions of perturbation theory provide an adequate
description and exhaust all the visible effects.  In this context, it
will be certainly useful to test as many different processes as
possible.

We present below some examples of such processes and observables, which
can potentially be studied in the LHC experiments.  Besides testing
QCD, there exist other motivations to understand production properties;
for instance, as a control of the systematics in CP violation.  Double
$b$ pair production is also a background in some channels of Higgs
detection for LHC \cite{bprod:HIGGS}.  Measurements of the $b$ production
 by ATLAS and CMS in the initial years of low luminosity
running will also be used to optimise the trigger selections at high
luminosity for rare $B$ decays.

\subsection{Detector and trigger characteristics relevant for
$b$-production}
The ATLAS, CMS and LHCb detectors and triggers are described in detail
elsewhere \cite{bprod:ALLDETTRIGG}.  Even though the signal-to-noise ratio
for $b$ events is higher at LHC than at lower energy hadron machines,
only about $1\%$ of the  non diffractive inelastic  collisions will produce $b$-quark pairs.
Events with $B$ hadrons can be distinguished from other inelastic $pp$
interactions by the presence of leptons, of secondary vertices and
particles with high $p_T$.  Each of the three experiments will have
several levels of triggers to efficiently select the interesting
events containing $B$ hadrons while maintaining manageable trigger
rates.  The information from the muon detectors and the
electromagnetic and hadronic calorimeters will be used by the lowest
level trigger in all the three experiments.  In LHCb the lowest level
trigger performs a pile-up veto followed by soft cuts on first level
trigger objects like muons ($p_T>1$~GeV), electrons ($E_T>2.1$~GeV) or
hadron clusters ($E_{T}>2.4$~GeV) reducing the trigger rate to 1 MHz. The
more time-consuming operations, like vertex reconstruction and using
information from the RICH for particle identification, will be
performed by the higher level triggers. The final event rate expected
from LHCb is $\sim$ 200 Hz.  ATLAS and CMS are central detectors for high
$p_T$ physics designed to  operate at high luminosities. The low-level
trigger objects have higher $p_T$ limits than in  LHCb: single muons
$p_T>6(7)$~GeV in ATLAS(CMS) or dimuon triggers with a minimal $p_T$ of
each muon in the interval $(3-6)$~GeV in ATLAS and $(2-4)$~GeV in
CMS \cite{bprod:ATLASTRIGG} \cite{bprod:CMSTRIGG}.  However, thanks to the higher
luminosity, despite the higher $p_T$ thresholds, they will have
statistics comparable to LHCb in many exclusive channels. Simulations
done on both experiments have demonstrated that at a luminosity of
$10^{33}cm^{-2}s^{-1}$, in spite of 2-3 pileup events on the average
accompanying the $b$ event in the same bunch crossing, $B$-decays
can be triggered on and further cleanly separated from background in
off-line reconstruction \cite{bprod:ATLASMINBIAS}
\cite{bprod:CMSTP}.

\subsection{Kinematic ranges}
\begin{figure}[htb]
\centering
\mbox{
\subfigure \ \ \ (a)
{ \epsfig{file=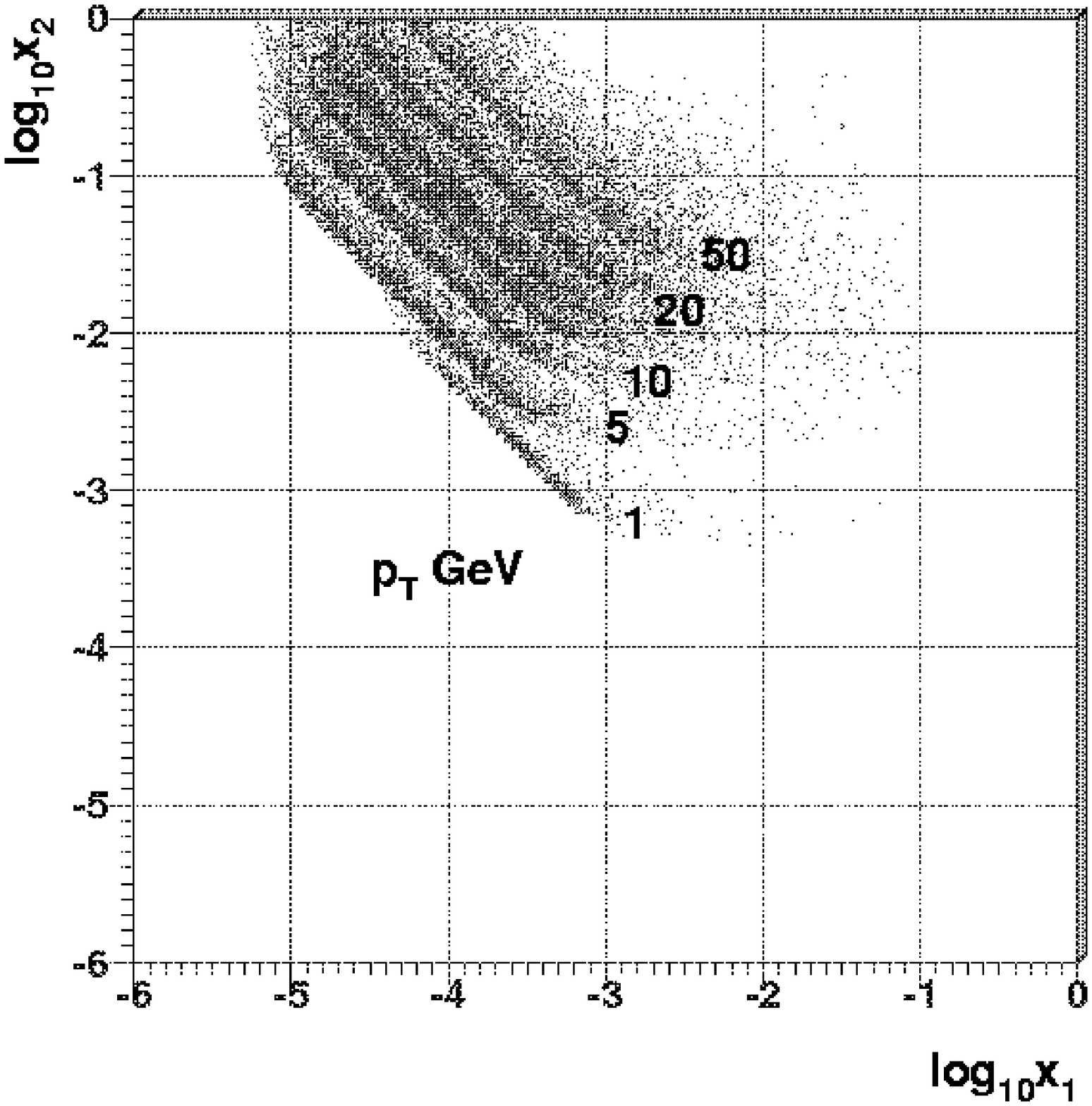,width=6cm,}} \quad
\subfigure \ \ \ (b)
{ \epsfig{file=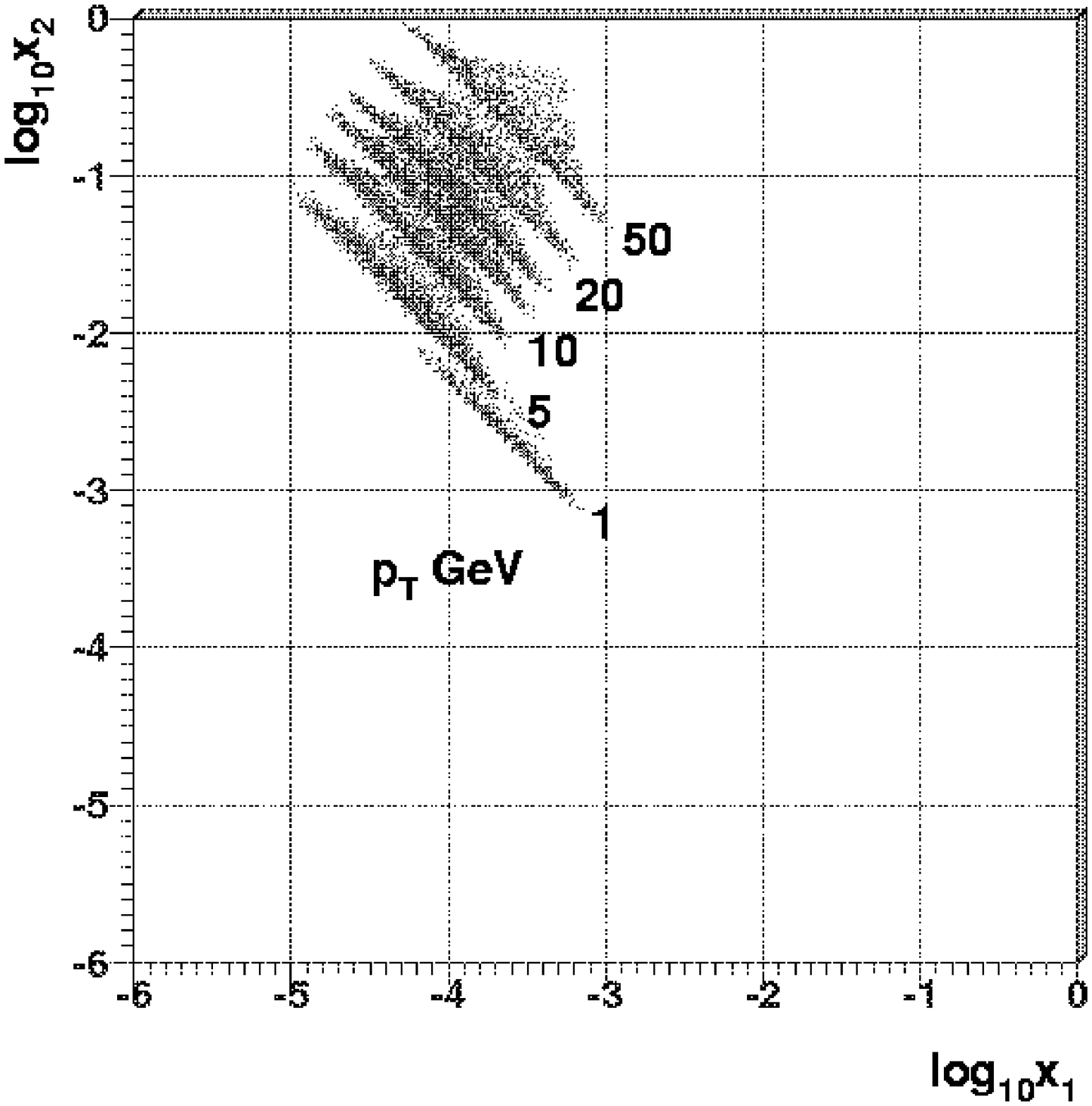,width=6cm}}
}
\mbox{
\subfigure \ \ \ (c)
{ \epsfig{file=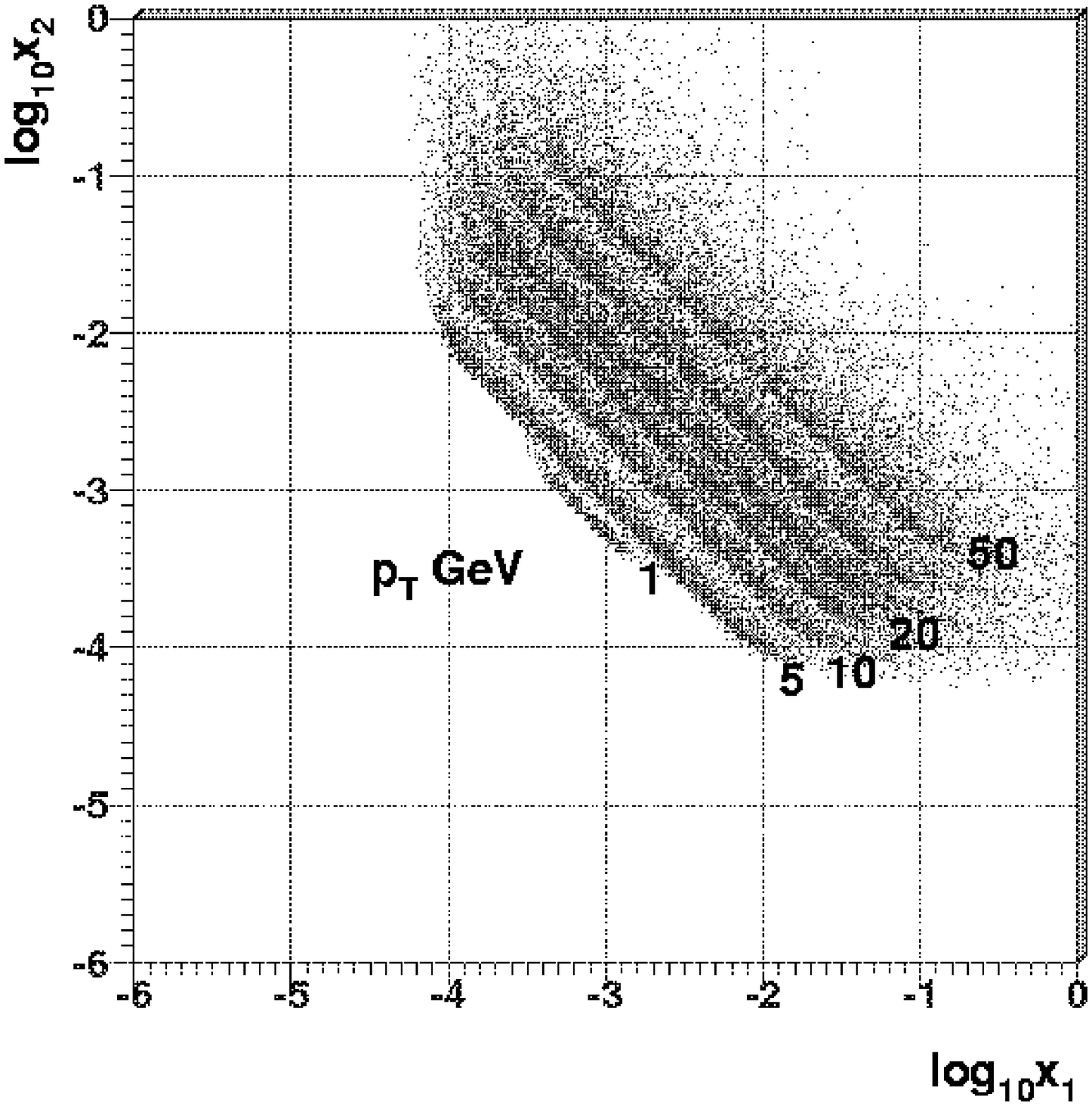,width=6cm}} \quad
\subfigure \ \ \ (d)
{ \epsfig{file=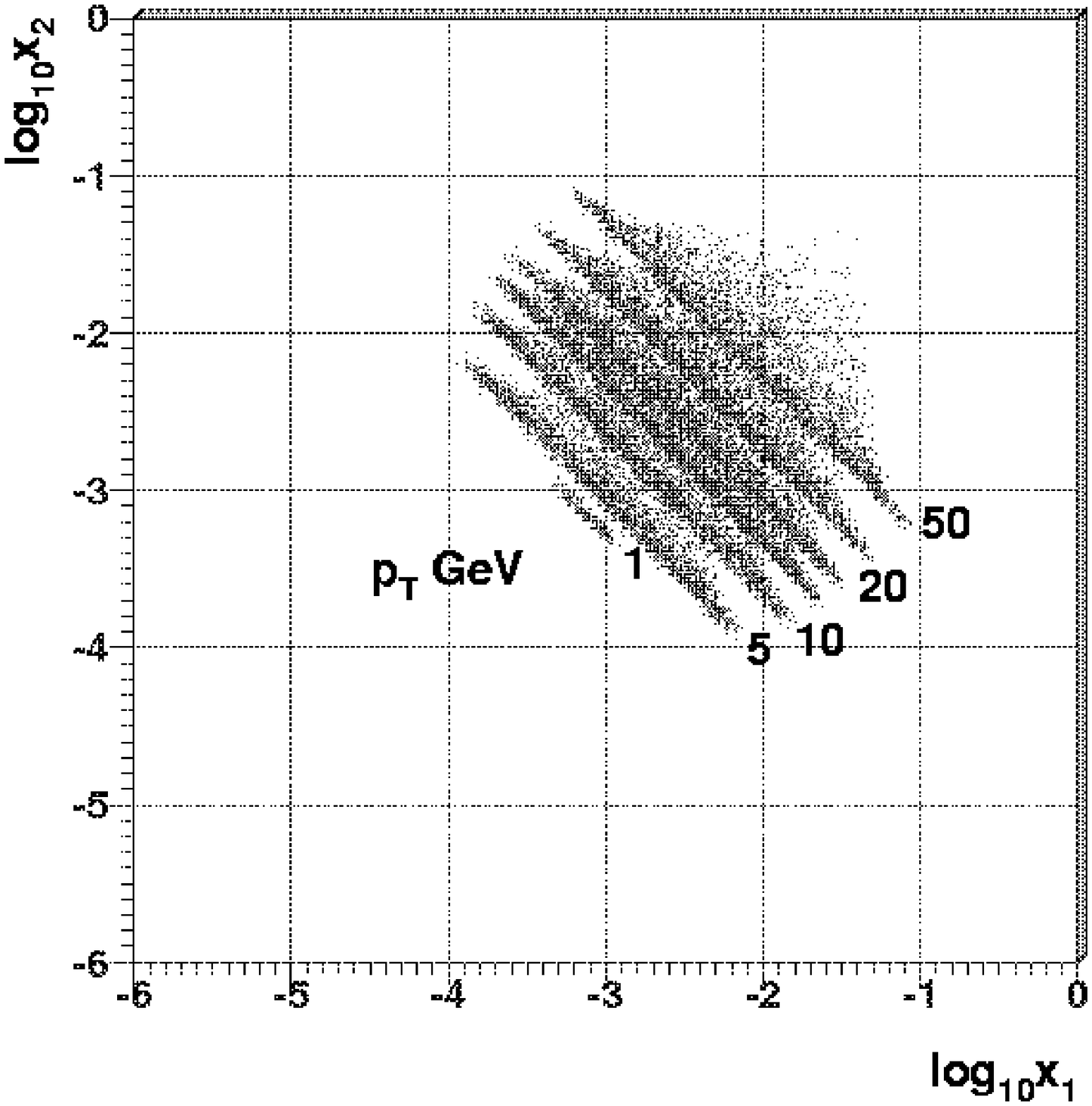,width=6cm}}
}
\caption{Bjorken $x$ region of LHCb for different values of the $b$
$p_T$ for the situation when one of the quarks is in the detector volume (a),
both $b$ and $\overline{b}$ are in the detector volume (b).
In (c) and (d), analogous distributions are given for ATLAS/CMS.}
\label{fig:xlhc}
\end{figure}

\begin{figure}[htb]
\centering
\mbox{
\subfigure \ \ \ (a)
{ \epsfig{file=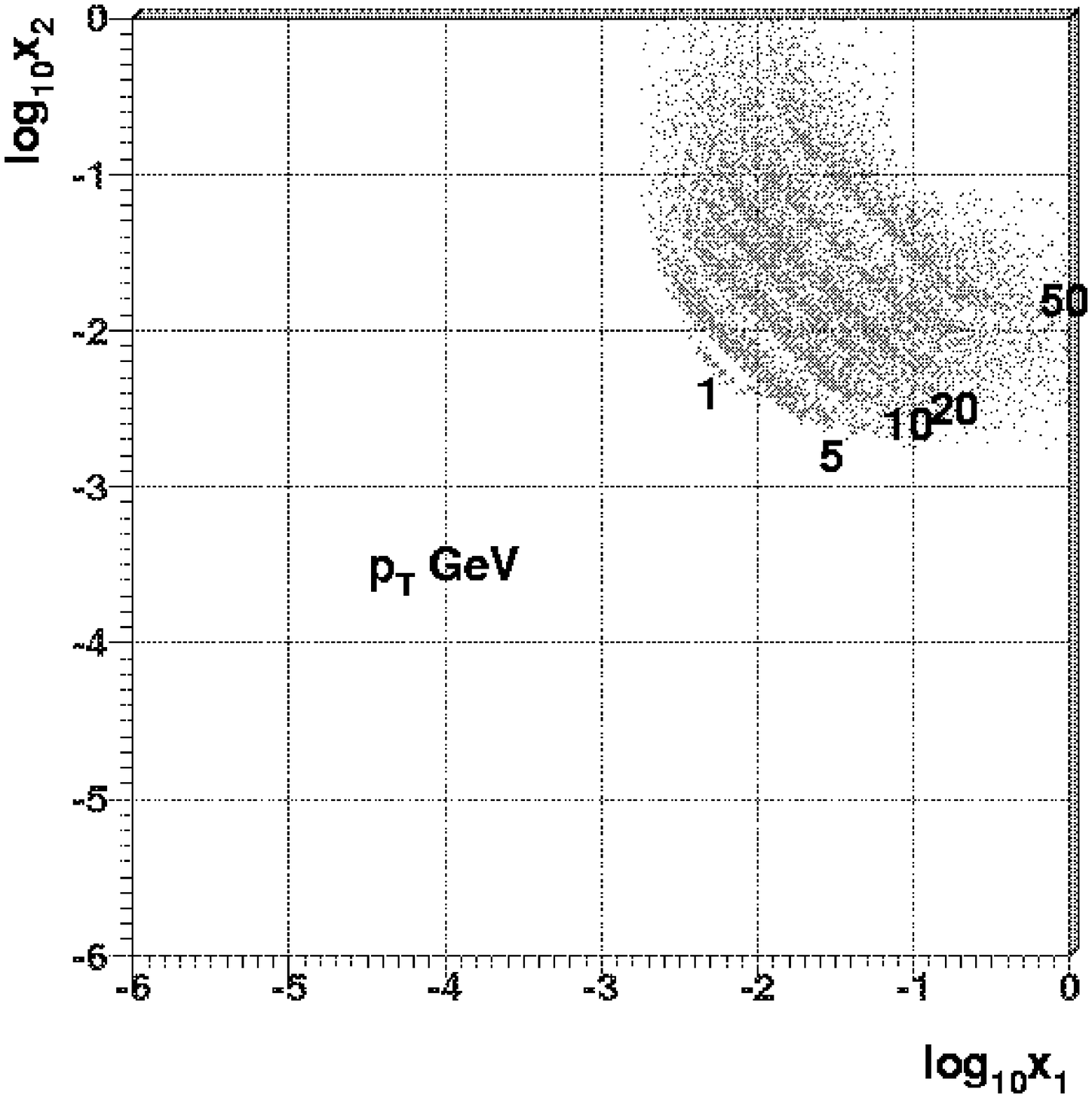,width=6cm}} \quad
\subfigure \ \ \ (b)
{ \epsfig{file=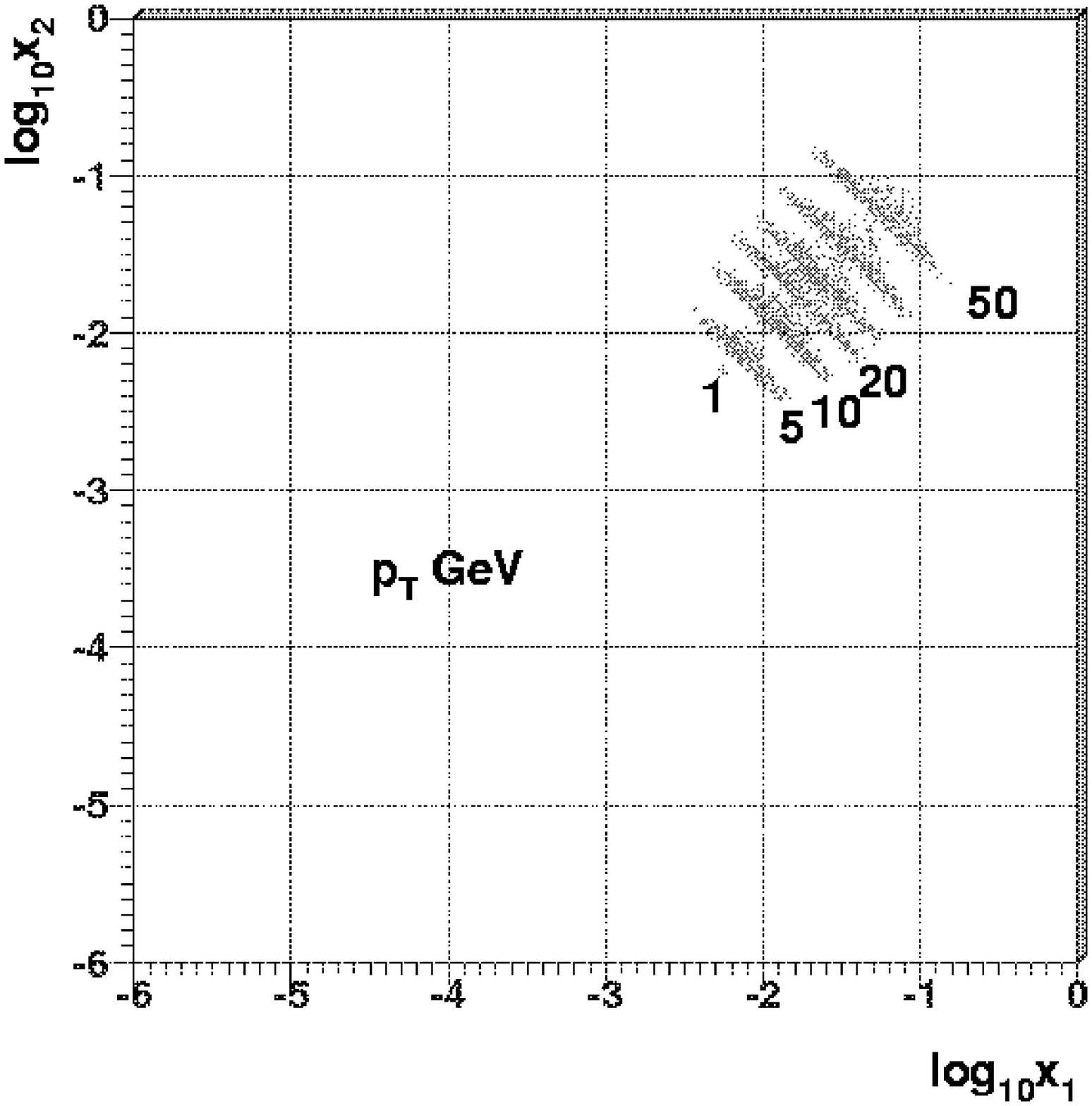,width=6cm}}
}

\caption{ Bjorken $x$ region of  CDF for different values of $b$ $p_T$
for the situation when  one of the quarks is in the detector volume (a),
both $b$ and $\overline{b}$ quarks are in the detector volume (b). }
\label{fig:xfnal}
\end{figure}

The central detectors ATLAS and CMS will cover the pseudorapidity
region $|\eta| < 2.5$; the more forward LHCb is optimised for $1.8 < \eta <
4.9\ $. The overlap between the experiments is less then a unit of
pseudorapidity, in the region $1.8 <\eta < 2.5$.  The low transverse
momentum cutoffs in each experiment are limited mainly by the admissible 
low-level trigger rates.  In the statistically dominant channels, ATLAS and CMS
will be efficient for $B$-hadrons with $p_T\gtrsim10$~GeV and LHCb for
$p_T>2$~GeV.  The domains of the Bjorken $ x$ variable for different
values of the $b$ quark transverse momentum $p_T$
are given in fig.~\ref{fig:xlhc} for two situations: when both the $b$
and $\overline{b}$ are in a fiducial volume of a detector; and when
only one of them is there.  It is clear that in all three LHC
experiments the sampled range of $x$ is contained within the region already
covered by HERA \cite{bprod:HERA}. For comparison, the analogous
distribution is calculated for CDF conditions (fig.~\ref{fig:xfnal}).

\subsection{Single  $b$ quark production }
\subsubsection{Theoretical motivations}

The inclusive differential cross section $d\sigma/dp_T d\eta$, where
$p_T$ and $\eta$ are the transverse momentum and the pseudorapidity of
the $b$ or $\overline{b}$ quarks, provide the basic information on
$b$ production. As discussed in the previous section,
next-to-leading order (NLO) calculations give a
cross section lower than CDF and D0 data by a factor of $\sim 2.4$
\cite{bprod:Buras2}. However, the shape of the $p_T$ distribution is well
reproduced by LO+NLO predictions, by a semihard model of the BFKL type
\cite{bprod:BFKL} and also by \Py\ \cite{bprod:my}. In the region of high
$p_T$, the effects of higher order contributions are taken into
account by means of the resummation technique~\cite{bprod:resumCG}
\cite{bprod:CacciariGrecoNason} \cite{bprod:resumOST}.
In ref.~\cite{bprod:CacciariGrecoNason} LO+NLO contributions
are included together with the resummation of all
terms of order $\alpha_s^k\ln^k(p_T/m_b)$
and $\alpha_s^{k+)}\ln^k(p_T/m_b)$.
These contributions change the shape of the $p_T$ spectrum. 
Thus measurements of  high
$p_T$ single $b$-spectra may be considered as a dedicated test for the QCD
resummation technique.

\subsubsection{Measurement possibilities}

Experiments can measure the doubly differential cross section
$d\sigma/dp_T d\eta$, where $p_T$ and $\eta$ are the transverse
momentum and the pseudorapidity of a $B$ hadron, or of a jet
associated with a $B$ hadron, or only one of the decay products of a
$B$ hadron (for example $J/\psi$ or $\mu$).
From these experimentally measured quantities, the
$d\sigma/dp_T d\eta$ of the parent $b$-quark can be extracted, using
appropriate models of hadronization and decay.

The determination of the absolute value of the cross section is also important.
Three independent measurements (ATLAS, CMS and LHCb) can be done at the same energy.
The determination of the absolute cross sections is always difficult,
since it requires a  precise understanding of the  luminosity, of the 
trigger and reconstruction efficiency and  of the background contributions.
Several techniques of  luminosity measurement are under study. It appears that
precisions of  $\sim 3 \%$ could be achieved \cite{bprod:LUMIATLAS}.
The overlap in the detection  phase space of  ATLAS,  CMS and LHCb, in the region 
$1.8<\eta<2.5$ and $p_T>10$~GeV,  can be used  for  cross-checks.

\subsubsection{Exclusive channels}
From trigger and offline studies and the present experience with CDF it is known that the three LHC
experiments can provide high statistics samples of some exclusive $B$-decay
channels cleanly separated from the background.   The statistically
dominant channels are those containing \Jpsitomm\ ($B_{d} \rightarrow
J/\psi K^{0}$, $B_{d} \rightarrow J/\psi K^{*}$, $B^{\pm} \rightarrow
J/\psi K^{\pm} $ and $B^{0}_s \rightarrow J/\psi \phi$), which are
also needed for CP violation studies. 
Moreover, LHCb will cleanly
separate large statistics of purely hadronic exclusive decays, where
the dominant ones are   $B_{d} \rightarrow D^{*-} \pi^{+}$ and 
 $B_{d} \rightarrow D^{*-} a_{1}^{+}$.  
 With these processes one can cover the differential $p_T$ cross
section measurements starting approximately from $p_T > 10$~GeV for ATLAS and CMS and
$p_T > 2$~GeV for  LHCb respectively.  The numbers of
these events after three years of run at luminosities of
$10^{33}cm^{-2}s^{-1}$ for ATLAS and CMS and five years at $2 \cdot
10^{32}cm^{-2}s^{-1}$ for LHCb,  are shown in
fig.~\ref{fig:exclusive} as a function of a minimal transverse
momentum of the $B$-hadron $p_T$.

\subsubsection{Inclusive \bJpsi channels}
\begin{figure}[htb]
\centering
\mbox{
\subfigure
{ \epsfig{file=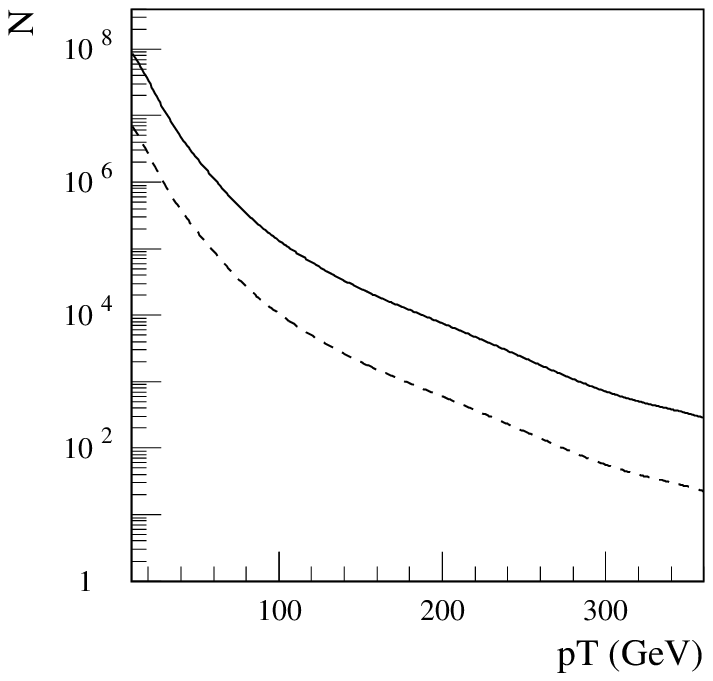,width=6cm}} \quad
\subfigure
{ \epsfig{file=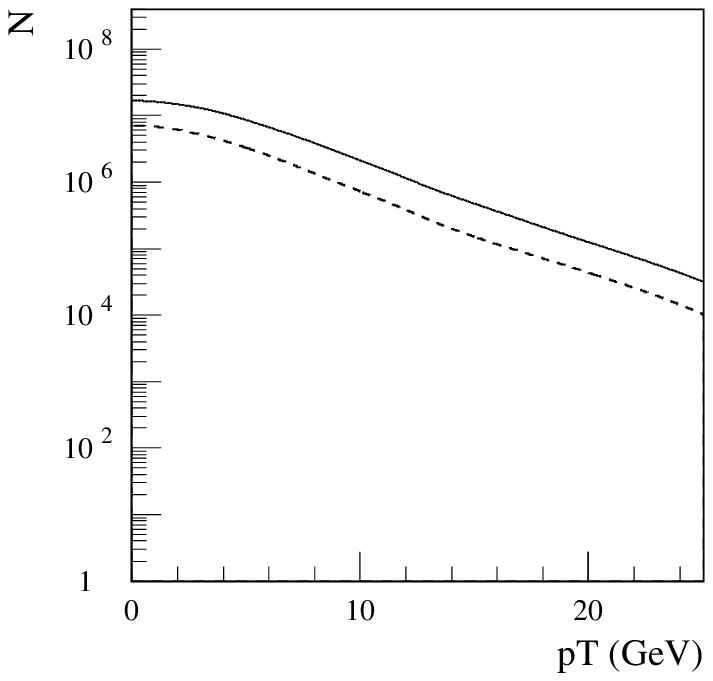,width=6cm}}}
\caption{Number of triggered and reconstructed  $B$-hadrons  as function of 
a lower cut on the $B$-hadron transverse momentum $p_T$.  Figure~(a)
shows the ATLAS expectations for inclusive and exclusive $B$-hadrons
decays to $J/\psi$ after 3 years, with an integrated luminosity of $
30 \mathrm fb^{-1}$.  The full line corresponds to inclusive events,
the dashed line to the sum of all exclusive channels. Fig.~(b) shows
the LHCb expectations for exclusive $B$-hadrons decays after 5 years.
The solid line is for all statistically dominant channels, the dashed
line shows only the channels with a $J/\psi$ in the final state.}
\label{fig:exclusive}
\end{figure}

The inclusive channels \bJpsi\ can be used to extend the available statistics for
production measurements to high transverse momenta
(fig.~\ref{fig:exclusive}).

A preliminary study from CMS~\cite{bprod:CMSJPSI} shows that
for $p_{T}^{J/\psi} \sim 300$~GeV, 
which corresponds to $p_{T}^{b} \sim 550$~GeV, a b-tagging efficiency
of $\sim 50\%$ can be achieved with a $J/\psi$ mass and decay length
reconstruction. This will give a signal-to-noise ratio of $\sim 2.5$ taking into account the
prediction for prompt $J/\psi$ production of 
ref.~\cite{bprod:Cano-Coloma:1997rn}.

In ATLAS a study has been done \cite{bprod:SIMON} for  events
\bJpsimm in which the $p_T$ of the $b$ quark was chosen larger than
50~GeV.  In particular, it was shown that the mass resolution of the
$J/\psi$ will not be degraded due to events in which a signal
reconstructed in the muon system is wrongly associated to a non-muon
track in the inner detector.

\subsubsection{Inclusive $b$-jet production}

Another method for $b$ production studies discussed at the workshop
was based on inclusive $b$-jet reconstruction. In both ATLAS and CMS this
technique was developed for the Higgs search \cite{bprod:ATLASPHYSTDR}
\cite{bprod:CMSTP}.  The $b$-jet cross section is expected to be a small
fraction (close to or larger than 2\% for jets with $E_T$ larger
than about 20~GeV)
of the single-jet cross section \cite{bprod:TRIGTPATLAS}
\cite{bprod:MFjets}. If
this method is to be used for  single $b$ quark production, it will
require prescaling of the trigger for the lower $p_T$ region or a cut
on very high transverse momenta ($p_T > 150$~GeV), to reduce the huge
rate of non-$b$ QCD background \cite{bprod:TRIGMENU}.

\begin{figure}[htb]
\begin{center}
\mbox{\epsfig{file=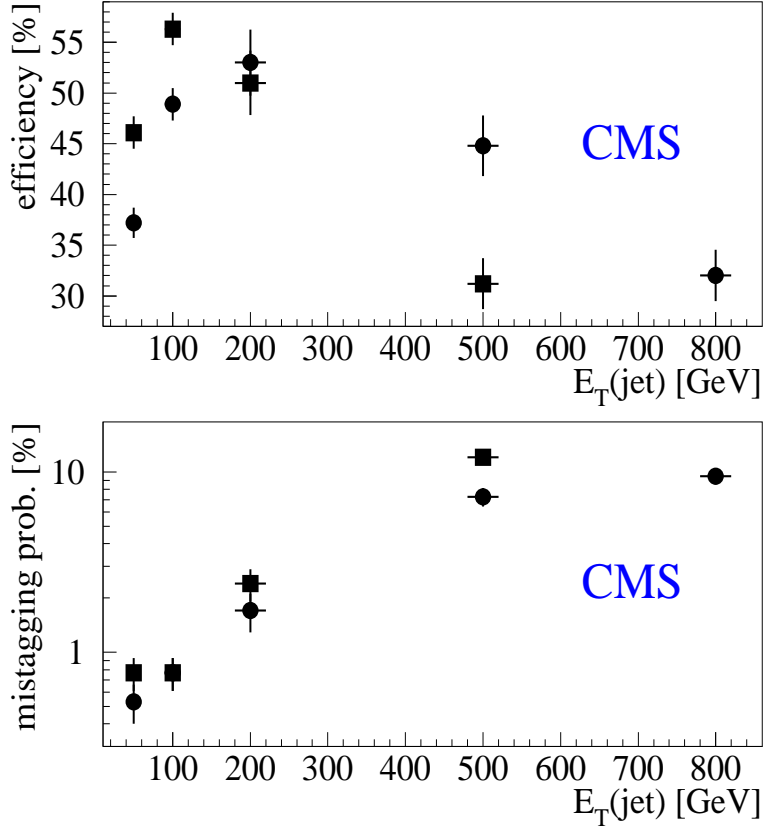,width=10cm, height=11cm}}
\caption[]{The results of CMS study of $b$-tagging efficiency and 
mistagging probability as a function of the jet $E_T$. The squares
represent the results of the phase-1 tracker, the circles those of the
phase-2 tracker.   } 
\label{fig:cmsvtx}
\end{center}
\end{figure}

Figure~\ref{fig:cmsvtx} shows the preliminary results of the CMS
$b$-tagging efficiency and mistagging probability for high $ E_{T}$
jets using the technique described in ~\cite{bprod:BJET}. The study
demonstrates that for tagging efficiencies of $35\%-55\%$ the
mistagging probability is better than $2\%$ up to $E_T\sim 200$~GeV.
Beyond that, the $b$-tagging efficiency and mistagging probability
deteriorate significantly. The algorithm will be further optimised,
possibly including lepton identification.

The method of $b$ cross section determination based on inclusive
$b$-jet identification will be heavily dependent on the precise
understanding of the non $b$-jet rejection factors.  Further
feasibility studies on this method are necessary.

\subsection{Correlations in $b$  production}
\subsubsection{Theoretical motivations}

As discussed in Section~\ref{bprod:theory}, the overall normalisation of the
production cross section, as well as the normalisation of the
inclusive $b$ spectra, remain uncertain within a factor $\sim 2$
because of inherent theoretical uncertainties.  Therefore the
measurement of these values does not provide a sringent test of NLO
contributions.  The expected correlations between the $b$ and the
$\overline{b}$ quarks can be computed in leading and next-to-leading
order \cite{bprod:MNR}. The shapes of two-particle distributions are
sensitive to the NLO contribution, and thus can be used for these tests.
In particular, distributions in the following quantities involving
both the $b$ and $\overline{b}$ quark can be considered: the relative azimuthal
distance $\Delta\phi(b\bar{b})<1$, the pair invariant mass, the pair transverse momentum and
the pair rapidity.

\subsubsection{Measurement possibilities}

\begin{figure}[htb]
\begin{center}
\mbox{\epsfig{file=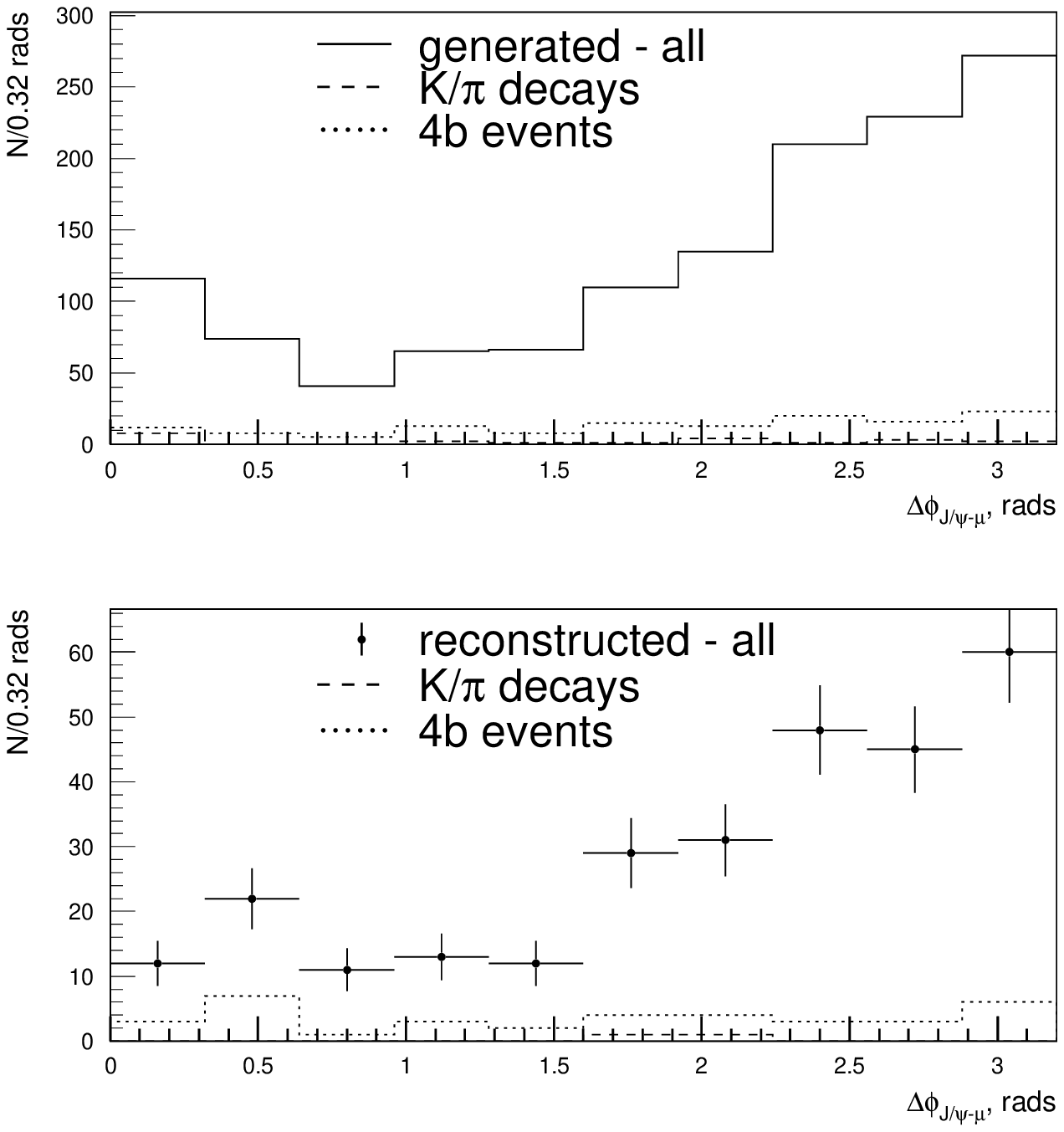,width=9cm, height=11cm}}
\caption[]{Reconstruction of $\bbJpsimm + \mu $ using combined Muon-Inner detector
off-line reconstruction in ATLAS. }
\label{simon_3mu}
\end{center}
\end{figure}

The choice of the decay channels is driven by the requirement that the
acceptance should not vanish when the $b$ and the $\overline{b}$ are close
in phase space.  The goal is to avoid isolation cuts in both
trigger and offline algorithms requiring a large separation between
the decay products of the $B$ and of the $\overline{B}$.  The
processes under consideration are based on the reconstruction of a
$J/\psi$ originating from the displaced vertex of a $B$-hadron, and of
an additional lepton coming from the semileptonic decay of the
associated $ \overline{B}$ hadron.  For example, in the ATLAS
experiment, for an integrated luminosity of $ 30\; \mathrm fb^{-1}$,
approximately $\sim 5 \cdot 10^5$ such events are expected, with the
exclusively reconstructed $B$-decays containing the $J/\psi$  (Table
\ref{table:a1}).  CDF and D0 measurements showed that  $b\bar{b}$
pairs are mostly produced back-to-back \cite{bprod:CDFCOREL}. However
the region most sensitive to  differences between the models is
$\Delta\phi(b\bar{b})<1$~rad, where only $\sim 14\%$ of the events are
expected \cite{bprod:my}.  The statistics may possibly be increased
using the semi-inclusive decays \bbJpsi accompanied by a lepton (Table
\ref{table:a1}).  As an example we quote recent studies in ATLAS
\cite{bprod:SIMON}, performed using simulated events with $B_{d}
\rightarrow J/\psi K^{0}$, \Jpsitomm.  They indicate that the signal
events can indeed be reconstructed in cases when the difference of
azimuthal angles between the $J/\psi$ and the other muon is small
(Fig.~\ref{simon_3mu}).  It is important to note that no selection cuts
requiring model dependent corrections were necessary.

The study can be extended to events with \Jpsitomm accompanied by an
electron and for \Jpsitoee\ combined with a muon or an electron.
Using all these combinations of leptons will allow the measurement of
the same variables by different detectors, leading to an improved
control of systematic errors.

\begin{table}[htb]
\begin{center}
\caption{Semi-inclusive and exclusive channels candidates for $b\bar{b}$
studies. Statistics are given for ATLAS after 3 years, with an
integrated luminosity of $ 30\; \mathrm fb^{-1}$.  }
\label{table:a1}
\begin{tabular}{ |c|c|c|c|}
\hline
Inclusive channels &
Statistics &
Exclusive channels &
Statistics \\
  & &with the same   & \\
  & & lepton content & \\
\hline
$\bbJpsimm + \mu $&
$2.8 \cdot 10^{6}$ &
$ b \overline{b} \rightarrow had \Jpsitomm + \mu$ &
$2.1 \cdot 10^{5}$ \\
$\bbJpsimm + e $&
$3.6 \cdot 10^{6}$ &
$ b \overline{b} \rightarrow had \Jpsitomm + e $&
$2.1 \cdot 10^{5} $\\
$\bbJpsiee + \mu $&
$0.6 \cdot 10^{6} $&
$ b \overline{b} \rightarrow had \Jpsitoee + \mu $&
$0.9 \cdot 10^{5}$ \\
\hline
\end{tabular}
\end{center}
\end{table}

\subsection{Multiple heavy quark pair production}
\subsubsection{Theoretical motivations}

At present, only the leading order calculation, $O(\alpha_s^4)$, is
available \cite{bprod:Likh4b} for the $bb\bar{b}\bar{b}$ production
cross section. The
effects of higher order corrections can only be estimated using the
event generator \Py\ 5.7. Since the predictions of \Py\  appear to
be about a factor of 10 above the leading order analytical
calculations \cite{bprod:BARADOUBLE}, further theoretical studies are
needed.

\subsubsection{Measurement possibilities}

Events with four $b$ quarks can be identified in several ways,
the most appropriate one depending upon the context.  As a background
to Higgs search, the requirement is four $b$ jets in the fiducial
volume. For the purpose of testing QCD predictions on double $b$
production, it may be sufficient to reconstruct events with three
$b$ quarks in the fiducial volume.  For three $b$ quarks with
$p_T>10$~GeV and $|\eta|<2.5$, \Py\ gives a cross section of
$140$~nb, which corresponds to 140 events produced per second. Despite
this large number, it will be necessary to define features allowing
on-line selection of these events in the presence of huge non-$b$
and single $b$ backgrounds.

As a source of an incorrect tag in CP violation measurements, the
relevant $bb\bar{b}\bar{b}$ events are those with two $b$ hadrons,
produced with the same flavour charge, identified in the fiducial
volume, while two other $B$-hadrons, produced with the opposite flavour,
are not detected. A direct measurement of this case could use
reconstructed charged mesons or baryons, which are
self-tagging. However the expected statistics of these events is
insufficient.  In fact, double $b$ production is expected to be
only a minor source of  wrong tags \cite{bprod:PHYSTDR}.  Techniques
exist to determine the wrong-tag rate from all processes regardless of
its origin, which does not need to be identified \cite{bprod:PHYSTDR}.

Similar to the case of double $b$ production, the production of
doubly heavy hadrons, such as the $B_c+b+c$, $\Xi_{bc}+b+c$, etc.,
refers to an $O(\alpha_s^4)$ lowest order QCD process, and also
provides a test of perturbative QCD calculations. The question of
higher order QCD contributions and, probably, nonperturbative
contributions is still open \cite{bprod:allBc}.  The total production
rate in this case will not be indicative enough to establish the role
of different production mechanisms, and the measurement should instead
concentrate on the specific event topologies. In particular,
various correlations 
between particles carrying charm and bottom may be of importance. 

Measurement possibilities are under investigation for the channel
$B_c^{(*)}\rightarrow J/\psi \pi$ \cite{bprod:Bcmu} .  A list of
possible semileptonic and nonleptonic $B_c$ decays can be found 
for instance in\cite{bprod:Anisimov}.  The
decays $B_c^{(*)}\rightarrow J/\psi \mu \nu $, $B_c^{(*)}\rightarrow
J/\psi \rho^{+}$, $B_c^{(*)}\rightarrow J/\psi K^{+*}$,
$B_c^{(*)}\rightarrow J/\psi D_{s}^{+}$ are other potentially
interesting modes.

\subsection{Other measurements}

$B$ hadrons with non-zero spin can be polarized perpendicular to
their production plane.  Polarization measurements of $b$ hadrons
produced in nucleon fragmentation could clarify the problems of
different polarization models \cite{bprod:polarth} that failed to
reproduce the existing data on strange hyperon production
\cite{bprod:polarexp}. In particular, information about the quark mass
dependence of polarization effects could be obtained.  For symmetry
reasons, in $pp$ collisions this polarization vanishes at zero rapidity,
so that the expected observed polarization in ATLAS and CMS will be
smaller than in the more forward LHCb.  Using the method of helicity
analyses of cascade decay $\Lambda^{0}_{b} \rightarrow \Lambda^0
J/\psi$ the $\Lambda^{0}_{b}$ polarization can be measured in ATLAS
with a precision better than 0.016 \cite{bprod:JALAMBDAb}.
Another approach to $\Lambda^{0}_{b}$ polarisation measurement,
using the same decay channel can be found in \cite{bprod:FRIDMAN}.

In proton-proton collisions a charge production asymmetry of $b$
hadrons is expected.  The asymmetry is defined as the  difference of
production probabilities of a $B$ hadron and its antiparticle.  From
the theoretical point of view, the asymmetries can provide information
on the effects of soft dynamics during the fragmentation and
hadronization (i.e., on the soft interactions between the produced $b$
quark and the remnants of the disrupted proton). The relevant physical
effects are expected to be unimportant \cite{bprod:VB:NPB478} \cite{bprod:Hwa}
\cite{bprod:Asymobs} in the central rapidity region covered by ATLAS and
CMS. In the more forward region of LHCb the asymmetry may rise to a
few percent. A detailed theoretical discussion of this issues are given
in Section~\ref{bprod:asymmetry}.

Any production asymmetry is always measured in the presence of a CP
violation asymmetry originating from $B$-hadron decays. In some cases
these two effects are expected to be of the same order. This is the
case, for instance, in the channels $B_{d} \rightarrow J/\psi
K^{*}(K^{+}\pi^{-})$, $B^{+} \rightarrow J/\psi K^{+} \ $ and
$\Lambda^{0}_{b} \rightarrow \Lambda^0 J/\psi$, which are expected to
have a small CP violation ($<1\%$). A way of estimating the relative
size of these two effects may be based on the fact that the production
asymmetry varies with the transverse momentum and the rapidity of
produced $b$-quark, while the decay asymmetry should remain the same.
Measurements of such small effects will require good understanding of
the possible instrumental detection asymmetries.

\subsection{Conclusions}

The properties of $b$ production at the LHC can be measured by the three
experiments, which are complementary in phase space. The small overlap
region will allow a cross check on the cross section
normalization. The kinematic conditions are such that Bjorken~$x$
values sampled in b-production are above 10$^{-5}$, a region lower than at the Tevatron, but
already covered by HERA. Differential cross section  measurements using exclusive $B$ hadron
decays will be most important at small $p_T$ values. At high $p_T$
values and  for correlations and multiple heavy flavour production
measurements, the statistics can be increased by semi-inclusive $B$
decays containing $J/\psi$.  Possible methods using $b$-jets require
further study, to control the non-$b$ QCD background. The enormous LHC
statistics will also allow to study the production polarization and
charge production asymmetries.
\section{TUNING OF MULTIPLE INTERACTIONS GENERATED BY 
PYTHIA\protect\footnote{Section coordinators: P.~Bartalini, O.~Schneider.}}
\labelsection{bprod:bartalini}
\subsection{Introduction}
\label{Introduction}
The track multiplicity distribution as well as the transverse momentum
distribution of charged particles in proton-proton interactions (the
so-called minimum bias events) affect the performances of the low
level triggers and the detector occupancy of the LHCb
experiment~\cite{bprod:TP}.  They should therefore be modelled
reliably in Monte Carlo programs.  In particular, at LHC energies,
multiple interactions play an important role, and should not be
neglected.

In Section~\ref{sec:mult_int} we examine the multiple interaction
models available in \Py~\cite{bprod:Pythia:ref} to describe the event
structure in hadron-hadron collisions.  In Section~\ref{sec:data} we
select a compilation of homogeneous data at different energies
suitable to tune the multiple interactions parameters of \Py; the
tuning procedure is presented in Section~\ref{sec:tuning}.  We use the
phenomenological extrapolations at LHC energy in order to get the
predictions for the track multiplicity and the transverse momentum
distributions in minimum bias and $b\overline{b}$ events; these are
reported in Section~\ref{sec:lhc}.
\subsection{Multiple interaction models}
\label{sec:mult_int}
The multiple interactions scenario is needed to describe the
multiplicity observables at hadron
colliders~\cite{bprod:multiparton_interactions} and is also supported
by direct observation~\cite{bprod:afs_direct} \cite{bprod:ua2_direct}
\cite{bprod:cdf_direct}.
The basic assumption is that several parton-parton interactions can
occur within a single hadron-hadron collision.  Four different models
are available in \Py. The main parameter of these models,
$P_{T_{min}}$, is the minimum transverse momentum of the parton-parton
collisions; it effectively controls the average number of
parton-parton interactions and hence the average track multiplicity.
The differences between the four models, which mainly affect the shape
of the multiplicity distribution, are the following:
\begin{itemize}
\item Model 1 (default in \Py)
\subitem All the hadron collisions are equivalent (as opposed to
model 3 and 4 below) and all the parton-parton interactions are
independent; the $P_{T_{min}}$ parameter represents an abrupt cut-off.

\item Model 2
\subitem Same as Model~1 but with a continuous turn-off of 
the cross section at $P_{T_{min}}$.

\item Model 3
\subitem Same as Model~2, but
hadronic matter in the colliding hadrons
is distributed according to a Gaussian shape, and a varying impact parameter 
between the two hadrons is assumed.

\item Model 4
\subitem Same as Model~3 but the matter distribution is described by two 
concentric Gaussian distributions.
\end{itemize}

The varying impact parameter models (Models~3 and 4) were
developed~\cite{bprod:multiparton_interactions} to fit the UA5
data~\cite{bprod:distr_ua5_old}.  A recent study performed by the CDF
collaboration~\cite{bprod:CDF_tuning} concludes that a varying impact
parameter model (Model~3) is also preferred to describe the underlying
tracks in $b$ events produced at the Tevatron.

\begin{figure}[htb]
\begin{center}
\mbox{\includegraphics[width=0.6\textwidth,clip]{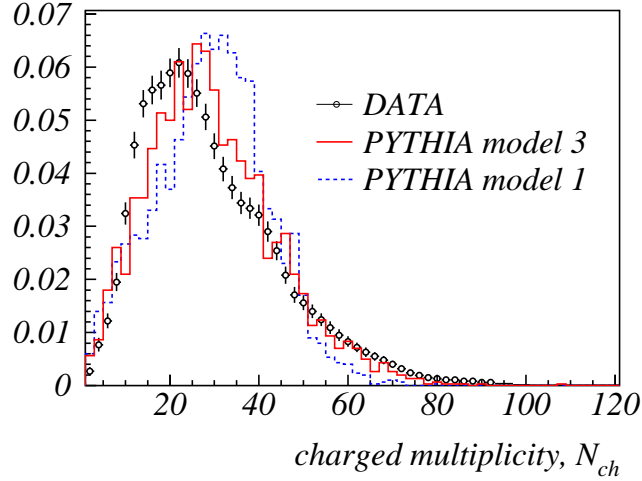}}\\
\caption{\label{fig:distr_ua5} Charged multiplicity distribution
for non single-diffractive events in $\overline{p}p$ collisions 
at $\sqrt{s} = 546$~GeV as measured by UA5~\cite{bprod:distr_ua5}
compared with \Py\ predictions using the CTEQ4L parton distribution 
functions and either Model~1 (solid) or 3 (dashed) for multiple interactions.
In each case the $P_{T_{min}}$ parameter has been tuned to reproduce 
the mean multiplicity measured in the data.}
\end{center}
\end{figure}

In the absence of published results on multiplicity distributions in
minimum bias events at the Tevatron, we compare again the predictions
of Model~1 and Model~3 with the UA5 data, using the final charged
multiplicity distribution from the full $\overline{p}p$ data sample
collected by UA5 at $\sqrt{s} = 546$~GeV~\cite{bprod:distr_ua5}, a
recent version of \Py\footnote{Version 6.125~\cite{bprod:Pythia6}
was used for all the studies reported here.}, and a modern set of
parton distribution functions, CTEQ4L~\cite{bprod:cteq4}, tuned on
both HERA and Tevatron data.  For this comparison, the main multiple
interactions parameter ($P_{T_{min}}$) is tuned in each model to
reproduce the mean value of the measured charged multiplicity
distribution in not single diffractive events\footnote{In this paper
we define as ``non single-diffractive event'' any inelastic
hadron-hadron interaction that cannot be regarded as a single
diffractive event; in the framework of the \Py\ hadronic
interactions, the ``non single-diffractive'' sample includes the $2
\rightarrow 2$ partonic processes and the double diffractive
hadron-hadron interactions.}  ($<N_{ch}> = 29.4 \pm 0.3 \pm 0.9$).  We
obtain $P_{T_{min}}=1.63 \pm 0.02$ for Model 1 and $P_{T_{min}}=1.97
\pm 0.03$ for Model~3.  The shapes of the multiplicity distributions
are compared in fig.~\ref{fig:distr_ua5}.  It is clear that Model~3 is
preferred over Model~1 to describe the UA5 data.  In particular the
shape of the tail at high multiplicities is reproduced much better by
Model~3.  The UA5 results are corrected for the lower efficiency
expected on double diffractive events. Therefore the simulated samples
include the generation of all kind of non single-diffractive events.
The uncertainty in the diffractive cross sections relative to the
partonic ones can affect the observed discrepancies between the data
and the \Py\ predictions in the low multiplicity
region\footnote{The $\overline{p}p$ cross sections predicted by
\Py\ at $\sqrt{s}=546$~GeV are $30.7$~mb for the partonic
processes and $5.3$~mb for the double diffractive processes.}.
\subsection{Mean charged multiplicity at \boldmath $\eta=0$}
\label{sec:data}
In order to produce realistic \Py\ predictions for the
multiplicity observables in the LHC environment, it is necessary to
take into account the energy dependence of the $P_{T_{min}}$
parameter.  Unfortunately there are not many published data concerning
the charged multiplicity distribution in minimum bias events at hadron
colliders. On the other hand there are some data available relative to
the average charged multiplicity in non single-diffractive events, in
particular for the central pseudo-rapidity region. Therefore, to study
the energy dependence of the $P_{T_{min}}$ parameter at generator
level, we consider an homogeneous sample of corrected average charged
multiplicity measurements at six different center-of-mass energies
($\sqrt{s} =$ $50$, $200$, $546$, $630$, $900$ and $1800$~GeV) in the
pseudo-rapidity region $|\eta|< 0.25$~\cite{bprod:average_ua5}
\cite{bprod:average_cdf}.
The energy dependence of $dN_{ch}/d\eta$ at $\eta=0$ is shown in
fig.~\ref{fig:average_data}a together with the fit of a quadratic
function of $\ln(s)$ proposed in Reference~\cite{bprod:average_cdf};
using this fit to extrapolate at LHC energy would predict
$dN_{ch}/d\eta = 6.11 \pm 0.29$ at $\eta=0$.
\subsection{Tuning of the multiple interaction
            parameter \boldmath $P_{T_{min}}$}
\label{sec:tuning}
The average charged multiplicity measurements performed on non
single-diffractive data in $\overline{p}p$ collisions and described in
Section~\ref{sec:data} are used to tune the main multiple interaction
parameter in \Py, $P_{T_{min}}$. We generate non
single-diffractive events.  At each value of $\sqrt{s}$, the
$P_{T_{min}}$ parameter is adjusted to reproduce the average
multiplicity measured in the data.  The uncertainty on the tuned value
of $P_{T_{min}}$ reflects the uncertainty on the data.  However, the
tuned parameters depend on other aspects of the
\Py\ simulation: in particular the effects of various choices for
the multiple interaction model and the parton distribution 
functions are investigated. For simplicity, the results of these studies are
shown only for some representative settings:
\begin{itemize}
\item
 as an example of pre-HERA parton
 distribution functions we consider the CTEQ2L~\cite{bprod:Tung:1994ua}
 set used by default in
 \Py\ versions 5.7, but recently retracted by their authors;
\item
 as an example of
 post-HERA parton distribution functions we consider the
 CTEQ4L~\cite{bprod:cteq4}
 and GRV94L~\cite{bprod:grv94l} sets, 
 the latter being the new default in \Py\ versions 6.1.
\end{itemize}
This study is restricted to Models 1 and 3 for multiple interactions
(see Section~\ref{sec:mult_int}).

\begin{figure}[htb]
$ 
\begin{array}{cc}
\mbox{\includegraphics[width=0.5\textwidth,clip]{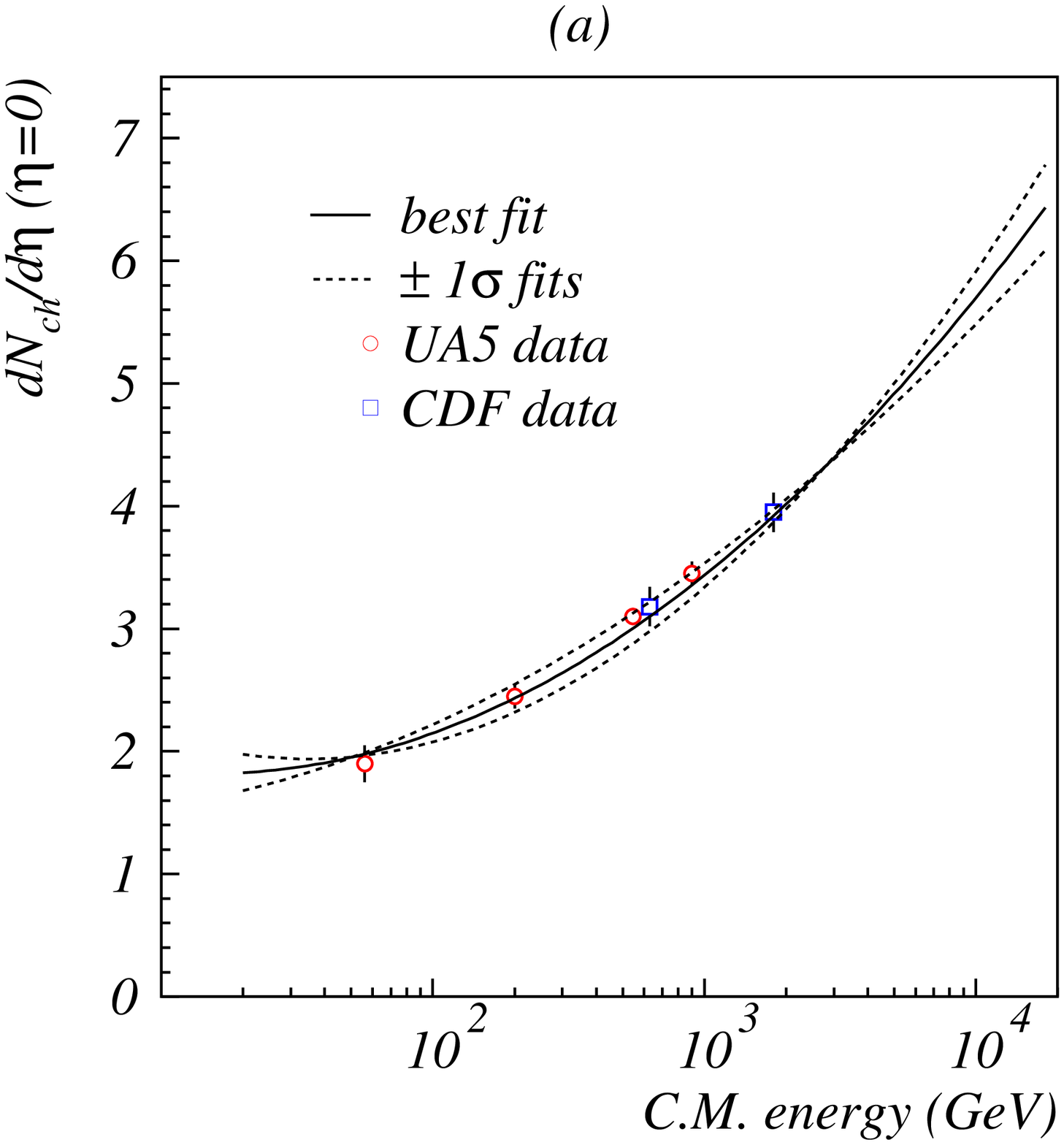}}
& \mbox{\includegraphics[width=0.5\textwidth,clip]{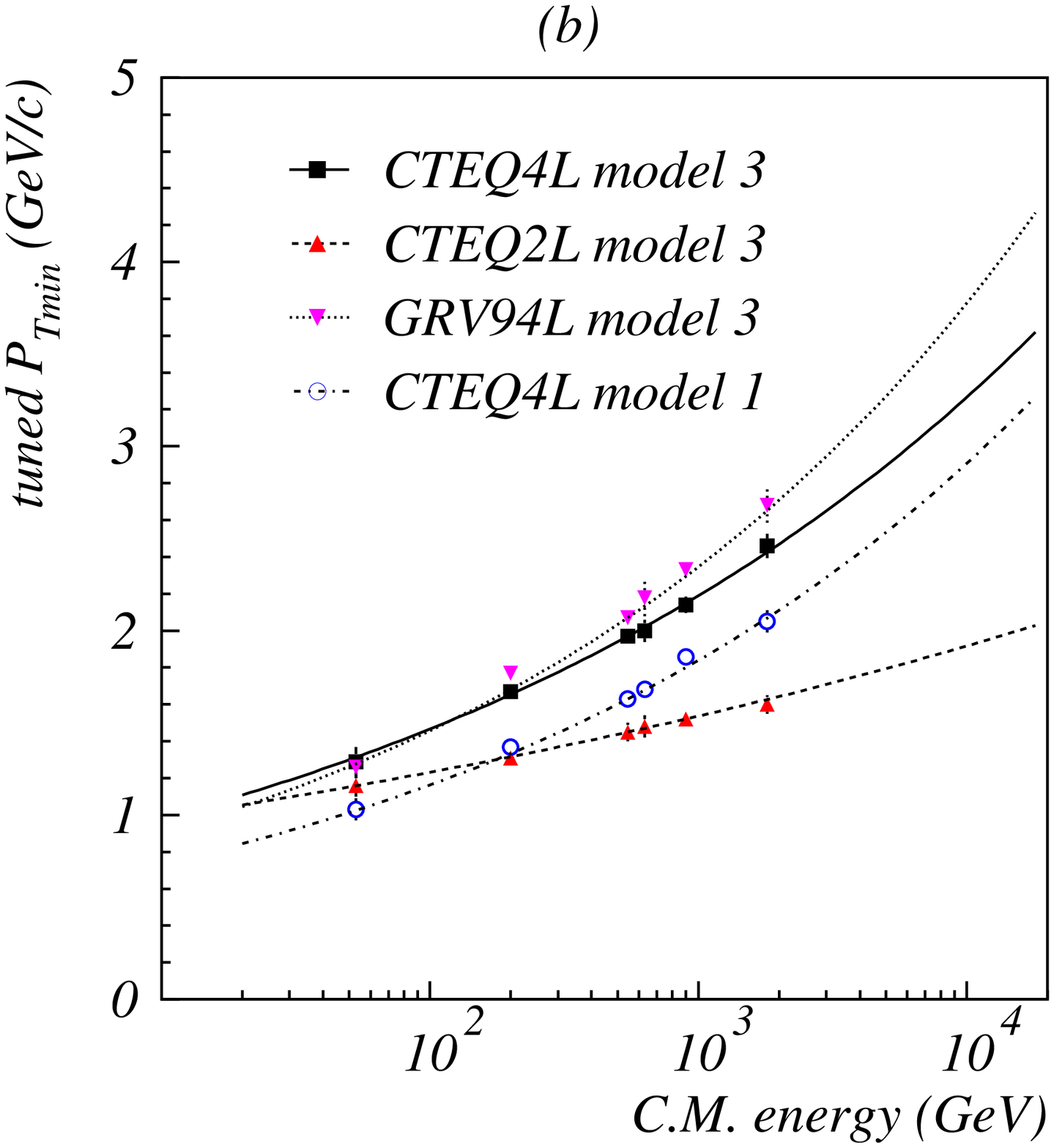}}
\end{array}
$
\caption{\label{fig:average_data} a) $dN_{ch}/d\eta$ at $\eta=0$
as a function of $\sqrt{s}$. The solid curve represent a phenomenological fit 
performed by CDF~\cite{bprod:average_cdf} with the formula
${dN_{ch}}/{d\eta}(s) = (0.023 \pm 0.008) \ln^2(s) - (0.25 \pm 0.19) \ln(s) +
(2.5 \pm 1.0)$.
The two dashed curves represent the 1 sigma variations of the fitted
parameters. \newline b) $\sqrt{s}$-dependence of the $P_{T_{min}}$
parameter for various parton distribution functions and multiple
interactions models.  The points with error bars are the results of
the tuning procedure on the data. The curves are the results of the
fits through the points assuming the functional form of
Equation~\ref{eq:ptcut_funct} and are characterized by the parameters
given in Table~\ref{tab:ptcut_fit}.}
\end{figure}

The results of the tuning procedure are shown in
fig.~\ref{fig:average_data}b: in each case, $P_{T_{min}}$ appears to
be monotonically increasing as a function of $\sqrt{s}$.  This
dependence is much more pronounced for the post-HERA parton
distribution functions regardless of the choice for the multiple
interactions model.

It was shown in Reference~\cite{bprod:fermi_workshop} that the
post-HERA parton distribution functions imply an energy-dependent
$P_T$ cut-off.  This is heuristically motivated by the Lipatov-like
dependence of the gluonic parton distribution function in the
small-$x$ limit:
\begin{equation}
xg(x,Q^2) \rightarrow {\rm constant} \times x^{-\epsilon} ~~  
{\rm for} ~ x \rightarrow 0
\label{lipatov}
\end{equation}
with $\epsilon \simeq 0.08$, while the pre-HERA parton distribution
functions give a reduced charge screening effect and consequently a
less sensitive running of the $P_T$ cut-off. This is heuristically
motivated by the Regge-like dependence of the gluonic parton
distribution function in the small-$x$ limit:
\begin{equation}
xg(x,Q^2) \rightarrow {\rm constant} ~~  {\rm for} ~ x \rightarrow 0\;.
\label{regge}
\end{equation}
In order to extrapolate $P_{T_{min}}$ at LHC energy, one needs to find
a reasonable function to fit the tuned $P_{T_{min}}$ values as a
function of $\sqrt{s}$ for the different parton distribution functions
and multiple interactions models; a four degree of freedom fit is
performed using the following exponential form, inspired by the recent
implementations added in \Py\ since
version~6.120~\cite{bprod:Pythia6}:
\begin{equation}
\displaystyle
P_{T_{min}}(\sqrt{s}) ~ = ~ P_{T_{min}}^{\rm LHC}
\left(\frac{\sqrt{s}}{14~{\rm TeV}}\right)^{2\epsilon}\;.
\label{eq:ptcut_funct}
\end{equation}
The fitted functions are superimposed on fig.~\ref{fig:average_data}b
and the results obtained for the fitted parameters $\epsilon$ and
$P_{T_{min}}^{\rm LHC}$ are given in Table~\ref{tab:ptcut_fit}.
This quantitative analysis demonstrates that the power law expressed 
in Equation~\ref{eq:ptcut_funct} holds for values of $\epsilon$ between 
$\simeq 0.08$ and $\simeq 0.10$ if post-HERA parton distribution functions 
are used, and for somewhat smaller values of $\epsilon$ ($\simeq 0.05$) 
for the pre-HERA parton distribution functions.
\begin{table}
\begin{center}
\caption{\label{tab:ptcut_fit} Results of the fits describing the exponential
running of the \Py\ multiple interactions parameter $P_{T_{min}}$ for 
different
parton distribution functions and multiple interactions models.}
\vskip 0.3cm
\begin{tabular}{|c|c|c|c|}
\hline\hline
Mult.\ int.\ model & PDF & $ P_{T_{min}}^{\rm LHC}$ (GeV/c) & $\epsilon$ \\
\hline\hline
 3  & CTEQ2L  & $1.99 \pm 0.11$  & $0.048 \pm 0.006$  \\
 3  & GRV94L  & $4.06 \pm 0.24$  & $0.103 \pm 0.006$  \\
 3  & CTEQ4L  & $3.47 \pm 0.17$  & $0.087 \pm 0.005$  \\
 1  & CTEQ4L  & $3.12 \pm 0.16$  & $0.100 \pm 0.005$  \\
\hline
\end{tabular}
\end{center}
\end{table}
\subsection{\Py\ predictions at LHC energy}
\label{sec:lhc}

\begin{figure}[htb]
\begin{center}
$ 
\begin{array}{cc} 
\mbox{\includegraphics[width=0.5\textwidth,clip]{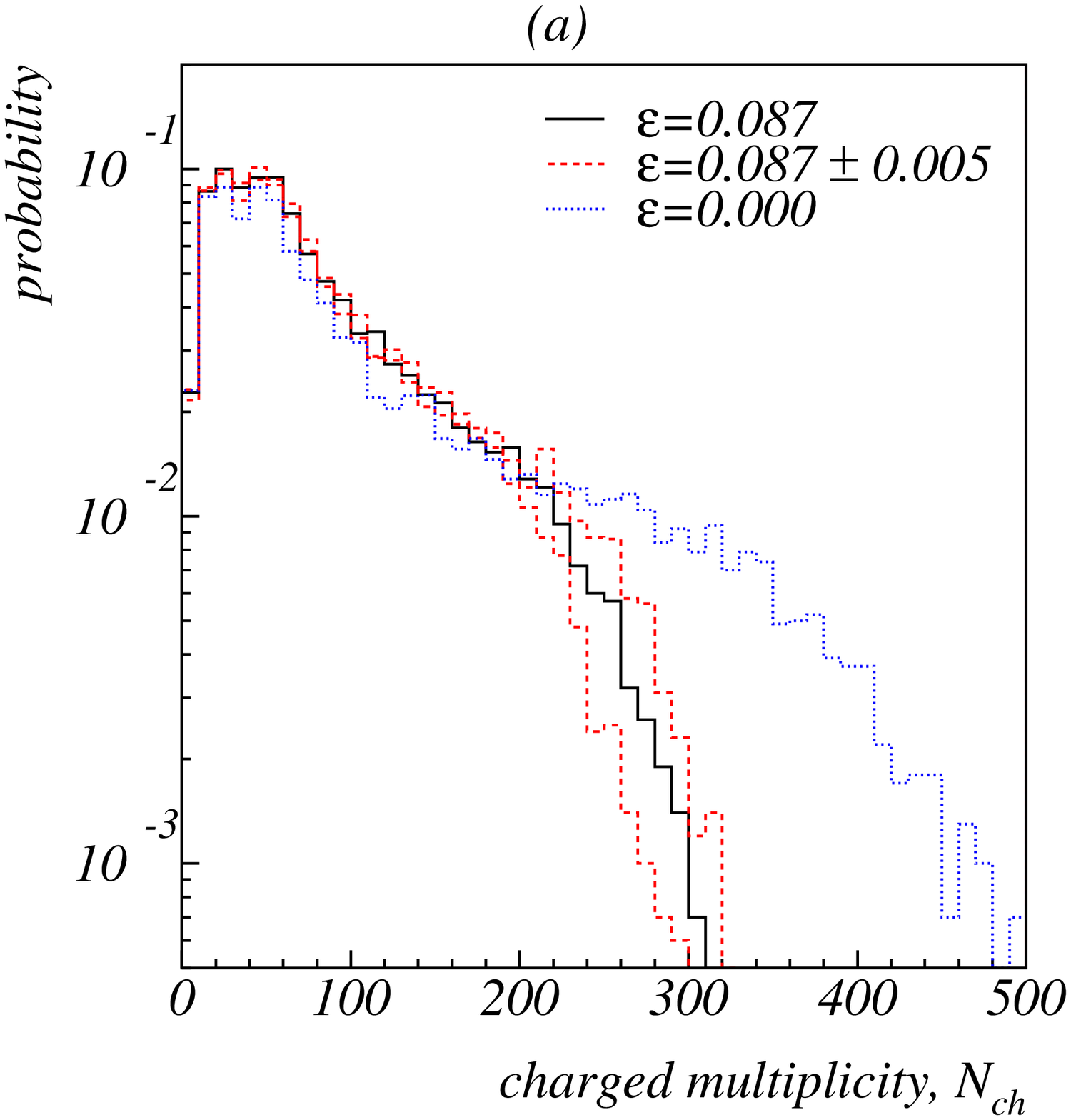}}
& \mbox{\includegraphics[width=0.5\textwidth,clip]{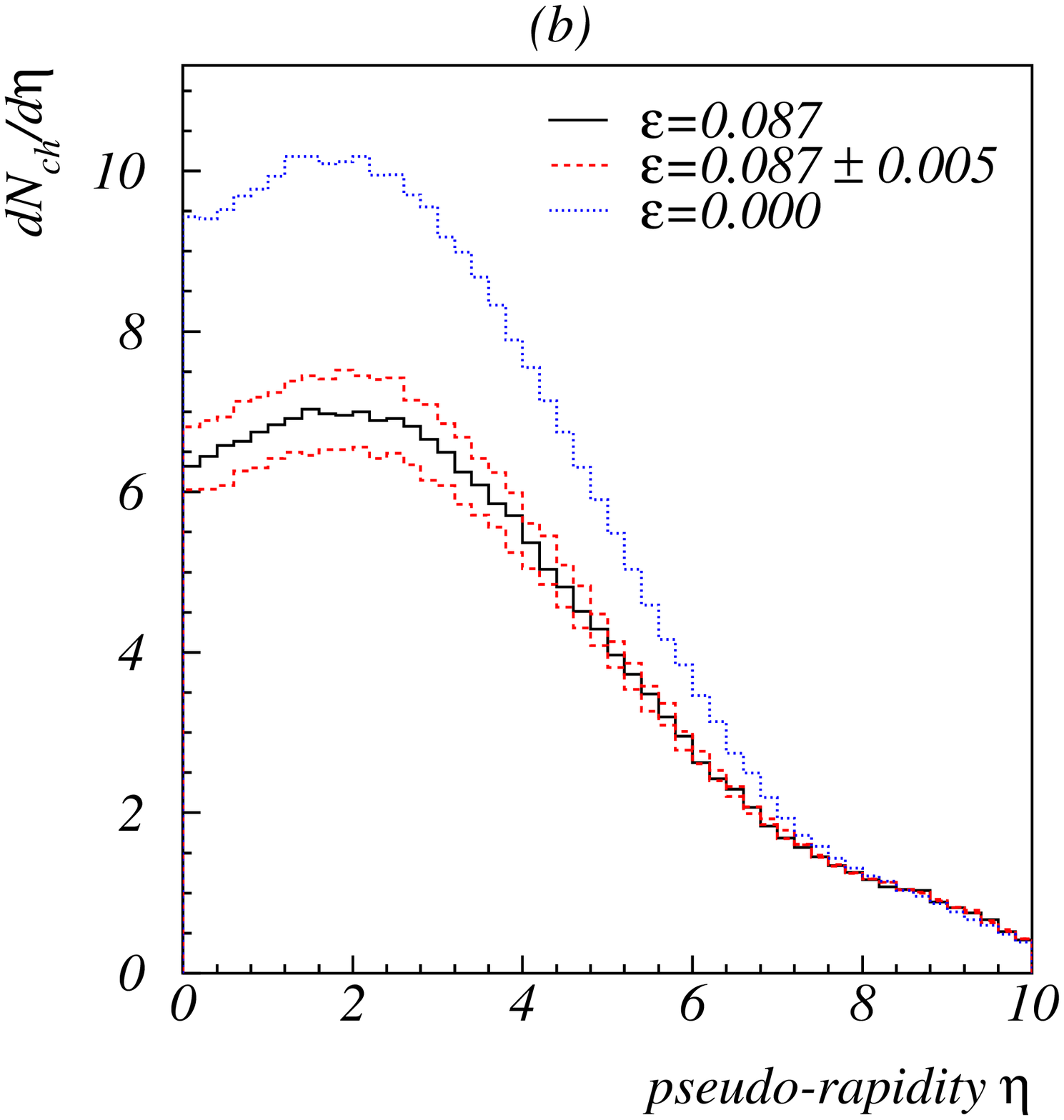}} 
\end{array}
$
\caption{\label{fig:epsilon} 
\Py\ predictions at LHC energy, using the CTEQ4L parton distribution 
functions and Model~3 for multiple interactions with $P_{T_{min}}$
given by Equation~\ref{eq:ptcut_funct2}: the solid and dashed
distributions correspond to the central value and $\pm 1\,\sigma$
uncertainties of the fitted value $\epsilon = 0.087\pm 0.005$. The
dotted histogram is obtained with $\epsilon=0$, i.e.\ using the
$P_{T_{min}}$ value tuned on Tevatron data and ignoring its energy
dependence.  a) Charged track multiplicity in the entire $4\pi$ solid
angle.  b) Average charged track multiplicity per unit pseudorapidity,
as a function of pseudorapidity.}
\end{center}
\end{figure}

Figures~\ref{fig:epsilon}a and b show multiplicity and pseudorapidity distributions 
for charged particles predicted by \Py\ at LHC with CTEQ4L and 
Model~3 for multiple interactions. 
The value of $P_{T_{min}}$ at LHC is obtained by an extrapolation
\begin{equation}
\displaystyle
 P_{T_{min}}^{\rm LHC} ~ = ~ P_{T_{min}}^{\rm Tevatron}
\left(\frac{14~{\rm TeV}}{1.8~{\rm TeV}}\right)^{2\epsilon}
\label{eq:ptcut_funct2}
\end{equation}
where $P_{T_{min}}^{\rm Tevatron}$ is the $P_{T_{min}}$ value tuned at
the Tevatron energy of $1.8~{\rm TeV}$. For the parameter $\epsilon$,
we adopt the results given in Table~\ref{tab:ptcut_fit}.  It is
important to note that the predictions $dN_{ch}/d\eta=6.30 \pm 0.42$
(for $\epsilon=0.087 \mp 0.005$) at $\eta=0$ are consistent with the
phenomenological fit displayed in fig.~\ref{fig:average_data}a
($dN_{ch}/d\eta=6.11 \pm 0.29$).

In order to demonstrate the importance of the correct $P_{T_{min}}$
extrapolation, figs.~\ref{fig:epsilon}a and b also show results
obtained by assuming $P_{T_{min}}^{\rm LHC} = P_{T_{min}}^{\rm
Tevatron}$, i.e.\ $\epsilon = 0$ not supported by the data as
demonstrated in Section~\ref{sec:tuning}. The multiplicity
distribution has a tail at high multiplicities and $dN_{ch}/d\eta$ at
$\eta=0$ is not consistent with that obtained from the
phenomenological fit.

\begin{figure}[htb]
\begin{center}
$ 
\begin{array}{cc} 
\mbox{\includegraphics[width=0.5\textwidth,clip]{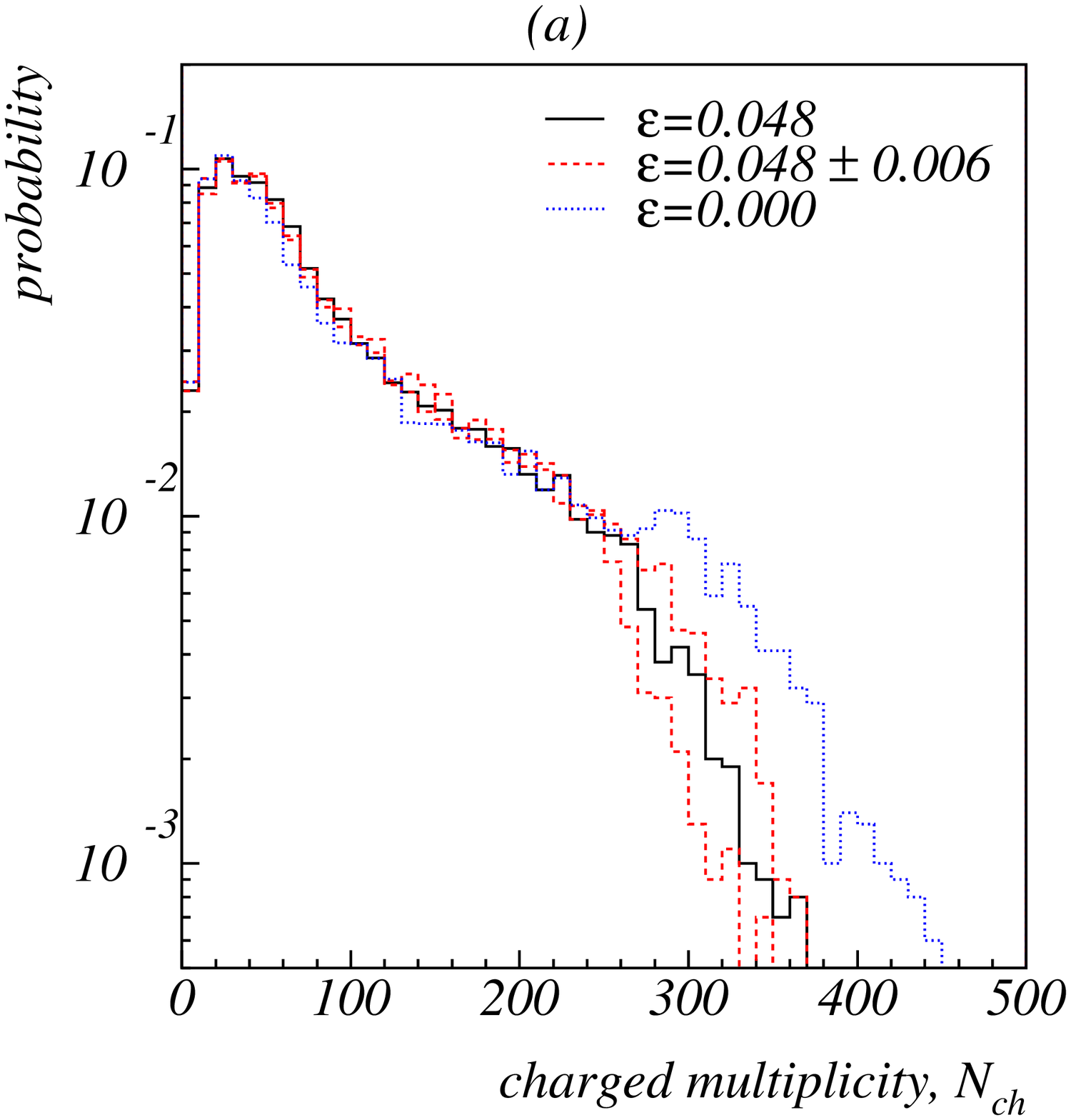}}
& \mbox{\includegraphics[width=0.5\textwidth,clip]{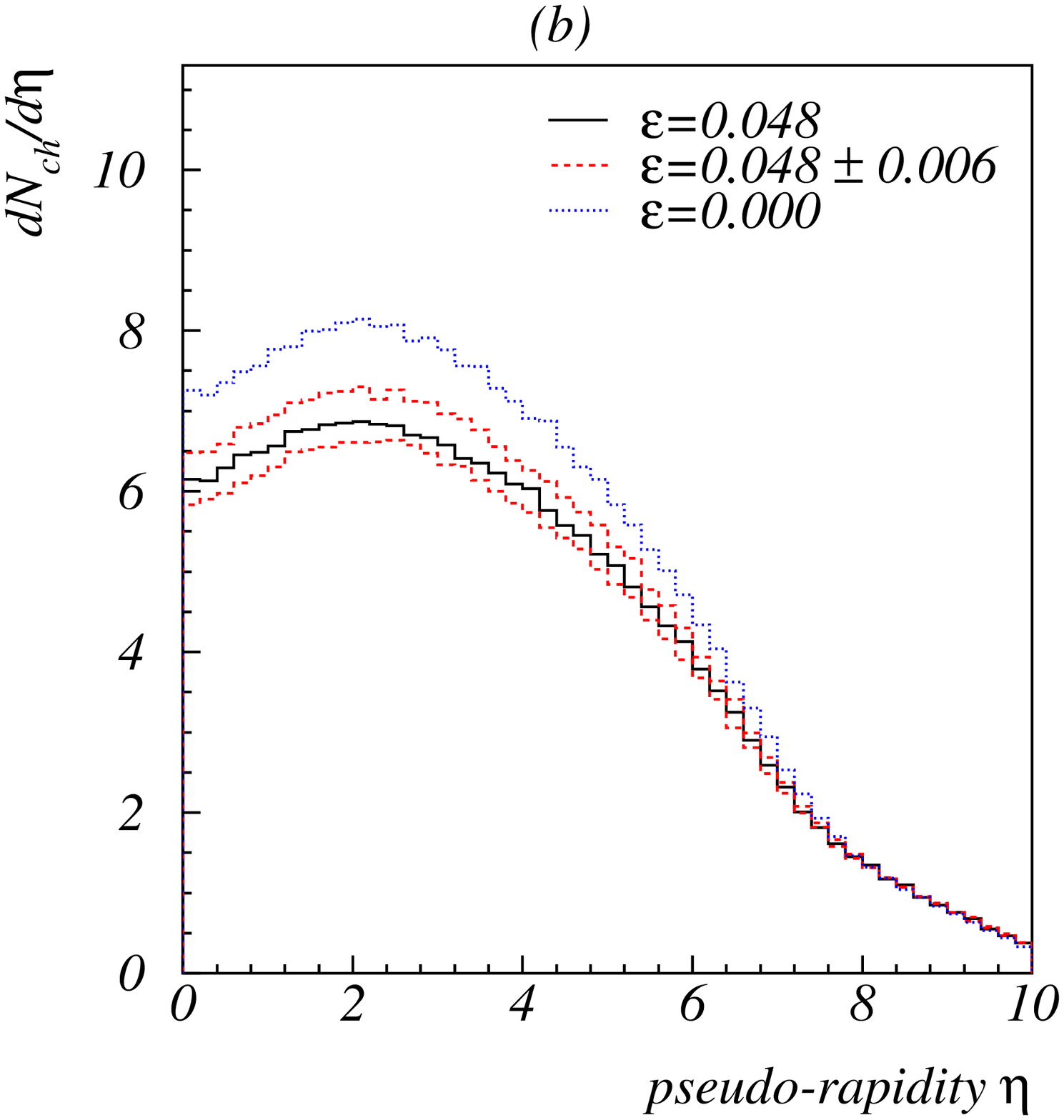}} 
\end{array}
$
\caption{\label{fig:epsilon_cteq2l}
\Py\ predictions at LHC energy, using the CTEQ2L parton distribution 
functions and Model~ 3 for multiple interactions with $P_{T_{min}}$
given by Equation~\ref{eq:ptcut_funct2}: the solid and dashed
distributions correspond to the central value and $\pm 1\,\sigma$
uncertainties of the fitted value $\epsilon = 0.048\pm 0.006$. The
dotted histogram is obtained with $\epsilon=0$, i.e.\ using the
$P_{T_{min}}$ value tuned on Tevatron data and ignoring its energy
dependence.  a) Charged track multiplicity in the entire $4\pi$ solid
angle.  b) Average charged track multiplicity per unit pseudorapidity,
as a function of pseudorapidity.}
\end{center}
\end{figure}

Figure~\ref{fig:epsilon_cteq2l} shows the same distributions as
fig.~\ref{fig:epsilon}, but for the CTEQ2L parton distribution
functions. It is interesting to note that, once the extrapolation of
$P_{T_{min}}$ is properly done, there is no large difference between
the multiplicity and pseudorapidity distributions obtained with
different structure functions.

\begin{figure}[htb]
\begin{center}
$ 
\begin{array}{cc} 
\mbox{\includegraphics[width=0.5\textwidth,clip]{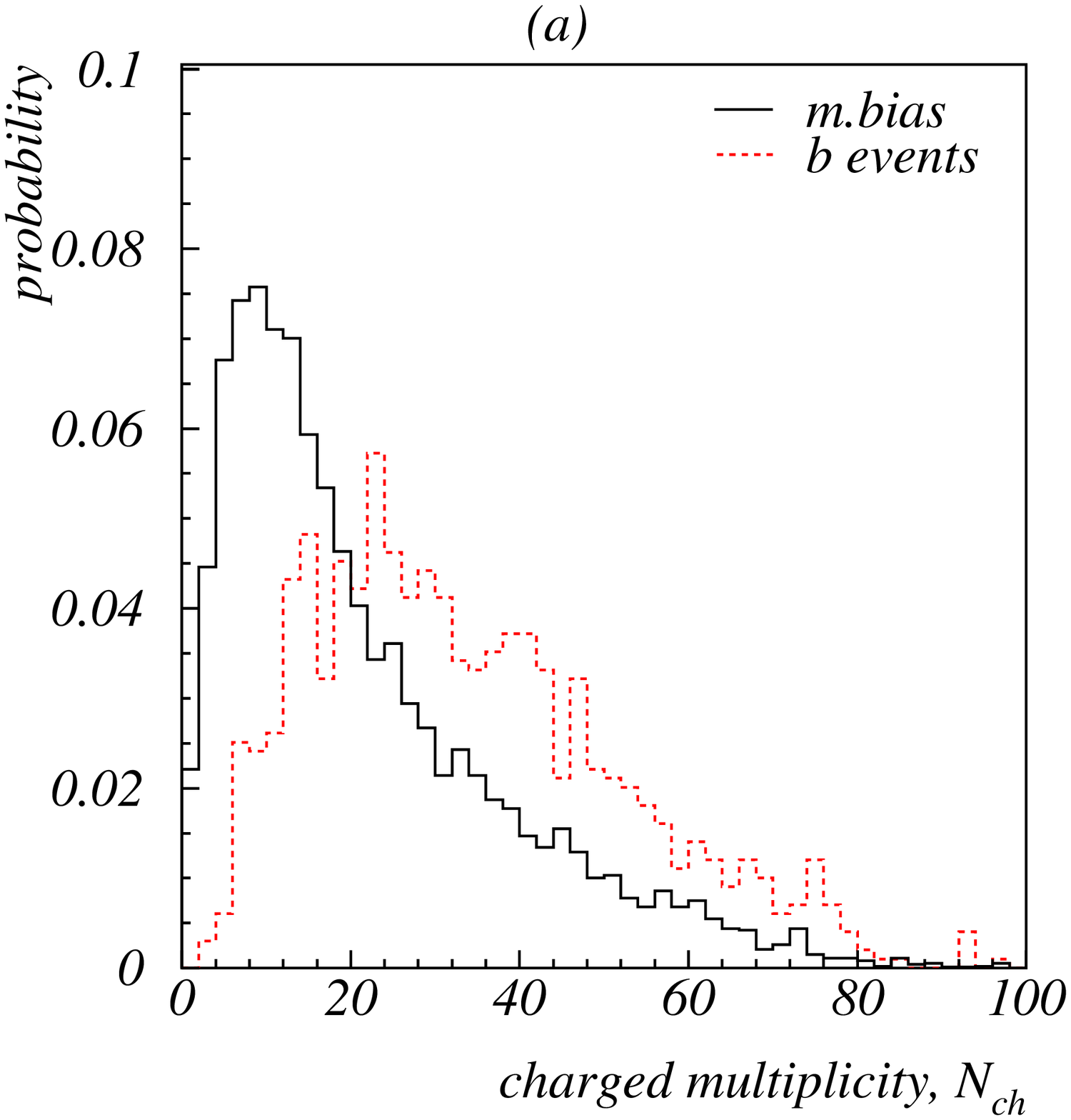}}
& 
\mbox{\includegraphics[width=0.5\textwidth,clip]{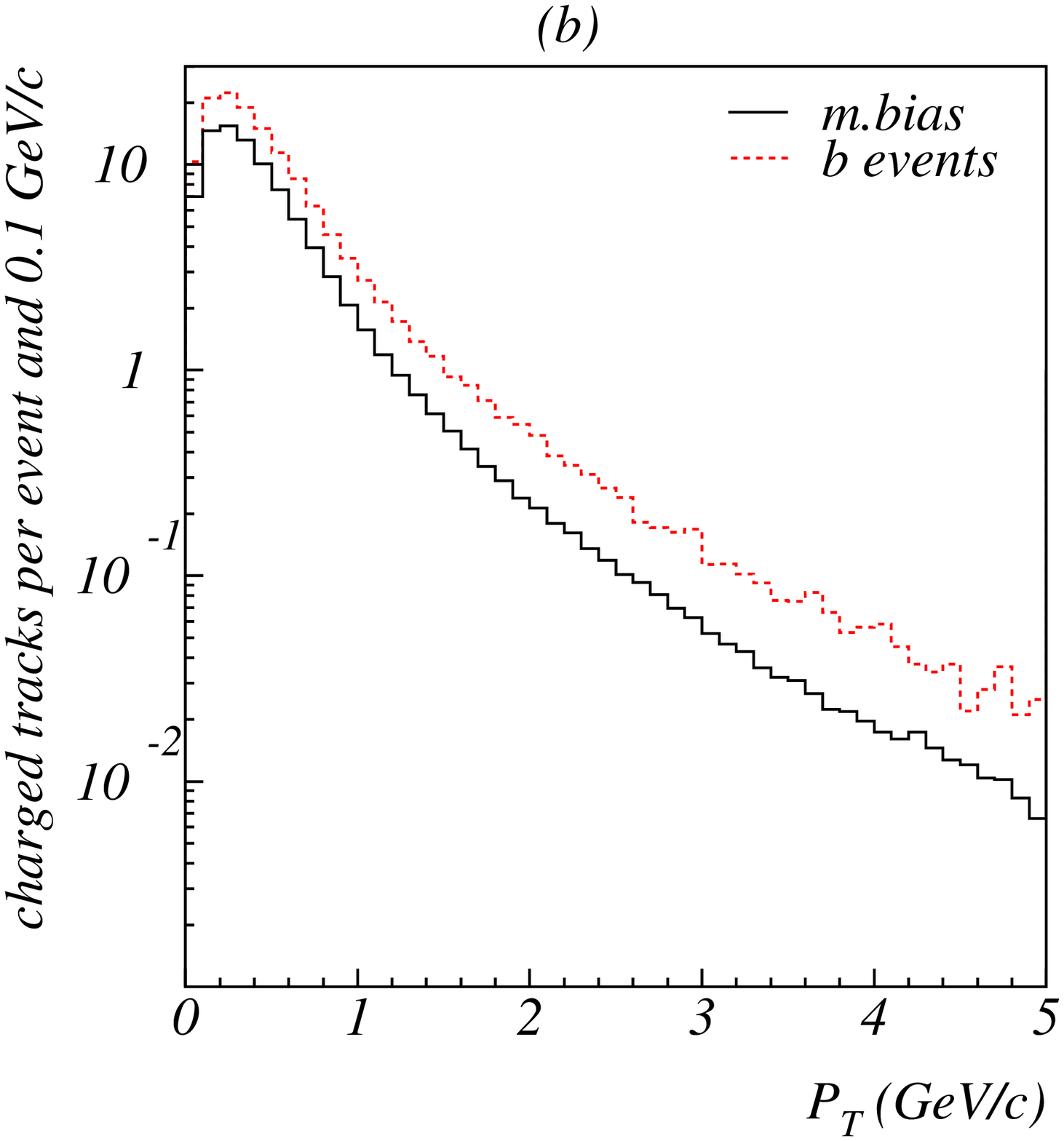}}
\\
\mbox{\includegraphics[width=0.5\textwidth,clip]{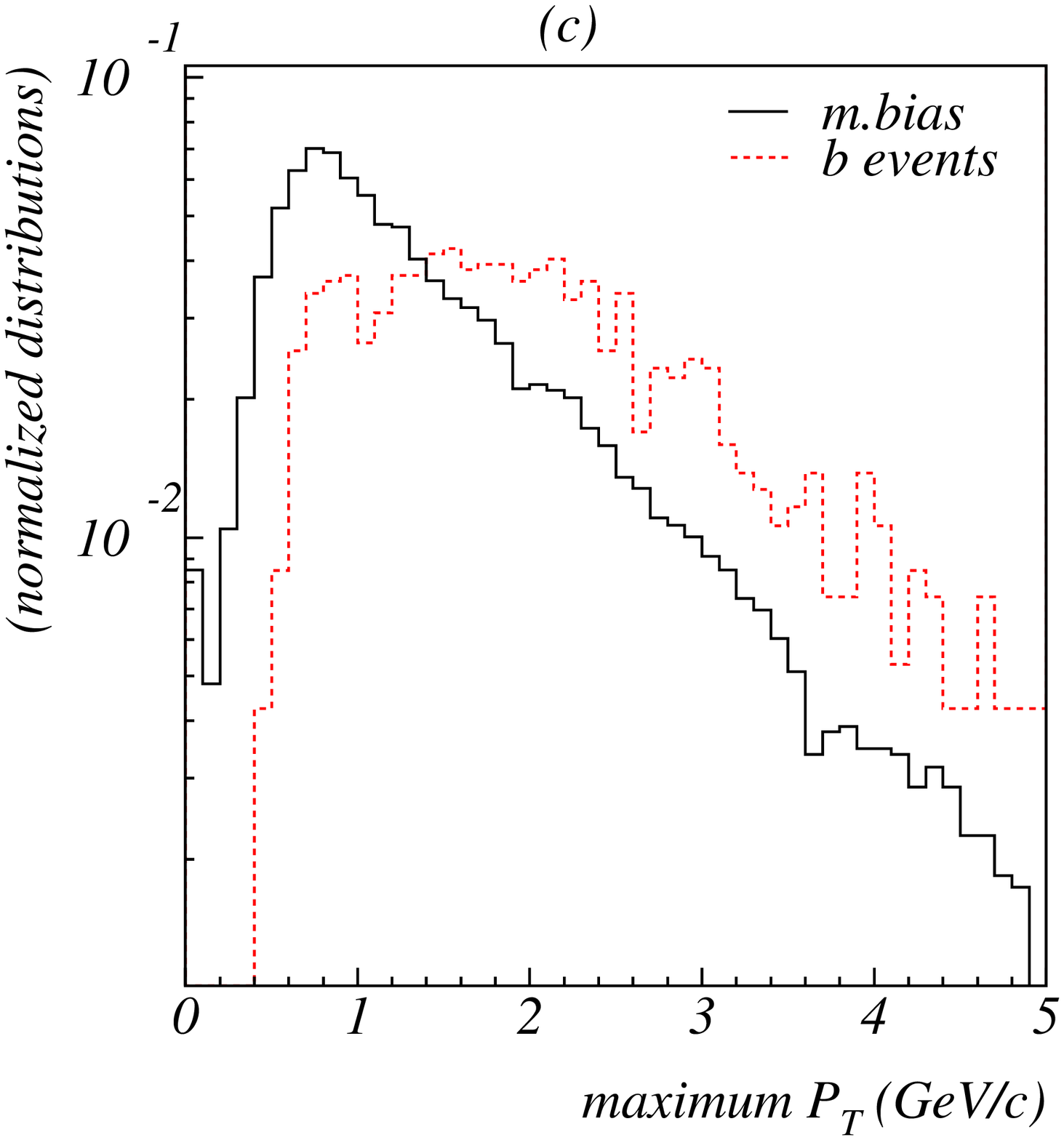}}
& 
\mbox{\includegraphics[width=0.5\textwidth,clip]{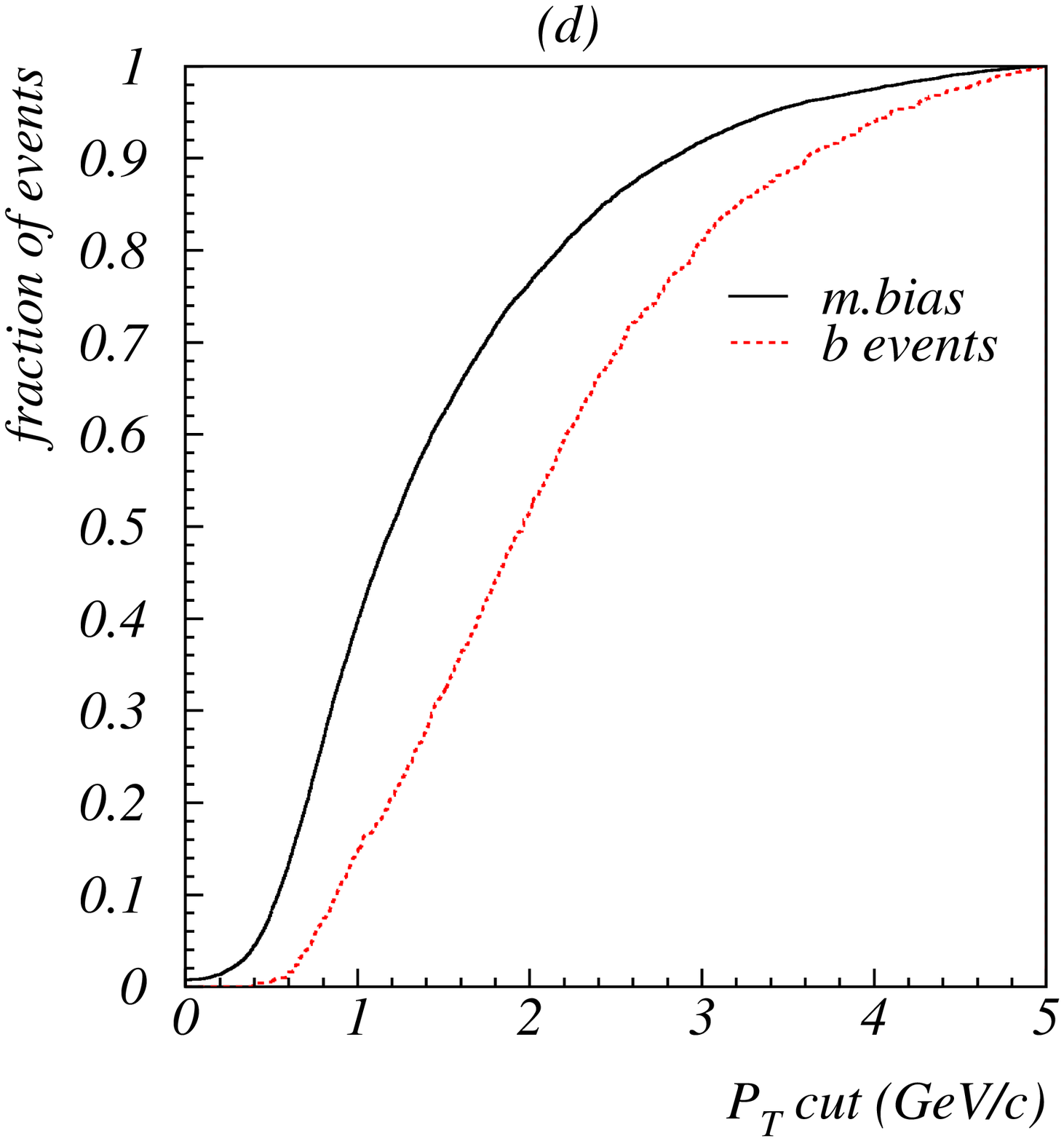}}
\\
\end{array}
$
\caption{\label{fig:summary3} 
\Py\ predictions 
for charged tracks in the LHCb acceptance using the CTEQ4L parton
distribution functions and Model~3 for multiple interactions with
proper $P_{T_{min}}$; the normalized predictions for $b\overline{b}$
events (dashed curve) and minimum bias events (solid curve) are
superimposed.  a) Charged track multiplicity distribution.  b)
Transverse momentum distribution of charged tracks.  c) Distribution
of the maximum transverse momentum of charged tracks.  d) Rejection
efficiency as a function of a traverse momentum cut (an event is
rejected if all the charged tracks have a transverse momentum below
the cut).}
\end{center}
\end{figure}

Figure~\ref{fig:summary3}a-d compare the \Py\ predictions for the
multiplicity and transverse momentum distributions in the LHCb angular
acceptance ($1.8<\eta<4.9$) between minimum bias and $b\overline{b}$
events\footnote{The $b\overline{b}$ events are selected among the
minimum bias events.}. These predictions are obtained with CTEQ4L,
multiple interactions Model~3 and the proper $P_{T_{min}}$
extrapolation. They show clear differences between minimum bias and
$b\overline{b}$ events, in particular higher average multiplicity and
transverse momentum for $b\overline{b}$ events.

\begin{figure}[htb]
\begin{center}
$ 
\begin{array}{cc} 
\mbox{\includegraphics[width=0.5\textwidth,clip]{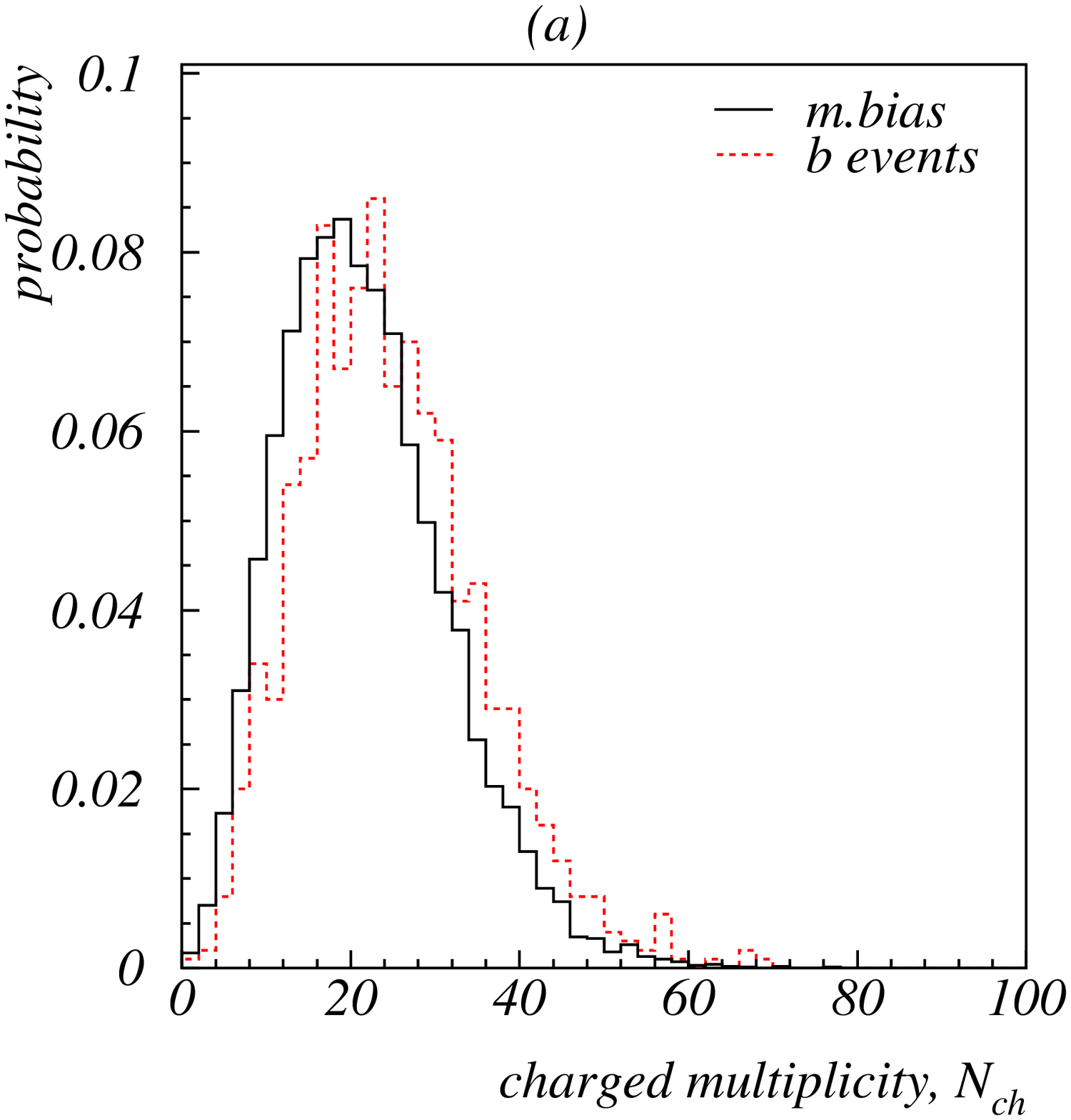}}
& 
\mbox{\includegraphics[width=0.5\textwidth,clip]{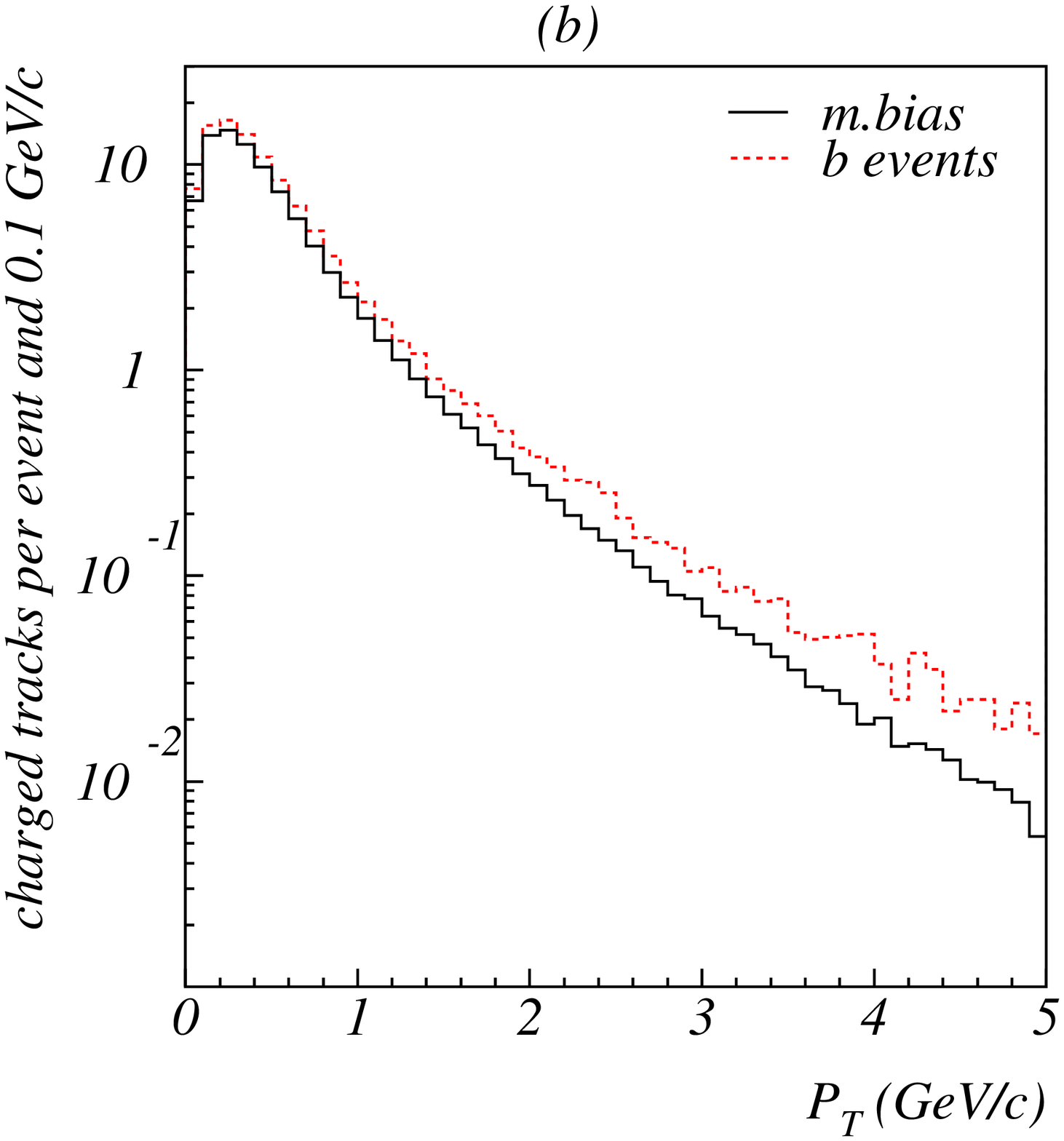}}
\end{array}
$
\caption{\label{fig:summary1} \Py\ predictions 
for charged tracks in the LHCb acceptance
using the CTEQ4L parton 
distribution functions and Model~1 for multiple interactions
with proper $P_{T_{min}}$;
the normalized predictions for $b\overline{b}$ events (dashed curve) 
and minimum bias events (solid curve) are superimposed.
a) Charged track multiplicity distribution. 
b) Transverse momentum distribution of charged tracks.
}
\end{center}
\end{figure}

\Py\
predictions with multiple interactions Model~1 for the multiplicity
and transverse momentum distributions are shown in
fig.~\ref{fig:summary1} for minimum bias and $b\overline{b}$
events. Compared to the results obtained with Model~3, less
significant differences between minimum bias and $b\overline{b}$
events is observed.  In Section~\ref{sec:mult_int} we have stressed
that a varying impact parameter model for multiple interactions
(i.e. Model~3) is needed to describe the charged track multiplicity in
hadron-hadron interactions.  There are arguments in favour of adopting
a multiple interactions model with varying impact parameter to
describe the heavy flavour production at hadron
colliders~\cite{bprod:Abe:1995dv}, though there are no experimental
data at low transverse momentum and high pseudorapidity (i.e.\ in the
LHCb acceptance region).  A more detailed discussion on the effect of
multiparton interactions in $b\overline{b}$ events at LHCb can be
found in Reference~\cite{bprod:galumian}.
\subsection{Conclusions}
\label{sec:conc}
Comparisons between \Py\ and experimental data demonstrate that,
in order to reproduce the charged track multiplicity spectrum, a
varying impact parameter model has to be adopted.

The varying impact parameter models predict 
sensitive differences in multiplicity and $P_T$ distribution
between light and heavy flavour events.

The running of the $P_T$ cut-off parameter in \Py\ multiple
interactions is mandatory.  Predictions made at LHC energy with a
fixed $P_T$ cut-off tuned at lower energies overestimate the
multiplicity observables.  Taking into account the running of the
$P_T$ parameter is even more important if post-HERA parton
distribution functions are used.
\newpage

\end{document}